\documentclass[11pt,twoside,a4paper,cmspaper,final,collab]{cms-tdr}
\def\svnVersion{4ec50a4}\def\svnDate{2026/11/11}\def\cmsCernNoTag{CERN-EP-2026-243}\def\cmsCernDate{\today}\def\cmsMessage{Published in the Journal of Instrumentation as \href{http://dx.doi.org/10.1088/1748-0221/21/11/P11012}{\doi{10.1088/1748-0221/21/11/P11012}.}}
\begin{document}\cmsNoteHeader{TAU-25-001}

\newcommand{\PNfix}{\hspace{-.04em}}
\newcommand{\Wjets}{\ensuremath{\PW+\text{jets}}\xspace}
\newcommand{\Zjets}{\ensuremath{\PZ/\PGgst+\text{jets}}\xspace}
\newcommand{\CP}{\ensuremath{C\!P}\xspace}
\newcommand{\mutau}{{\ensuremath{\PGt_{\PGm}\PNfix\tauh}}\xspace}
\newcommand{\tautau}{{\ensuremath{\tauh\tauh}}\xspace}
\newcommand{\ltau}{{\ensuremath{\PGt_\Pell\PNfix\tauh}}\xspace}
\newcommand{\FF}{\ensuremath{F_{\mathrm{F}}}\xspace}
\newcommand{\FFs}{\ensuremath{F_{\mathrm{F}}}\text{s}\xspace}
\newcommand{\jtf}{\ensuremath{\text{jet}\to\tauh}\xspace}
\newcommand{\pttauh}{\ensuremath{\pt^{\tauh}}\xspace}
\newcommand{\ptmu}{\ensuremath{\pt^{\PGm}}\xspace}
\newcommand{\mvis}{\ensuremath{m_{\text{vis}}}\xspace}
\newlength\cmsFigWidth\setlength\cmsFigWidth{0.45\textwidth}
\newlength\cmsTabSkip\setlength{\cmsTabSkip}{1ex}

\cmsNoteHeader{TAU-25-001}
\title{A machine-learning method for the estimation of the background from jets misidentified as hadronic tau lepton decays}

\date{\today}

\abstract{Backgrounds arising from quark- and gluon-initiated jets misidentified as hadronically decaying tau leptons constitute a major experimental challenge for analyses involving tau leptons. The MUltivariate Fake-Factor INference (MUFFIN) approach, a machine-learning-based method for estimating this background directly from observed events, is presented. Boosted decision tree reweighting is used to learn multidimensional, per-object correction factors directly from observed data, thereby modelling the jet to hadronic tau lepton misidentification probability across the kinematic and topological phase space without relying on low-dimensional parametrisations. Performance studies are presented using proton-proton collision data collected by the CMS experiment at the CERN LHC at $\sqrt{s}=13.6\TeV$, corresponding to an integrated luminosity of 62.4\fbinv. Compared to low-dimensional parametrisation methods, improved modelling of key kinematic observables is achieved, particularly in regions of high \pt. The improved modelling leads to a reduction of up to 70\% in the systematic uncertainty associated with jets being misidentified as hadronically decaying tau leptons. The MUFFIN method provides a robust and scalable method for background estimation in precision measurements and searches for beyond-the-standard-model physics involving tau lepton final states.}

\hypersetup{
pdfauthor={CMS Collaboration},
pdftitle={A machine-learning method for the estimation of the background from jets misidentified as hadronic tau lepton decays},
pdfsubject={CMS},
pdfkeywords={CMS, taus, fake factors, machine learning, BDT reweighting}}

\maketitle 
\section{Introduction}\label{sec:introduction}

In the standard model (SM) of particle physics, tau (\PGt) leptons are the heaviest leptons and play a crucial role in measurements involving the Higgs boson (\PH), as well as in searches for physics beyond the SM (BSM). 
Measurements involving \PGt leptons provide a sensitive probe of Higgs boson properties, including its couplings to fermions, its charge-parity (\CP) properties, and its self-coupling via the Higgs boson pair production mechanism~\cite{Sirunyan:2017khh,ATLAS:2018ynr,CMS:2022hgz,cp_paper,CMS:2022kdi,CMS:2022uox,ATLAS:2022akr,ATLAS:2024pov,ATLAS:2024wfv,ATLAS:2025bts,CMS:2024nmz}. 
The \PGt leptons are also key signatures in many BSM scenarios, including models with extended Higgs sectors, supersymmetry, and heavy gauge bosons~\cite{HIG-21-001, ATLAS:2020zms, CMS:2023qdw, ATLAS:2022wcu, CMS:2020wzx, CMS:2026wpe, ATLAS:2019gti, ATLAS:2020tre}. 
The \PGt leptons can decay either leptonically or hadronically, with the latter mode, denoted as \tauh, being the most abundant and occurring in approximately 65\% of the \PGt lepton decays. 
Since \tauh decays consist of charged and neutral hadrons and are accompanied by neutrinos, they represent a complex experimental signature that requires dedicated reconstruction and identification techniques. 
Despite significant advances in \PGt lepton reconstruction and identification at the CERN LHC, analyses involving \tauh candidates remain challenging due to substantial backgrounds from quark- and gluon-initiated jets that can mimic \tauh decays (\jtf). 
An accurate estimation of these backgrounds is therefore essential for precision measurements and searches involving \PGt leptons.

These backgrounds, primarily originating from quantum chromodynamics (QCD) multijet production, \PW boson production in association with jets (\Wjets), and top quark pair production (\ttbar) processes, can mimic the signature of \tauh decays and thus complicate the extraction of signal processes. Additional processes such as \Zjets or diboson production can also contribute depending on the final state.  
The dominant background from \jtf misidentification is often estimated using data-driven methods, among which the ``Fake Factor'' (\FF) method is widely used~\cite{HIG-21-001, ATLAS_UFF, FF_ZTT, FF_MSSM, cp_paper}. 
This method relies on the definition of determination regions (DRs), enriched in misidentified \tauh candidates, and application regions (ARs), which are kinematically similar to the signal regions (SRs) but differ in the \tauh identification requirements. 
Transfer factors (TFs) are measured in the DRs and applied in the ARs to estimate the misidentified background in the SRs.

While the \FF method is robust and effective, it relies on a number of limiting assumptions. 
In particular, \FFs are typically parametrised using fixed binning in a limited number of kinematic variables, which constrains the achievable parametrisation and leads to both statistical limitations from sparsely populated bins and systematic uncertainties associated with missing variable dependencies.
The method assumes that the misidentification probability factorises between these variables and a separate set of variables used to define the analysis categories, implicitly neglecting potential correlations with other event-level or object-level quantities. 
As analyses increasingly exploit multivariate information, these assumptions can become limiting factors in achieving an optimal background modelling accuracy. 
Residual mismodelling may arise in regions where the assumptions of the \FF method do not hold. 
In such cases, additional non-closure (\ie discrepancies between predicted and observed background in control regions) or extrapolation corrections are often required, together with associated systematic uncertainties. 
Moreover, the need to define multiple parametrisations and correction steps can lead to complex workflows that are difficult to maintain, reproduce, and validate consistently. 
This method can lead to reduced precision and, in some cases, limit the sensitivity of analyses involving \tauh candidates.

Advances in machine learning (ML) techniques provide an opportunity to generalise the \FF paradigm by learning multidimensional reweighting functions directly from observed data and simulation. 
In this paper, MUltivariate Fake-Factor INference (MUFFIN), an ML-based alternative to the \FF method based on boosted decision tree (BDT) reweighting~\cite{BDTReweighting}, is presented. 
The MUFFIN method replaces explicit low-dimensional parametrisations with per-object event weights that capture correlations between multiple kinematic and event-level variables, reducing the reliance on factorisation assumptions. 
At the same time, this novel procedure retains the fully data-driven nature of the background estimation strategy and provides a more unified and reproducible background estimation workflow, developed and demonstrated here using proton-proton (pp) collision data collected with the CMS detector.
It is applicable to all analysis channels involving \tauh final states, and is demonstrated explicitly in the fully hadronic \tautau and in the semileptonic \mutau channel, where $\PGt_{\PGm}$ denotes a \PGt lepton decaying to a muon.
Although the method is demonstrated using the \textsc{DeepTau}~\cite{DeepTau_v2p1,CMS:2025kgf} algorithm for \tauh identification, the approach does not depend on the specific choice of identification algorithm and can be readily adapted as identification methods evolve.
The \jtf background constitutes a significantly larger fraction of the total background in the \tautau channel than in the \mutau channel. For the inclusive SR selections considered in this paper, defined in Section~\ref{sec:selections}, it accounts for approximately 75\% of the total background in the \tautau channel, compared to approximately 35\% in the \mutau channel. Consequently, improvements in the \jtf background modelling are expected to have a larger impact on the overall background estimation in the \tautau channel, while still providing benefits in the \mutau channel.
Throughout this paper, the \FF method is used as the reference baseline against which the MUFFIN method is compared.

This paper is organised as follows. Section~\ref{sec:detector} describes the CMS detector and the event reconstruction algorithms.
Sections~\ref{sec:samples} and~\ref{sec:event_reconstruction} outline the data and simulated samples used in this analysis, and the event reconstruction process, respectively. Section~\ref{sec:selections} details the event selection criteria applied to define the SRs for the \tautau and \mutau channels.
Section~\ref{sec:bdt_ff} introduces the MUFFIN method and its implementation in the context of the \jtf background estimation.
Section~\ref{sec:performance} compares the performance of the MUFFIN method with that of the \FF method, as implemented in Ref.~\cite{HIG-21-001}. This is followed by a description of the systematic uncertainties associated with the MUFFIN method in Section~\ref{sec:uncertainties}.
Section~\ref{sec:physics_impact} presents the application of the method to events with two \PGt candidates selected with typical SR selections used in analyses targeting Higgs boson decays to \PGt lepton pairs.
Finally, Section~\ref{sec:summary} summarises the findings and discusses potential future applications of the MUFFIN method.

\section{The CMS detector}\label{sec:detector}

The CMS apparatus~\cite{CMS:2008xjf,CMS:2023gfb} is a multipurpose, nearly hermetic detector, designed to trigger on~\cite{CMS:2020cmk,CMS:2016ngn,CMS:2024aqx} and identify~\cite{CMS:2020uim,CMS:2018rym,CMS:2014pgm} electrons, muons, photons, and (charged and neutral) hadrons. Its central feature is a superconducting solenoid of 6\unit{m} internal diameter, providing a magnetic field of 3.8\unit{T}. Within the solenoid volume are a silicon pixel and strip tracker, a lead tungstate crystal electromagnetic calorimeter (ECAL), and a brass and scintillator hadron calorimeter (HCAL), each composed of a barrel and two endcap sections. Forward calorimeters extend the pseudorapidity ($\eta$) coverage of the barrel and endcap detectors. Muons are reconstructed using gas-ionisation detectors embedded in the steel flux-return yoke outside the solenoid.

Events of interest are selected using a two-tiered trigger system. The first level, composed of custom hardware processors, uses information from the calorimeters and muon detectors to select events at a rate of around 100\unit{kHz} within a fixed latency of about 4\mus~\cite{CMS:2020cmk}. The second level, known as the high-level trigger, consists of a farm of processors running a version of the full event reconstruction software optimised for fast processing, and reduces the event rate to a few kHz before data storage~\cite{CMS:2016ngn,CMS:2024aqx}.

A more detailed description of the CMS detector, together with a definition of the coordinate system used and the relevant kinematic variables, can be found in Refs.~\cite{CMS:2008xjf,CMS:2023gfb}.

\section{Data and simulated samples} \label{sec:samples}

The data set analysed in this work was recorded with the CMS detector during 2022--2023 and corresponds to an integrated luminosity of 62.4\fbinv of pp collisions at a centre-of-mass energy of $13.6~\TeV$~\cite{CMS-PAS-LUM-22-001}.

Production of \Zjets and \Wjets events is simulated with \MGvATNLO (Version~2.9.13)~\cite{Alwall:2014hca}. For the \Zjets process, next-to-leading order (NLO) QCD matrix elements with up to two additional partons are employed, with different parton multiplicities merged via the FxFx prescription~\cite{FxFx}. The \Wjets process is modelled at leading order (LO) in QCD, retaining matrix elements with up to four additional partons, combined through the MLM jet-matching scheme~\cite{Alwall:2007fs}. Diboson production is simulated at LO using \PYTHIA (Version~8.306)~\cite{PYTHIA}. The predicted yields for the \Zjets, \Wjets, and diboson processes are normalised to theoretical cross sections computed at next-to-next-to-leading order (NNLO) in QCD with NLO electroweak corrections applied~\cite{Grazzini:2017mhc,Grazzini:2019jkl}.

Top quark pair (\ttbar) and single top quark production, including the $t$-channel and $\PQt\PW$ associated production modes, are modelled at NLO in QCD using the \POWHEG 2.0 generator~\cite{Alioli:2010xd,Alioli:2011as,Re:2010bp,Frederix:2012dh}. The \ttbar and $t$-channel single top quark processes are normalised to cross sections from NNLO QCD calculations~\cite{Czakon:2011xx,Campbell:2020fhf}, while the $\PQt\PW$ process is normalised to an NLO QCD cross section~\cite{Kidonakis:2021vob}.

All simulated samples use the NNPDF3.1 NNLO parton distribution functions~\cite{Ball_2017}. Parton showering, hadronisation, and \PGt lepton decays are modelled by \PYTHIA, with the underlying event tuned to the CP5 parameter set~\cite{Sirunyan_2020_CP5}. The response of the CMS detector is evaluated with a full simulation based on \GEANTfour~\cite{Agostinelli:2002hh}, and all simulated events are reconstructed using the same algorithms as applied to data. The effects of additional pp interactions within the same or nearby bunch crossings (pileup) are included by overlaying minimum bias collisions simulated with \PYTHIA, and the pileup distribution in simulation is reweighted to match that observed in data.

\section{Event reconstruction}\label{sec:event_reconstruction}

Events recorded by the CMS detector are reconstructed using the particle-flow (PF) algorithm~\cite{PF2017}, which combines information from the tracking system, calorimeters, and muon detectors to identify and reconstruct individual particles produced in pp collisions. The PF algorithm classifies reconstructed particles as charged or neutral hadrons, photons, electrons, or muons, and provides a global and consistent description of the event.

The energy of photons is obtained directly from measurements in the ECAL. Electron momenta are reconstructed by combining the momentum measured in the silicon tracker with the energy deposited in the ECAL, including contributions from bremsstrahlung photons associated with the electron track. Muon momenta are determined from the curvature of tracks reconstructed using information from the silicon tracker and the muon system. The energies of charged hadrons are reconstructed from a combination of the track momentum and the associated calorimetric energy deposits, corrected for the calorimeter response to hadronic showers. Neutral hadron energies are obtained from corrected ECAL and HCAL measurements.

For each event, hadronic jets are clustered from these reconstructed particles using the infrared and collinear safe anti-\kt algorithm~\cite{Cacciari:2008gp, Cacciari:2011ma} with a distance parameter of 0.4.
Jet momentum is determined as the vectorial sum of all particle momenta in the jet, and is found from simulation to be, on average, within 5--10\% of the true momentum over the whole \pt spectrum and detector acceptance.
Jet energy corrections are derived from simulation to bring the measured response of jets to that of particle level jets on average.
In situ measurements of the momentum balance in dijet, $\text{photon} + \text{jet}$, $\PZ + \text{jet}$, and multijet events are used to account for any residual differences in the jet energy scale between data and simulation~\cite{CMS:2016lmd}.
The jet energy resolution amounts typically to 15--20\% at 30\GeV and 10\% at 100\GeV~\cite{CMS:2016lmd}.
Additional selection criteria are applied to each jet to remove jets potentially dominated by anomalous contributions from the readout of various detector components or reconstruction failures.
The \textsc{DeepJet} algorithm~\cite{Bols:2020bkb} is used for \PQb jet identification. The algorithm produces a continuous discriminant score for each jet, and a working point is defined by placing a fixed threshold on this score, yielding a binary \PQb jet selection. The medium working point is used, corresponding to an identification efficiency for \PQb jets of approximately 70\% and a misidentification rate for charm, light-flavour quark, and gluon jets of about 1\%~\cite{Bols:2020bkb}. Scale factors derived from data are applied to simulated events to correct for differences in \PQb-tagging efficiency between data and simulation. The \PQb-tagging requirements are used to define dedicated control regions enriched in \ttbar events, as described in Section~\ref{sec:bdt_ff}, and to suppress \ttbar contributions in the SRs by vetoing events containing \PQb-tagged jets.

Interaction vertices are reconstructed by clustering tracks along the beam axis and fitting each cluster with an adaptive vertex fitting algorithm~\cite{CMS:2014pgm}. The primary vertex (PV) is selected as the vertex with the largest value of the squared \pt sum of associated physics objects, constructed from anti-\kt track jets, remaining tracks including identified leptons, and a track-based transverse momentum imbalance~\cite{CMS:2014pgm}.

Pileup can contribute extra tracks and calorimetric energy depositions to the jet momentum.
The pileup-per-particle identification algorithm (PUPPI)~\cite{Sirunyan:2020foa,Bertolini:2014bba} is used to mitigate the effect of pileup at the reconstructed-particle level, making use of local shape information, event pileup properties, and tracking information.
A local shape variable is defined, which distinguishes between collinear and soft diffuse distributions of other particles surrounding the particle under consideration.
Collinear distributions are attributed to particles originating from the hard scatter, whereas soft diffuse distributions are attributed to particles originating from pileup interactions.
Charged particles identified to be originating from pileup vertices are discarded.
For each neutral particle, a local shape variable is computed from nearby particles. Within the tracker acceptance ($\abs{\eta} < 2.5$), only charged particles compatible with the PV are considered. Outside the tracker coverage, both charged and neutral particles are used.
The momenta of the neutral particles are then rescaled according to their probability to originate from the PV, deduced from the local shape variable, superseding the need for jet-based pileup corrections~\cite{Sirunyan:2020foa}.

The missing transverse momentum vector \ptvecmiss is computed as the negative vector \pt sum of all the PF candidates in an event, and its magnitude is denoted as \ptmiss~\cite{CMS:2019ctu}.
The \ptvecmiss is corrected for the energy scale of the reconstructed jets in the event.
The PUPPI algorithm is applied to reduce the pileup dependence of the \ptvecmiss observable.
The \ptvecmiss is computed from the PF candidates weighted by their probability of originating from the PV~\cite{CMS:2019ctu}.

Muons are reconstructed within $\abs{\eta}<2.4$ using information from the muon detectors and the silicon tracker. The reconstruction and identification efficiency for muons exceeds 96\% over the full acceptance, with \pt resolutions of about 1\% in the barrel and 3\% in the endcap regions for muons with $\pt < 100 \GeV$~\cite{CMS:2018rym}.

Electrons are reconstructed by matching tracks in the silicon tracker to energy clusters in the ECAL. Their momentum resolution depends on the detector region and the amount of brems\-strahlung radiation emitted in the tracker material, and is approximately 1.6--5.0\% for electrons with \pt around 45\GeV from $\PZ\to\Pe\Pe$ decays~\cite{CMS:2020uim,CMS-DP-2020-021}.

The reconstruction of \tauh leptons is performed from anti-\kt jets using the hadrons-plus-strips (HPS) algorithm~\cite{HPS2,DeepTau_v2p1}. The reconstruction is based on PF candidates and targets the characteristic low-multiplicity decay topologies of \tauh decays. Charged hadrons reconstructed by the PF algorithm are combined with electromagnetic PF candidates consistent with neutral pion decays. Neutral pions are identified through their $\Pgpz \to \PGg\PGg$ decay, including cases affected by photon conversions, by clustering nearby photons and electrons into elongated structures in the $\eta$-$\phi$ plane, referred to as strips. The strip size is dynamically adjusted as a function of the \pt of the associated electromagnetic candidates to account for the collimation of the decay products at high momentum.

The four-momentum of each reconstructed neutral pion candidate is constructed by summing the energies of all PF photons and electrons clustered into the strip as a scalar sum, exploiting the superior energy resolution of the ECAL. Its direction is defined by the leading electromagnetic constituent, which in most cases is a photon, as it carries the dominant fraction of the strip energy and therefore provides the most reliable directional estimate. The photon direction is reconstructed by extrapolating the associated ECAL clusters back to the PV, and the invariant mass of the strip is fixed to the known neutral pion mass. The \tauh candidates are then formed by combining these neutral pion strips with one or three charged hadrons within the jet, yielding decay mode hypotheses corresponding to the dominant \tauh decay channels.

Based on the reconstructed charged-particle and strip multiplicities, \tauh candidates are categorised into specific decay modes, including the 1-prong, 1-prong+1-strip, 3-prong, and 3-prong+1-strip topologies. These categories correspond to the most prevalent \tauh decay modes and are used throughout the analysis. To suppress backgrounds from quark- and gluon-initiated jets, as well as from electrons and muons misidentified as \tauh, the \textsc{DeepTau}~v2.5 algorithm is employed. This deep neural network combines information from individual PF candidates in the vicinity of the \tauh axis with the global properties of the \tauh candidate and the event. The combined HPS and \textsc{DeepTau} v2.5 reconstruction achieves an efficiency of approximately 45\% for genuine \tauh candidates with $\pt > 20\GeV$, while reducing the jet misidentification probability to the level of $\mathcal{O}(10^{-3})$~\cite{DeepTau_v2p1,CMS:2025kgf}.

\section{Event selection} \label{sec:selections}

In both channels, events with a pair of oppositely charged \PGt lepton candidates are selected. Events are required to pass the selection imposed by the CMS trigger system. In the \mutau channel, events are selected using both a single-muon trigger and a cross trigger. The cross trigger is specifically designed to identify events containing a muon in association with a \tauh candidate~\cite{CMS:2026zzq}. In the \tautau channel, a combination of a di-\tauh trigger, a di-\tauh plus jet trigger, and a single-\tauh trigger is employed. The thresholds applied during data acquisition (online) and those used in the subsequent event selection (offline) are set such that the selected events lie on the efficiency plateau of the trigger turn-on curve. The trigger configuration and thresholds remained unchanged throughout the full 2022--2023 data-taking period~\cite{CMS:2026zzq}. The trigger and analysis selection thresholds for each channel are summarised in Table~\ref{tab:triggertable}.

\begin{table}[htbp]
\renewcommand{\arraystretch}{1.5}
\centering
\topcaption{Online trigger and offline selection requirements applied in the 
\mutau and \tautau channels. All \pt values are given in \GeVns{}. Online 
trigger thresholds are indicated in parentheses.
\label{tab:triggertable}}
\begin{tabular}{llcc}
\hline
Channel & Trigger & \pt [\GeVns{}] & $\abs{\eta}$ \\
\hline
\multirow{3}{*}{\tautau} 
    & di-\tauh        
    & $\pttauh > 40\ (35)$, $\pttauh > 40\ (35)$                              
    & $< 2.1$ \\
    & di-\tauh+jet  
    & $\pttauh > 35\ (30)$, $\pttauh > 35\ (30)$, 
    & $< 2.1$ \\
    & 
    & $\pt^{\text{jet}} > 60\ (60)$
    &   \\
    & single-\tauh
    & $\pttauh > 190\ (180)$
    & $< 2.1$ \\[\cmsTabSkip]
\multirow{2}{*}{\mutau}
    & single-\PGm     
    & $\ptmu > 26\ (24)$                                                        
    & $\abs{\eta_{\PGm}} < 2.4$ \\
    & \PGm-\tauh cross trigger
    & $21\ (20) < \ptmu < 26$, $\pttauh > 32\ (27)$                         
    & $\abs{\eta_{\PGm}} < 2.1$, $\abs{\eta_{\tauh}} < 2.1$ \\
\hline
\end{tabular}
\end{table}

In both channels, a pair of oppositely charged \PGt leptons with a separation of $\DR > 0.5$ is required, where $\Delta R = \sqrt{\smash[b]{(\Delta\eta)^2 + (\Delta\phi)^2}}$ is the distance in the $\eta$--$\phi$ plane between the two candidates. Each offline-reconstructed object must be geometrically matched to the corresponding trigger object within $\DR < 0.5$. A veto on events containing additional loosely identified electrons or muons is applied in both channels to suppress contributions from processes with higher lepton multiplicities. Here, loosely identified electrons and muons refer to objects passing minimal preselection criteria: $\pt > 10\GeV$ and $\abs{\eta} < 2.5$ for electrons (with a loose identification requirement~\cite{CMS:2020uim}), and $\pt > 10\GeV$ and $\abs{\eta} < 2.4$ for muons (with a loose isolation requirement~\cite{CMS:2018rym}). When multiple \PGt lepton pair candidates are present in an event, candidates of each object type entering the pair are first ranked independently by decreasing isolation, using \pt as a tie-breaker; for \tauh candidates, the raw \textsc{DeepTau} $D_\text{jet}$ score is used as the ranking discriminant. All combinations satisfying the pair-level requirements described above are then formed from these ranked candidate collections, and the highest-ranked pair is retained.

\subsection{The \texorpdfstring{\tautau}{tautau} channel}\label{sec:tautau_selection}

Both \tauh candidates are required to be reconstructed in one of the decay modes described in Section~\ref{sec:event_reconstruction}, and to satisfy the kinematic and identification criteria summarised in Table~\ref{tab:sel_combined}. A minimum visible mass of $\mvis > 40\GeV$, defined as the invariant mass of the visible decay products of the lepton pair, is required for the \tautau pair. To suppress backgrounds from quark- and gluon-initiated jets, electrons, and muons misidentified as \tauh candidates, both \tauh candidates are required to pass the \textsc{DeepTau} working points also listed in Table~\ref{tab:sel_combined}. There, the labels $D_\text{jet}$, $D_\text{e}$, and $D_{\mu}$ denote the \textsc{DeepTau} discriminator decisions against jets, electrons, and muons, respectively, with the working point names (\eg VVLoose, Medium, Tight) corresponding to fixed thresholds on the discriminator score, as defined in Ref.~\cite{DeepTau_v2p1}.

\subsection{The \texorpdfstring{\mutau}{mutau} channel}\label{sec:sel_mutau}
The muon is required to satisfy a relative isolation $I_{\text{rel}} < 0.15$, defined as
\begin{linenomath}
\begin{equation}
    I_{\text{rel}} = \frac{\sum_{\text{charged}}\pt + 
    \max\left(\sum_{\text{neutral}}\ET + \sum_{\gamma}\ET - 
    \Delta\beta\sum_{\text{charged,PU}}\pt,\ 0\right)}{\pt^{\mu}},
    \label{eq:iso}
\end{equation}
\end{linenomath}
where the sums run over photons, charged and neutral hadrons within a cone of $\DR = 0.4$ around the muon direction. Pileup contributions are corrected for using the $\Delta\beta$ method~\cite{CMS:2018rym}, in which the neutral pileup component is estimated as $\Delta\beta = 0.5$ times the scalar \pt sum of charged pileup particles within the isolation cone, reflecting the assumed 2:1 ratio of charged to neutral pileup activity. A minimum visible mass of $\mvis > 40\GeV$ is also required for the \mutau pair. The \tauh candidate must satisfy the \textsc{DeepTau} working points summarised in Table~\ref{tab:sel_combined}. The selection requirements for both objects are summarised in Table~\ref{tab:sel_combined}.

To suppress the otherwise large \Wjets background, the transverse mass ($\mT$) of the muon-\ptvecmiss system is required to satisfy $\mT < 65\GeV$, where $\mT$ is defined as
\begin{equation}
    \mT \equiv \sqrt{2\,\pt^{\mu}\,\ptmiss\left[1 - \cos(\Delta\phi)\right]},
    \label{eq:mT}
\end{equation}
where $\Delta\phi$ is the azimuthal angle between the muon momentum direction and \ptvecmiss.

The longitudinal and transverse impact parameters of the muon are required to satisfy $\abs{d_z} < 0.2\cm$ and $\abs{d_{xy}} < 0.045\cm$, consistent with the muon originating from the PV. For the \tauh candidate, only the requirement $\abs{d_z} < 0.2\cm$ is imposed, in order to avoid a significant loss in the selection efficiency due to a relatively long \PGt lepton lifetime.

\begin{table}[htbp]
\renewcommand{\arraystretch}{1.35}
\setlength{\tabcolsep}{5pt}
\centering
\topcaption{Offline selection requirements in the \tautau and \mutau channels. A dash indicates that the requirement is not applied in that channel.}
\label{tab:sel_combined}
\begin{tabular}{lcc}
\hline
Variable & \tautau & \mutau \\
\hline
Muon \pt                                          & \NA               & ${>}26\GeV$ \\
Muon $\abs{\eta}$                                 & \NA               & ${<}2.4$ \\
Muon isolation $I_{\text{rel}}$                   & \NA               & ${<}0.15$ \\
Muon $\abs{d_{xy}}$                               & \NA               & ${<}0.045\cm$ \\
Muon $\abs{d_z}$                                  & \NA               & ${<}0.2\cm$ \\[\cmsTabSkip]
\tauh \pt                                       & ${>}35\GeV$       & ${>}32\GeV$ \\
\tauh $\abs{\eta}$                              & ${<}2.1$          & ${<}2.1$ \\
\tauh $\abs{d_z}$                               & ${<}0.2\cm$       & ${<}0.2\cm$ \\
\textsc{DeepTau} $D_\text{jet}\xspace$            & Medium            & Medium \\
\textsc{DeepTau} $D_\text{e}\xspace$              & VVLoose           & VVLoose \\
\textsc{DeepTau} $D_{\mu}\xspace$                & Tight             & Tight \\[\cmsTabSkip]
Pair \mvis                             & ${>}40\GeV$       & ${>}40\GeV$ \\
Pair $\mT$                                        & \NA               & ${<}65\GeV$ \\
\hline
\end{tabular}
\end{table}

\section{The MUFFIN method} \label{sec:bdt_ff}
 
\subsection{Conceptual overview of MUFFIN}\label{sec:bdt_ff_overview}

The \FF method~\cite{HIG-21-001} estimates the contribution of backgrounds due to quark- or gluon-initiated jets misidentified as \tauh decays using TFs defined as
\begin{equation}
  \FF(\mathbf{x}) = \frac{N_{\text{pass}}(\mathbf{x})}{N_{\text{fail}}(\mathbf{x})},
\end{equation}
where $N_{\text{pass}}$ and $N_{\text{fail}}$ denote the number of \tauh candidates passing and failing the nominal \tauh identification requirements, respectively, measured in DRs as functions of a limited set of
kinematic variables $\mathbf{x}$. These TFs are applied to events in the ARs to predict the misidentified background in the SRs.

The MUFFIN method generalises this approach with multidimensional per-object reweighting. A BDT is trained in background-enriched DRs to distinguish failing (fail $D_\text{jet}$ Medium but pass VVVLoose \textsc{DeepTau} working point) from passing (pass $D_\text{jet}$ Medium \textsc{DeepTau} working point) \tauh candidates. The calibrated outputs of the BDT are interpreted as class probabilities (see Section~\ref{sec:bdt_ff_calibration}), which are then used to construct continuous TFs in the full feature space.

\begin{figure}[htbp]
\centering
\includegraphics[width=\textwidth]{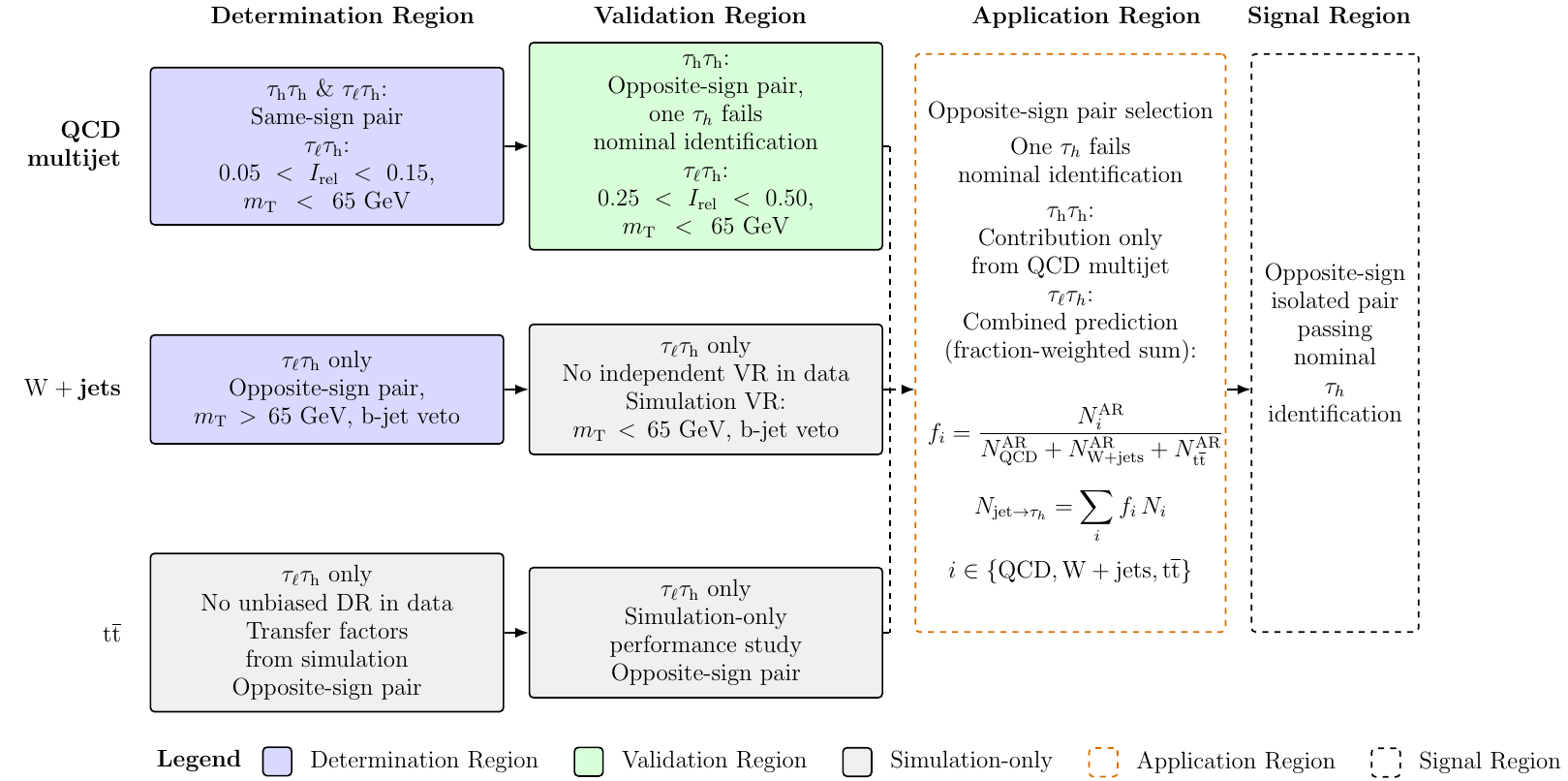}
\caption{
Schematic overview of the DRs, VRs, AR, and SR used in the estimation of the \jtf backgrounds.
For each process, TFs are derived in dedicated DRs enriched in jets misidentified as \tauh candidates.
The performance of the method is assessed in VRs, where available, or through simulation-based studies in cases without an unbiased DR.
The prediction is then applied in a common AR, defined by opposite-sign events with one \tauh candidate failing the nominal identification requirement.
In the \tautau channel, the AR is dominated by QCD multijet events, while in the \mutau channel a combined prediction is constructed from QCD multijet, \Wjets, and \ttbar contributions using relative fractions derived in the AR.
The final estimate is obtained in the SR, defined by opposite-sign isolated events passing the nominal \tauh identification requirements.
}
\label{fig:dr_vr_sr_schematic}
\end{figure}

After calibration, the model outputs $P_{\text{pass}}(\mathbf{z})$ and $P_{\text{fail}}(\mathbf{z})$ for input features $\mathbf{z}$. Their ratio,
\begin{equation}
  w_{\text{MUFFIN}}(\mathbf{z}) \equiv \frac{P_{\text{pass}}(\mathbf{z})}{P_{\text{fail}}(\mathbf{z})},
\end{equation}
plays the role of a continuous multidimensional TF, replacing binned $\FF(\mathbf{x})$. The precise definition of $P_{\text{pass}}$ and $P_{\text{fail}}$ from the calibrated four-class BDT outputs is given in Section~\ref{sec:bdt_ff_calibration}; the BDT is trained separately for each analysis channel and background process, as described in Sections~\ref{sec:bdt_ff_dr_ar}--\ref{sec:bdt_ff_calibration}. The DRs in Fig.~\ref{fig:dr_vr_sr_schematic} define where this mapping is learned, and the resulting per-object weights are applied in SR-like regions to predict the \jtf background. Dedicated validation regions (VRs), orthogonal to the DRs and SRs, are used to assess the modelling performance of the method. Figure~\ref{fig:dr_vr_sr_schematic} illustrates the region definitions and their role in the method for both the \tautau channel and, generically, semileptonic \ltau channels ($\Pell$ = electron or muon), although only the \mutau channel is shown in this paper.

\subsection{Construction of determination and application regions} \label{sec:bdt_ff_dr_ar}

The TFs are measured separately for each dominant source of jets misidentified as \tauh leptons: QCD multijet, \Wjets, and \ttbar production in the \mutau channel, and QCD multijet production in the \tautau channel. For each process and channel, dedicated DRs are defined to enrich the corresponding background contribution while suppressing processes with genuine \PGt leptons. Where appropriate, orthogonal VRs are constructed to assess the modelling performance of the method. In all cases, the DRs are defined by requiring one \tauh candidate to fail the nominal identification criterion, while passing a looser \tauh identification working point ($D_\text{jet}\xspace$ VVVLoose), as defined in Ref.~\cite{DeepTau_v2p1}. Additional selections on charge sign requirements, lepton isolation, $\mT$, and \PQb-tagged jet multiplicity are applied to enhance the purity of the targeted background process. The DRs and ARs are common to both the \FF and MUFFIN methods. The final background estimate in each channel is then obtained by combining the weighted contributions from the relevant sources of misidentified \tauh candidates. A schematic overview of the region definitions and their role in the method is shown in Fig.~\ref{fig:dr_vr_sr_schematic}. The resulting background estimate from the MUFFIN method in a signal-like region of phase space is discussed in Section~\ref{sec:signal_regions}.

For the estimation of the QCD multijet background, same-sign charge requirements are used in all channels to suppress electroweak and top quark processes, which are predominantly opposite-sign. In the \mutau channel, the nominal muon isolation requirement ($I_{\text{rel}} < 0.15$) and \mT selection ($\mT<65\GeV$) are retained to ensure the DR remains kinematically representative of the SR, while the same-sign requirement provides the primary QCD multijet enrichment by suppressing \Wjets and \Zjets contributions. 
In the \tautau channel, a VR is defined by requiring one of the \tauh candidates to fail the nominal \tauh identification requirement while passing a looser working point, and by requiring the two \tauh candidates to have opposite signs. 
In the \mutau channel the VR is defined by modifying the muon isolation requirement ($0.25 < I_{\text{rel}} < 0.5$). 
These VRs provide independent samples for testing the performance of the method, while maintaining a high purity of the targeted background process.

The \Wjets DR is defined in the \mutau channel by requiring opposite-sign events with $\mT>65\GeV$. 
To suppress contamination from \ttbar events, a veto on \PQb-tagged jets is applied. 
As the inversion of the large \mT requirement would lead to overlap with the SR, no independent VR is defined for this background; instead, the performance is assessed using a dedicated simulation-only training and validation procedure.
To this end, MUFFIN is trained on simulated \Wjets events using the same DR selections as for the data (\ie $\mT>65\GeV$) and this is validated by applying the methods to simulated events with $\mT<65\GeV$.
Likewise, the \FF method derives the TFs in high-\mT events and applies them in low-\mT events.

The \ttbar TFs are measured from simulated events.
This is necessary as it is not possible to define a DR in data that is sufficiently pure in \ttbar \jtf events without applying selections that bias the measured TFs.
In particular, selections that substantially enhance the \ttbar purity (\eg requirements on \PQb-tagged jets and/or specific lepton flavour configurations) can modify the flavour composition and kinematics of the jets that are misidentified as \tauh candidates, thereby distorting the $\jtf$ misidentification probability.
As a result, the performance of the method for \ttbar is assessed using simulation-only studies, where the true composition of the sample is known and the impact of the method can be reliably evaluated.
The uncertainty model for the \ttbar background estimation is designed to account for potential mismodelling of the \ttbar \jtf TFs in simulation.

\subsection{Training inputs}

The data are split into independent training (70\%) and evaluation (30\%) samples to ensure stable training and unbiased performance evaluation. The VR, defined in Section~\ref{sec:bdt_ff_dr_ar}, serves as a fully blind test set, as it is a separate physics region whose data never enters the training or hyperparameter optimisation. The training is performed separately for each analysis channel and background process to capture channel-specific kinematic properties and event topologies. For each targeted process, the training is performed on data events in the corresponding DR. However, the data sample contains small contributions ($\mathcal{O}(1\%)$) from other background processes with genuine \tauh candidates (\eg\ \ttbar, \Zjets). To isolate the \jtf misidentification TF, these genuine \tauh contributions enter the training as a source of contamination that is subtracted from the data components in the final weight definition (Section~\ref{sec:bdt_ff_calibration}). Crucially, no simulation of the targeted \jtf process itself is used at any stage: \eg when deriving the \Wjets TFs, no \Wjets simulation enters either the training or the subtraction, ensuring that the method is entirely data-driven for the targeted \jtf component. For dedicated simulation-only studies, a binary setup with two classes (pass and fail) is used.

The input feature set is chosen to capture both object-level properties of the \tauh candidate and aspects of the global event environment that influence the \jtf misidentification probability. 
Local variables include the reconstructed decay mode of the \tauh candidate (the specific decay channel of the \tauh as determined by the HPS algorithm), its \pt, $\eta$, and azimuthal angle $\phi$, as well as the ratio of the \pt of the jet from which the \tauh candidate is reconstructed to that of the \tauh candidate. 
This ratio provides sensitivity to differences in energy sharing between the reconstructed \tauh decay products and the other jet constituents.
In genuine \tauh decays, the visible decay products typically carry a large fraction of the momentum of the original \PGt lepton and are more collimated, leading to a reconstructed \tauh candidate whose \pt closely matches that of the associated jet. 
In contrast, jets misidentified as \tauh candidates often arise from broader hadronic activity, where only a subset of the jet constituents is selected by the \tauh reconstruction algorithm. 
As a result, the \pt of the reconstructed \tauh candidate tends to represent a smaller and more variable fraction of the full jet momentum, making this ratio a powerful discriminant between the two cases.

In addition to local quantities, a small number of global event-level variables are included in the training. These variables, such as the total number of reconstructed jets and the
number of \PQb-tagged jets in the event, encode information about the overall hadronic activity and event topology. Although they are not directly associated with the \tauh candidate, they can impact the composition and kinematic properties of the \jtf background and therefore improve the modelling of the misidentification rate. This is not dissimilar to how, in the \FF method, similar variables are used to define separate \FF parametrisations in different analysis categories. Lastly, an explicit data-taking period label is included as an input feature to account for potential variations in detector conditions and performance across different data-taking periods. The relative importance of the input variables, as determined by feature importance studies, can be ranked as follows:
\begin{enumerate}
    \item Decay mode of the \tauh candidate (most important);
    \item Ratio of jet $\pt$ to \tauh \pt;
    \item The \pt of the \tauh candidate;
    \item Total number of reconstructed jets in the event;
    \item Number of \PQb-tagged jets in the event;
    \item The $\eta$ of the \tauh candidate;
    \item The $\phi$ of the \tauh candidate;
    \item An explicit data-taking period label (subdominant but non-negligible).
\end{enumerate}
Feature importance is evaluated using the built-in \textsc{XGBoost}~\cite{XGBoost} metrics ``weight'', ``gain'', and ``cover'', which respectively quantify the number of times a feature is used to split the data, the average improvement in accuracy brought by a feature, and the relative coverage of samples affected by a feature. These metrics are extracted from the trained BDT and visualised to assess the contribution of each input variable. These studies help to validate the choice of input features and provide insights into the physical properties that govern the \jtf misidentification rates. Across all channels and processes, the BDT models consistently identify the decay mode and jet-to-\tauh kinematics such as the ratio of jet \pt to \tauh \pt as the most important features. Angular variables and event-level quantities are found to have a secondary but non-negligible impact, highlighting the benefit of including global event information in the training. Another important conclusion from these studies is that the relative importance of the explicit data-taking period label (as an input feature) is found to be non-negligible but subdominant compared to the main kinematic features, indicating that while there are some differences in the detector performance across data-taking periods that can lead to heightened/reduced misidentification rates, the core physics properties driving the misidentification rates remain consistent.

\subsection{The BDT architecture}

The ML models used for the reweighting are based on BDT classifiers implemented using the \textsc{XGBoost} framework. The model architecture and training hyperparameters are optimised using the \textsc{Optuna} toolkit~\cite{Optuna}, which performs an automated search of the hyperparameter space using a Bayesian optimisation strategy~\cite{BayesianOptimization}. The objective of the optimisation is to minimise the loss evaluated on an independent evaluation data set, thereby ensuring good generalisation performance and stable behaviour across the phase space relevant to the analysis.

The hyperparameter search includes variations of the maximum tree depth, the learning rate, the fraction of events and features used at each boosting iteration, and regularisation parameters controlling the complexity of the model. In particular, parameters governing tree depth and minimum node population are used to limit overfitting, while regularisation terms are introduced to penalise overly complex decision boundaries: a sparsity penalty that encourages the model to rely on fewer features per tree, and a weight-decay penalty that prevents individual node weights from becoming too large. This choice of hyperparameters allows the model to capture non-linear dependencies between input variables, while maintaining robustness against statistical fluctuations in the training data.

\begin{figure}[htbp]
  \centering
  \includegraphics[width=0.7\textwidth]{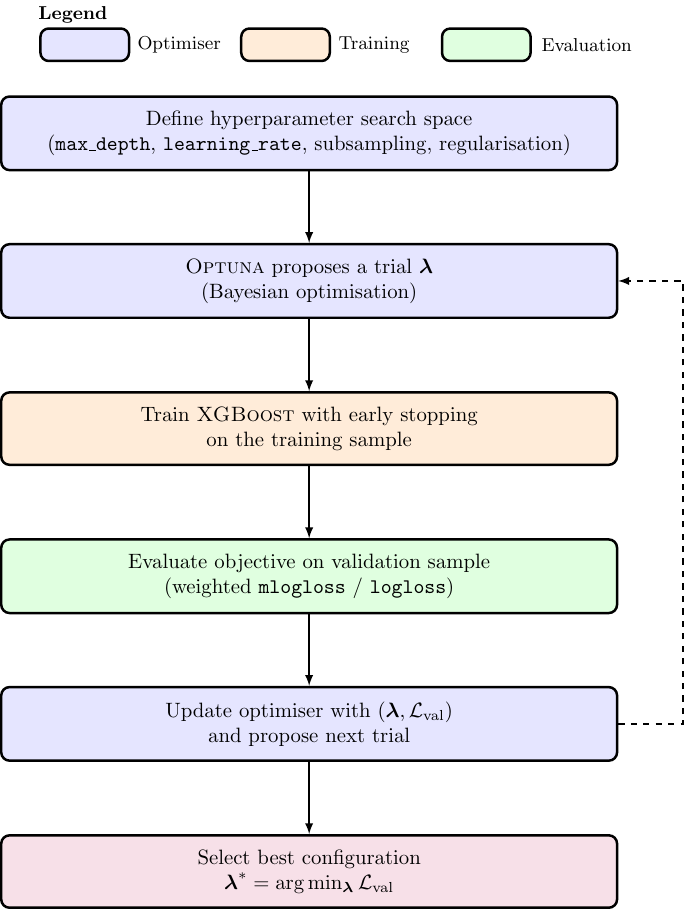}
  \caption{Schematic overview of the hyperparameter optimisation procedure used to configure the BDT models. For each trial, \textsc{Optuna} proposes a hyperparameter set, the model is trained with early stopping, and the evaluation loss is used as the optimisation objective.}
  \label{fig:hyperopt_schematic}
\end{figure}

For each optimisation trial, a BDT model is trained with a maximum number of boosting iterations of 1000, and early stopping is employed based on the evaluation loss to determine the optimal number of trees. Training stops when the evaluation loss fails to improve by more than 0.1\% in absolute value for 10 consecutive boosting iterations. The optimisation is performed over a fixed number of trials ($N=50$), after which the hyperparameter configuration yielding the best evaluation performance is selected for the final training. This procedure, illustrated in Fig.~\ref{fig:hyperopt_schematic}, provides a reproducible and data-driven approach to selecting the model architecture, reducing sensitivity to manual tuning and ensuring consistent performance across analysis channels and data-taking periods.

\subsection{Training performance}

The performance and stability of the BDT training are evaluated using independent training and evaluation data sets. The evolution of the loss function as a function of the boosting iteration is monitored to assess convergence and to identify potential overtraining. The training minimises the cross-entropy loss,
\begin{equation}
    \mathcal{L} = -\frac{1}{N} \sum_{i=1}^{N} \sum_{k=1}^{K} y_{ik} \log \hat{p}_{ik},
    \label{eqn:bce_loss}
\end{equation}
where $y_{ik} \in \{0, 1\}$ is the true-class indicator ($y_{ik} = 1$ if candidate $i$ belongs to class $k$, and $y_{ik} = 0$ otherwise), $\hat{p}_{ik}$ is the corresponding predicted class probability, and $K$ is the number of classes. In the nominal subtraction-based training, $K = 4$ (data pass, data fail, simulation pass, and simulation fail). In dedicated simulation-only studies, a binary setup with $K = 2$ is used. The initial loss therefore differs between configurations: $\ln 4 \approx 1.386$ for the four-class channels and $\ln 2 \approx 0.693$ for the binary simulation-only channels, explaining the lower starting loss observed for the simulation-based \Wjets and \ttbar models. The simulation-based \Wjets model requires more boosting iterations to converge, reflecting the greater complexity of the learned TF for this process. In all channels, the training and evaluation losses are observed to decrease smoothly and to stabilise at similar values, indicating good generalisation performance. 

\begin{figure}[htbp]
    \centering
    \includegraphics[width=0.7\textwidth]{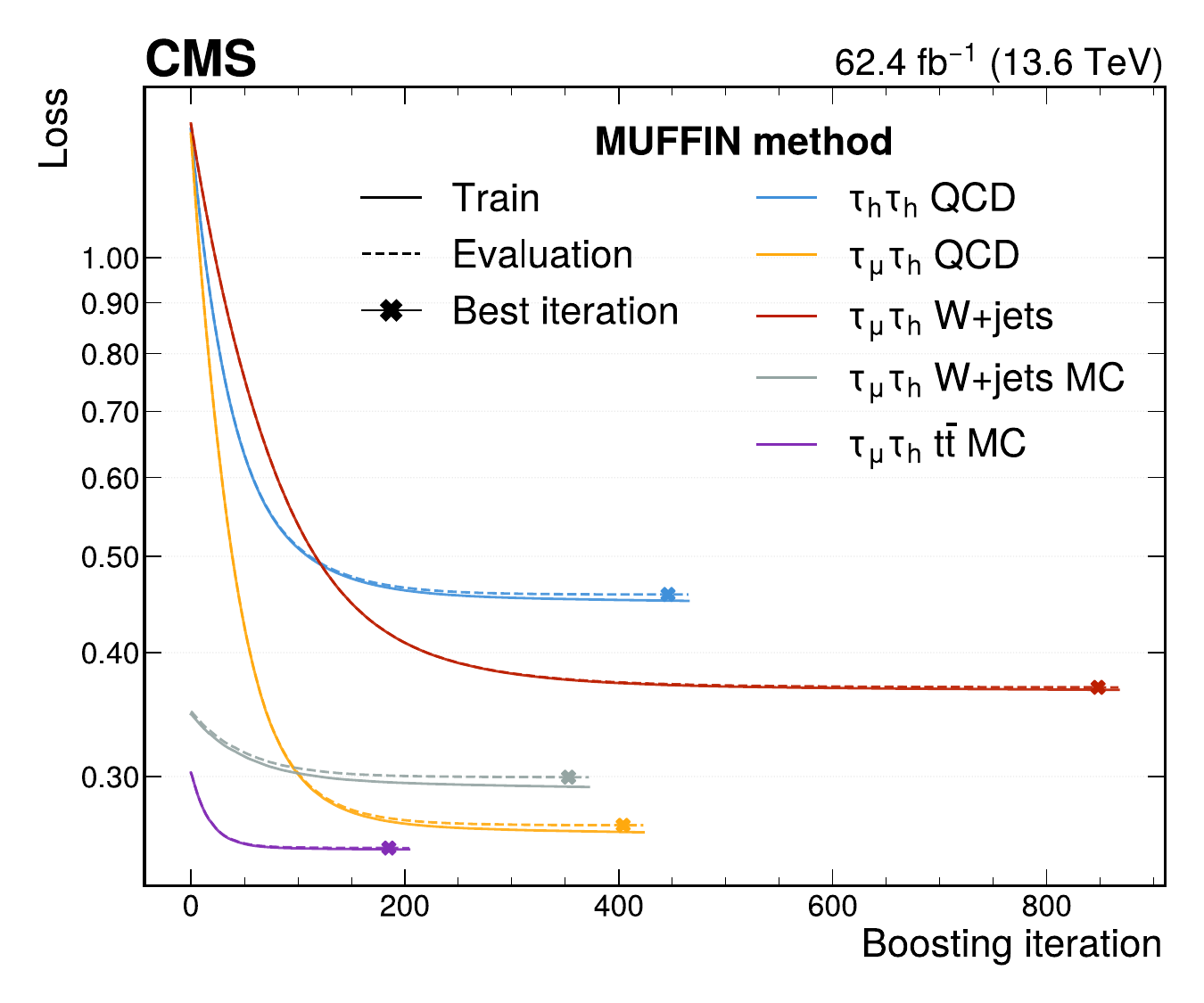}
    \caption{Cross-entropy loss (Eq.~(\ref{eqn:bce_loss})) versus boosting iteration for all BDT models in the \tautau and \mutau channels, with the loss axis shown on a logarithmic scale. Solid (dashed) lines correspond to the training (evaluation) loss. Cross markers indicate the best (saved) iteration; curves extend up to 10 iterations beyond this point, corresponding to the early stopping window. The lower initial loss of the simulation-only (\Wjets, \ttbar) models reflects the binary ($K=2$) classifier setup ($\ln 2 \approx 0.693$) compared to the four-class ($K=4$) data channels ($\ln 4 \approx 1.386$).}
    \label{fig:bdt_loss_curve}
\end{figure}

Figure~\ref{fig:bdt_loss_curve} shows representative loss curves for all channels and processes. The cross marker indicates the best iteration, at which the model is saved; the loss curves extend up to 10 iterations beyond this point, corresponding to the early stopping window over which no improvement exceeding 0.1\% was observed. A small gap between training and evaluation loss is expected, as the model has seen the training data. Overfitting would manifest as a rising evaluation loss while the training loss continues to fall. The absence of such divergence confirms that the model is not overfitting to statistical fluctuations in the training data, and that it is learning meaningful patterns that generalise well to unseen data. This behaviour is consistent across all channels and background processes, demonstrating the robustness of the training procedure and the suitability of the chosen hyperparameters for capturing the relevant features of the \jtf misidentification background.

\subsection{Calibration of BDT outputs} \label{sec:bdt_ff_calibration}

The BDT output scores are subsequently used to derive per-object weights for the \jtf background. For this purpose, it is important that the classifier output can be interpreted as a well-calibrated probability, such that the predicted score is consistent with the empirical class frequency observed in the evaluation sample. Calibration in this sense is a purely empirical requirement: it does not require prior knowledge of the true class proportions, but rather ensures that a predicted confidence of $p$ corresponds to an observed frequency of $p$ in the held-out data. A post-training calibration of the BDT outputs is therefore performed using temperature scaling~\cite{TemperatureScaling, Platt:1999}. In this approach, the logits $\mathbf{z}$ of the trained classifier are divided by a single global temperature parameter $T^*$, yielding calibrated class probabilities $\mathrm{softmax}(\mathbf{z}/T^*)$. The optimal $T^*$ is determined by minimising the cross-entropy loss of Eq.~(\ref{eqn:bce_loss}) on the independent 30\% held-out evaluation sample, ensuring the calibration is unbiased with respect to the training. This procedure preserves the ordering of classifier scores and does not alter the decision boundaries of the model, while correcting potential mismatches between predicted probabilities and observed frequencies.

The calibration performance is assessed using reliability diagrams, which verify that the predicted class probability (confidence) matches the empirical fraction of correctly classified events (accuracy) in bins of the classifier output. For a well-calibrated classifier, a predicted confidence of $p$ should correspond to an observed accuracy of $p$, lying along the diagonal. Here, the confidence is defined as the maximum predicted class probability for each event, while the accuracy corresponds to the event-weight-averaged fraction of correctly classified events within a given confidence bin. Uncertainties on the accuracy are estimated using Wilson score 68\% confidence intervals, with the effective sample size $n_\text{eff} = (\sum_i w_i)^2 / \sum_i w_i^2$ accounting for the non-uniform event weights.

\begin{figure}[htbp]
    \centering
    \includegraphics[width=0.7\textwidth]{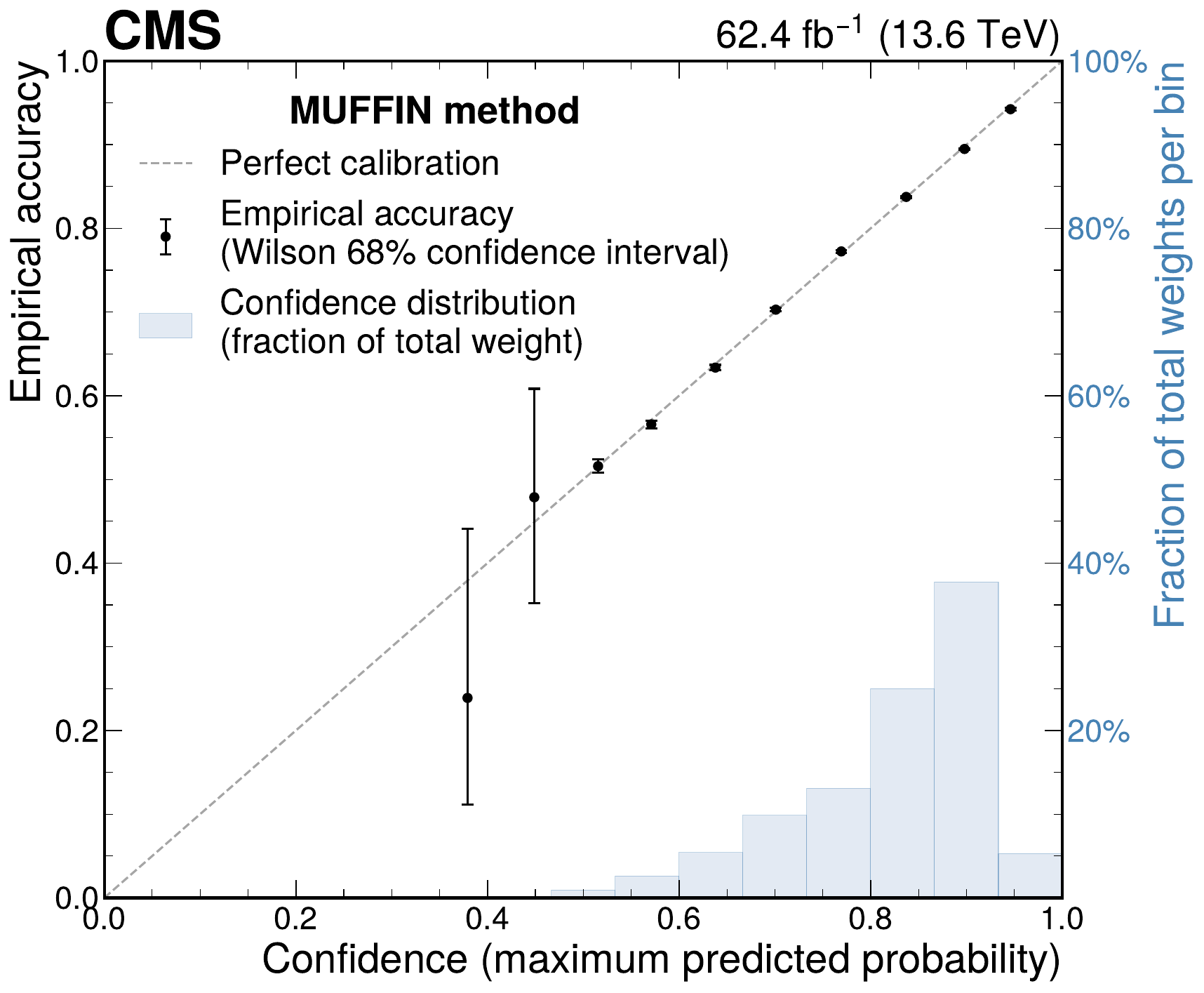}
     \caption{Reliability diagram for the QCD multijet BDT in the \tautau channel after temperature scaling ($T^\ast = 0.999$), shown on the evaluation data set. Each point is placed at the weighted mean confidence of events in the bin (not the geometric bin centre) and shows the event-weight-averaged fraction of correctly classified events (accuracy), with Wilson score 68\% confidence intervals using the effective sample size $n_\text{eff} = (\sum_i w_i)^2 / \sum_i w_i^2$. The dashed diagonal indicates perfect calibration. The shaded histogram (right axis) shows the weighted confidence-score distribution. The model is well calibrated across the full range, with most events (${>}30\%$ per bin) at high confidence (${>}0.8$); the larger uncertainties at low confidence reflect the correspondingly small event counts there.}
    \label{fig:bdt_reliability_diagram}
\end{figure}

For the QCD multijet BDT in the \tautau channel, the optimal temperature parameter is found to be $T^\ast = 0.999 \approx 1$, meaning the BDT outputs require negligible rescaling and are already well-calibrated class probabilities. This confirms the stability of the training and that the BDT class probabilities are directly suitable for use in the weight formula introduced below. Figure~\ref{fig:bdt_reliability_diagram} shows the reliability diagram after calibration for the QCD multijet BDT in the \tautau channel. Similar behaviour is observed across other channels and background processes ($T^\ast \in [0.990, 1.001]$). Given the small magnitude of the calibration correction in all cases, temperature scaling is primarily employed as a validation step to ensure probabilistic interpretability of the classifier output, rather than as a substantial modification of the model predictions.

The calibrated BDT output scores are then used to derive per-object weights for the background reweighting, following the same logic as the \FF method. The weights are defined as
\begin{equation}
w_{\text{MUFFIN}}(\mathbf{z}) =
\frac{
P_{\text{pass}}^{\text{data}}(\mathbf{z}) - P_{\text{pass}}^{\PGt_{\text{sim}}}(\mathbf{z})
}{
P_{\text{fail}}^{\text{data}}(\mathbf{z}) - P_{\text{fail}}^{\PGt_{\text{sim}}}(\mathbf{z})
}.
\label{eq:weights}
\end{equation}
Here, the calibrated BDT output provides four class probabilities for a given set of input features $\mathbf{z}$: $P_{\text{pass}}^{\text{data}}$ and $P_{\text{fail}}^{\text{data}}$ are the probabilities for an object in data to pass or fail the \tauh identification, containing contributions from both genuine \PGt leptons and \jtf; $P_{\text{pass}}^{\PGt_{\text{sim}}}$ and $P_{\text{fail}}^{\PGt_{\text{sim}}}$ are the corresponding probabilities for simulated genuine \PGt leptons only. The subtraction of the genuine \PGt simulation contributions serves to remove any residual genuine \PGt contamination in the DRs, ensuring that the weights capture only the \jtf misidentification rates. Importantly, no simulation of the targeted \jtf process itself (\eg QCD, \Wjets, or \ttbar when treated as the target) is used in this subtraction, preserving the fully data-driven nature of the method. This explicit use of the BDT output probabilities is a key feature of the ML-based method, enabling a continuous, multidimensional reweighting that generalises the \FF method. In the simulation-only configurations for \ttbar and \Wjets, the data class probabilities are set to zero, ($P_{\text{pass}}^{\text{data}}(\mathbf{z})=P_{\text{fail}}^{\text{data}}(\mathbf{z})=0)$,
so the subtraction-based expression simplifies to the ratio of the genuine \PGt simulation probabilities.
Since $w_{\text{MUFFIN}}$ depends on the ratio of class probabilities, any dependence on the class proportions $\pi_k$ assumed during training enters only as a global multiplicative constant, leaving the shape of the weight function invariant to assumptions about these proportions. The overall normalisation of the background estimate is independently constrained by the observed data yield in the AR, ensuring robustness with respect to the training class balance.

In the limit of reduced input dimensionality, this approach naturally converges to the \FF method. The incorporation of ML techniques enables a more flexible multidimensional reweighting that captures correlations between input variables that are not explicitly modelled in the \FF method. Owing to their per-object definition and evaluation in a high-dimensional feature space, the BDT-derived weights can be applied consistently across different analysis categories and regions of phase space. This reduces the reliance on additional corrections and uncertainties that account for missing variable dependencies, and simplifies the overall background estimation workflow.
\section{Performance comparison}\label{sec:performance}

The performance of the MUFFIN method is benchmarked directly against the \FF method using a series of closure tests performed in the respective VRs for each process. 
These regions are designed to be kinematically similar to the SRs while remaining orthogonal to the regions used for the derivation of the background estimates, thereby providing an unbiased assessment of the
predictive power of each method.
In the \mutau channel the performance is also assessed in a dedicated simulation-only validation procedure for the \Wjets background, as described in Section~\ref{sec:bdt_ff}.

For each method, the predicted distribution of \jtf background events is compared to the data in the VRs for a range of key observables. The level of agreement is quantified by the mean absolute per-bin Poisson pull, $\langle\abs{z}\rangle$~\cite{Cowan:1998, ParticleDataGroup:2026}, where lower values indicate improved agreement and larger values signal worse residual mismodelling. The metric is defined as:
\begin{equation}
\langle\abs{z}\rangle = \frac{1}{N}\sum_{i=1}^{N} \abs{z_i}, \qquad z_i = \frac{N_i^{\text{obs}} - N_i^{\text{pred}}}{\sqrt{N_i^{\text{pred}}}},
\label{eq:z_metric}
\end{equation}
where $N_i^{\text{obs}}$ and $N_i^{\text{pred}}$ are the observed and predicted yields in bin $i$, $N$ is the total number of bins, and $z_i$ is the per-bin Poisson pull. The metric has no free parameters, and its denominator is independent of the degree of closure, ensuring that methods with worse agreement are not artificially favoured.

The set of observables considered is representative of the kinematic and topological quantities commonly used in analyses involving \tauh decays.
These variables probe both the overall event kinematics and the properties of the reconstructed \tauh candidates, and are therefore sensitive to potential mismodelling of the \jtf background. 
The distributions examined include the total transverse mass~\cite{HIG-21-001, HIGG-2013-31},
\begin{equation}
    \mT^{\text{total}} = \sqrt{\mT^2(\PGt_1,\PGt_2) + \mT^2(\PGt_1,\ptvecmiss) + \mT^2(\PGt_2,\ptvecmiss)},
    \label{eqn:mt_tot}
\end{equation}
where $\PGt_1$ and $\PGt_2$ denote the two \PGt lepton candidates, \ptvecmiss is the missing transverse momentum vector, and $\mT(a,b)$ is the pairwise transverse mass defined in Eq.~(\ref{eq:mT}), and the visible invariant mass \mvis, both of which are key discriminating variables in measurements and searches and are particularly sensitive to mismodelling of the \jtf background. 
Kinematic properties of the reconstructed \tauh candidates are also probed through their \pt and $\eta$, as well as the angular separation between the decay products, quantified by $\Delta R$. These variables test the modelling of both the object-level properties of the \tauh candidates and their correlations with the surrounding event topology. Finally, the reconstructed decay mode of the \tauh candidates is examined to validate the modelling across different \tauh topologies. Together, this set of observables provides a broad test of the background estimation performance across the kinematic and topological phase space relevant for physics analyses, ensuring that improvements observed in closure tests translate into robust modelling in SRs.

A notable limitation of the reference \FF method is its treatment of observables not directly used in the derivation of the TFs, such as the \tauh $\eta$ and decay mode. In practice, the \FF TFs are derived as averages over observables such as $\eta$ and decay mode, effectively marginalising over any dependence these variables may have on the \jtf misidentification probability. The choice of variables to average over is often driven by practical considerations, and is dependent on the specific analysis and the available data sample size in the DRs. While this is a pragmatic choice that yields acceptable modelling of the primary fit observables, it introduces a residual mismodelling whenever the distributions of these averaged-over variables differ between the DR and the SR. This condition is generally not guaranteed by the region definitions and can vary across analysis categories. Since these marginalised observables enter downstream discriminants and feed directly into the signal-over-background separation, even small shape differences can translate into increased systematic uncertainties on the background estimate.
The MUFFIN method avoids this limitation by learning the full multidimensional dependence of the misidentification probability from data, incorporating $\eta$, decay mode, and other correlated observables as continuous inputs to the BDT. This allows the method to implicitly capture the dependence of the \jtf misidentification probability on these variables without requiring explicit binning, and therefore provides improved modelling even for observables that the \FF method cannot efficiently parametrise. The improvement is clearly visible in the closure tests presented below, where pronounced gains relative to the \FF method are consistently observed in the $\eta$ and decay mode distributions.

\subsection{Performance in the \texorpdfstring{\tautau}{fully hadronic} channel}

\begin{figure}[htbp]
  \centering
    \includegraphics[width=\cmsFigWidth]{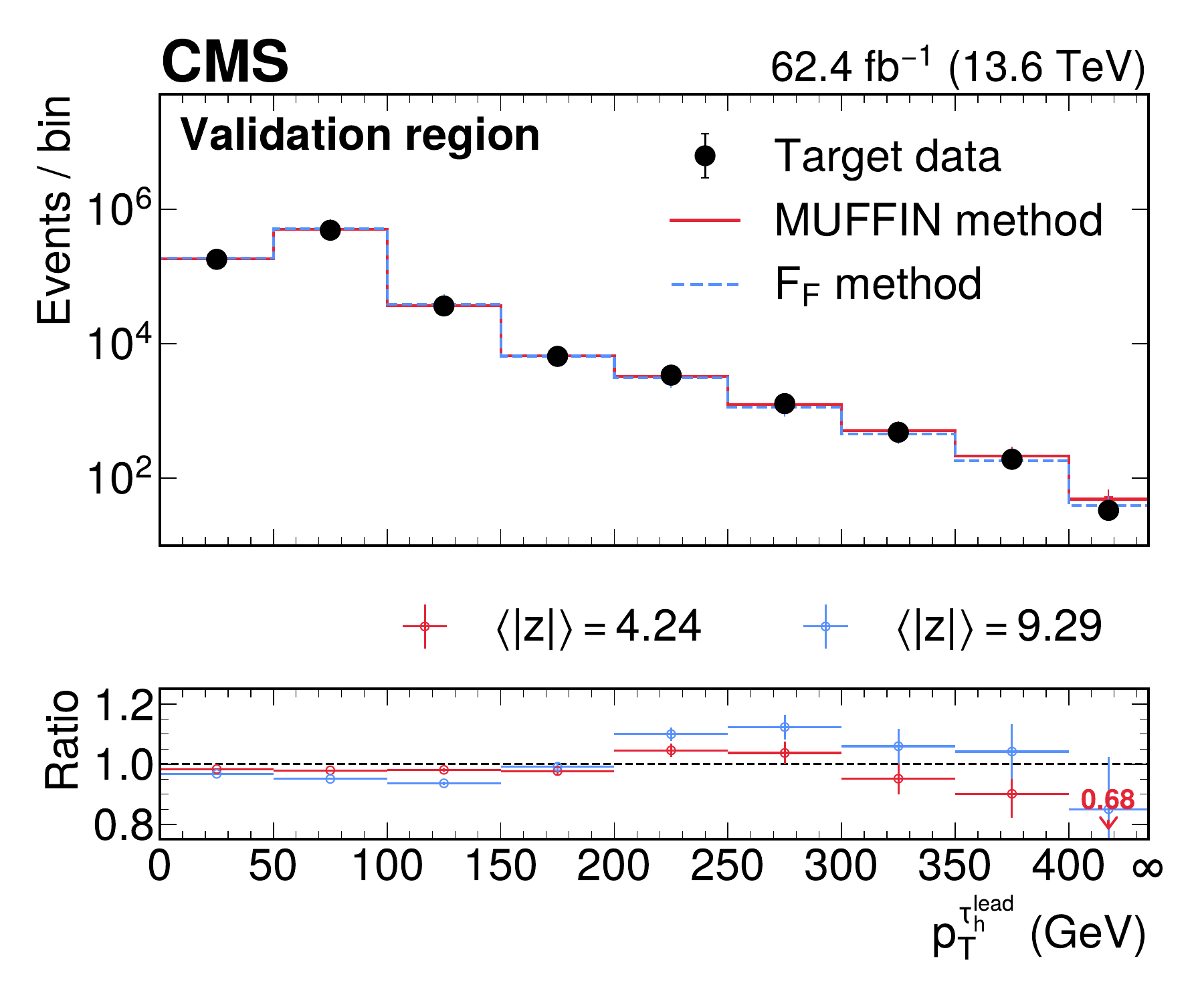}
    \includegraphics[width=\cmsFigWidth]{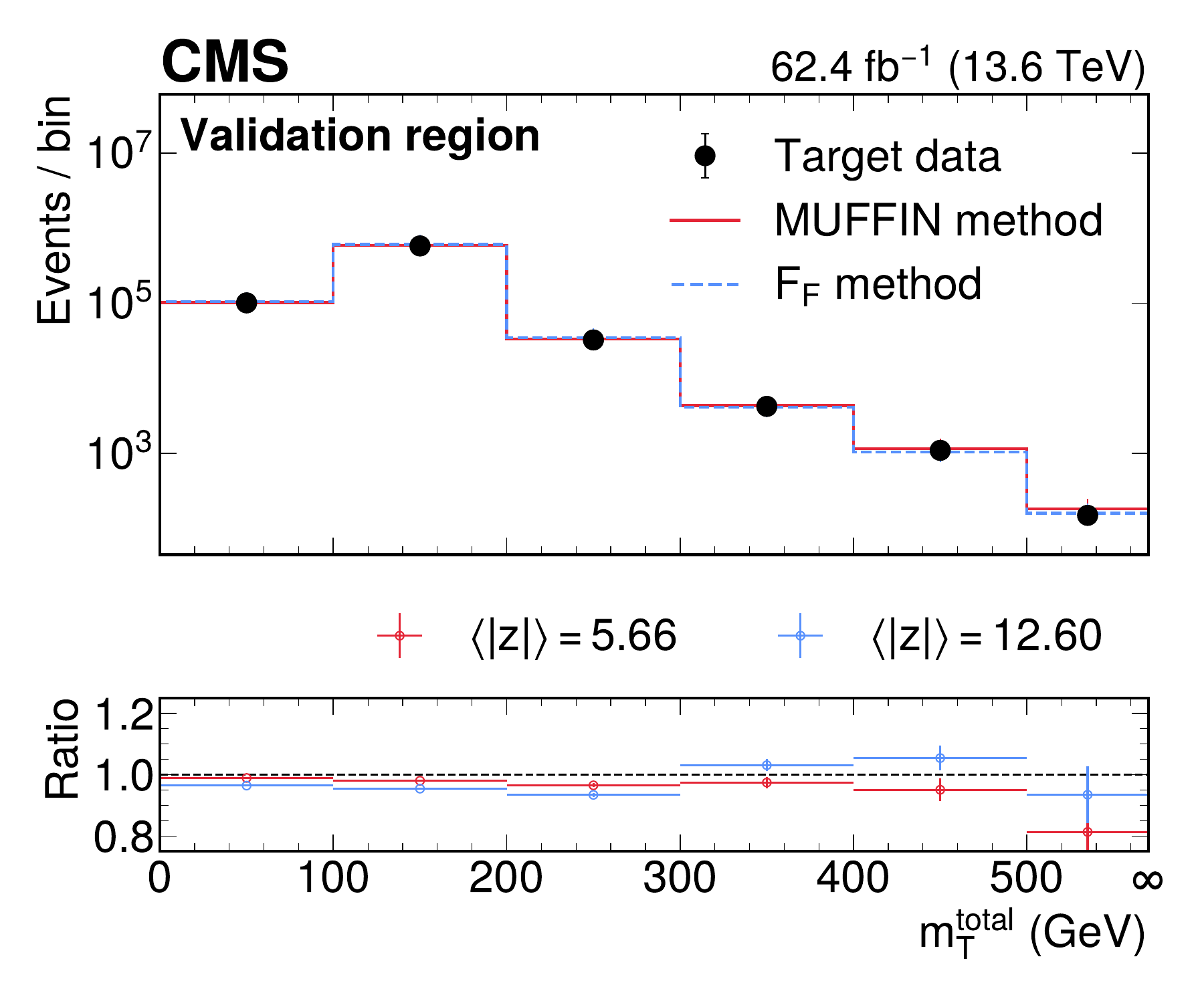}
    \includegraphics[width=\cmsFigWidth]{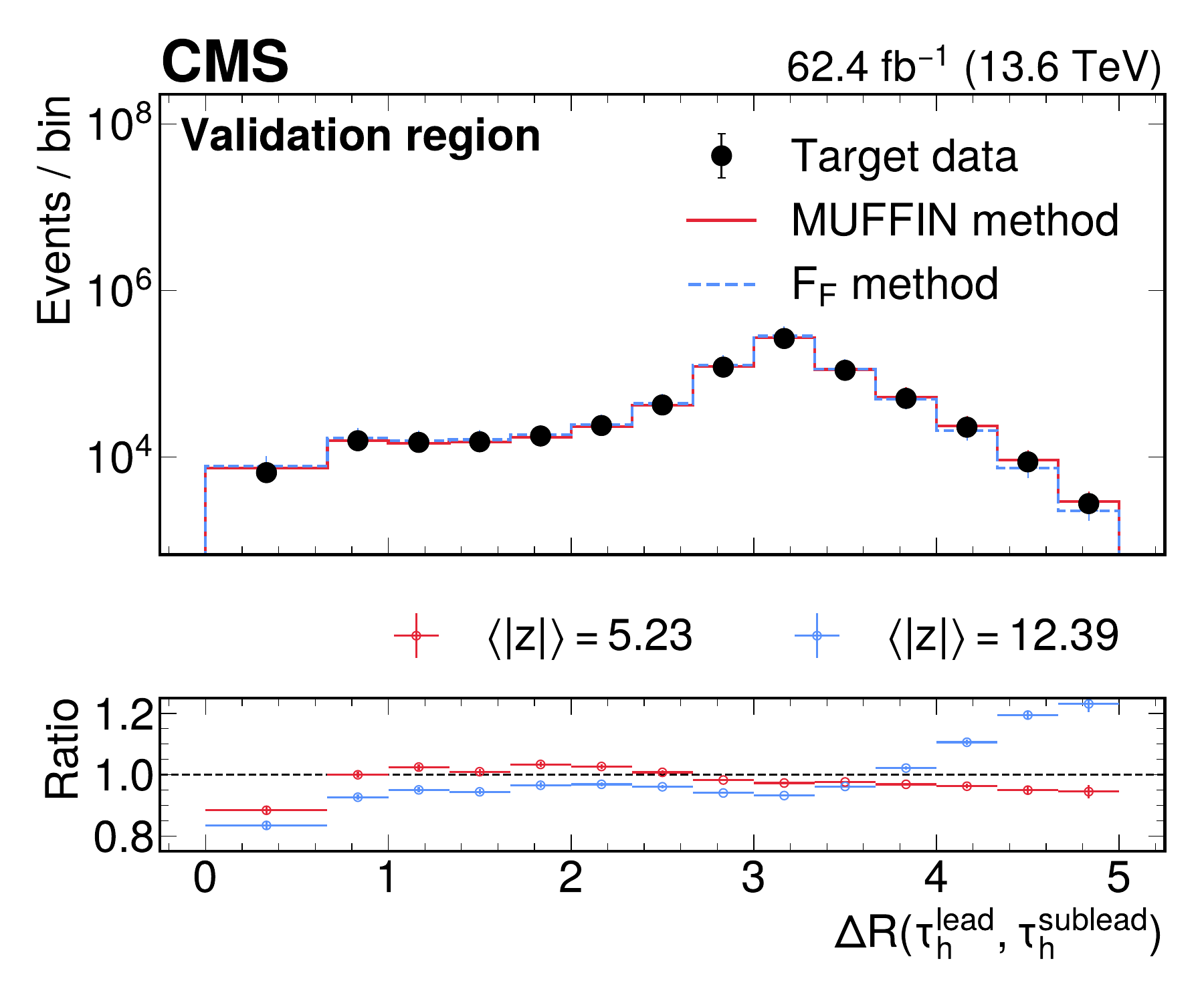}
    \includegraphics[width=\cmsFigWidth]{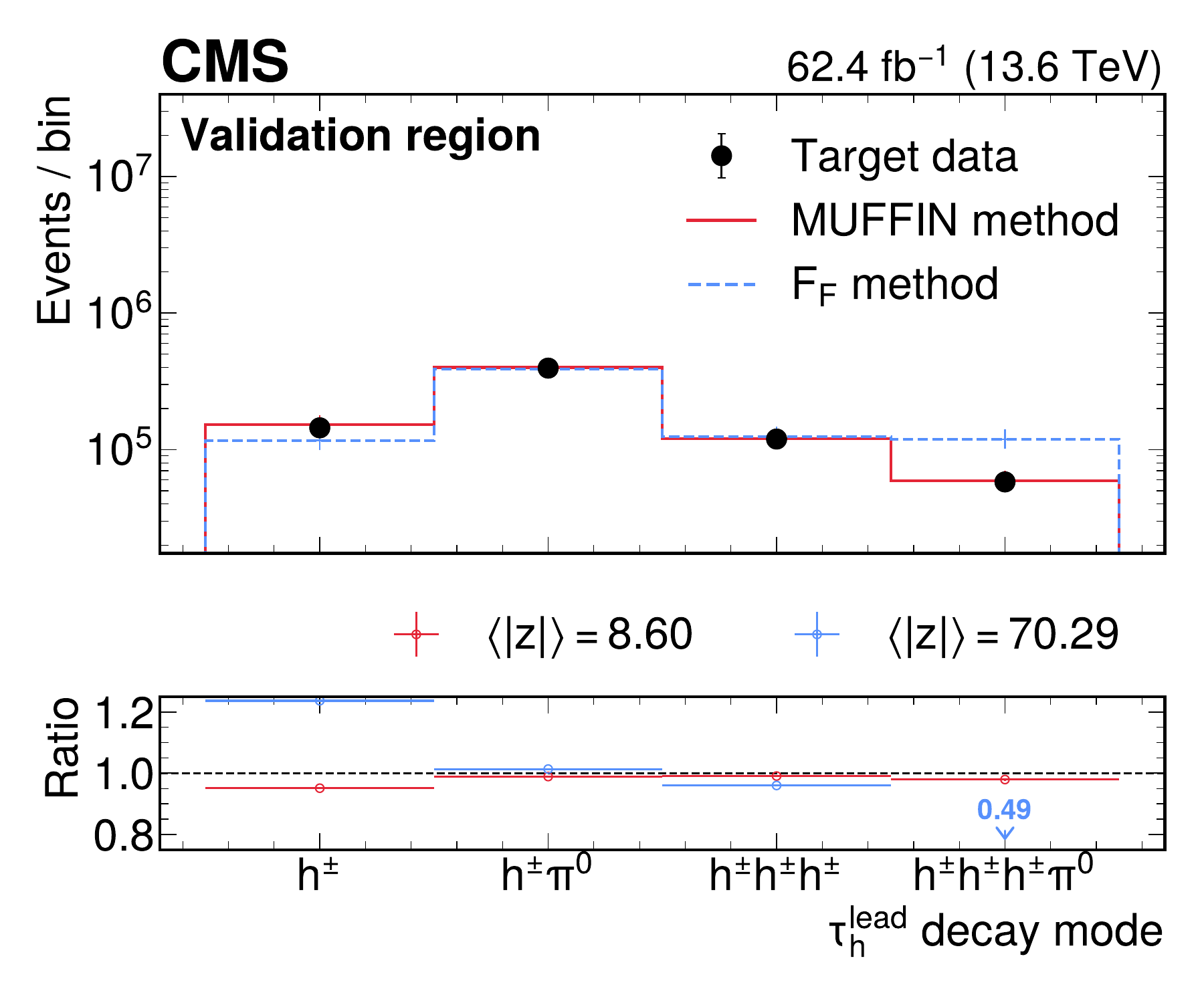}
  \caption{Distributions of representative observables in the \tautau AR comparing the MUFFIN prediction (red) and  \FF estimate (blue, dashed) to the target distribution (black markers) for \pttauh, $\mT^{\text{total}}$, $\Delta R$, and decay mode. Vertical bars on the points indicate the statistical uncertainty. The ratio panels display the target-to-prediction ratio, with the corresponding $\langle\abs{z}\rangle$ values quoted in each panel. Arrows indicate bins outside the plotted range. Overflow is included in the final bin ($\infty$), where applicable.}
  \label{fig:tautau_ar}
\end{figure}

\begin{figure}[htbp]
  \centering
    \includegraphics[width=\cmsFigWidth]{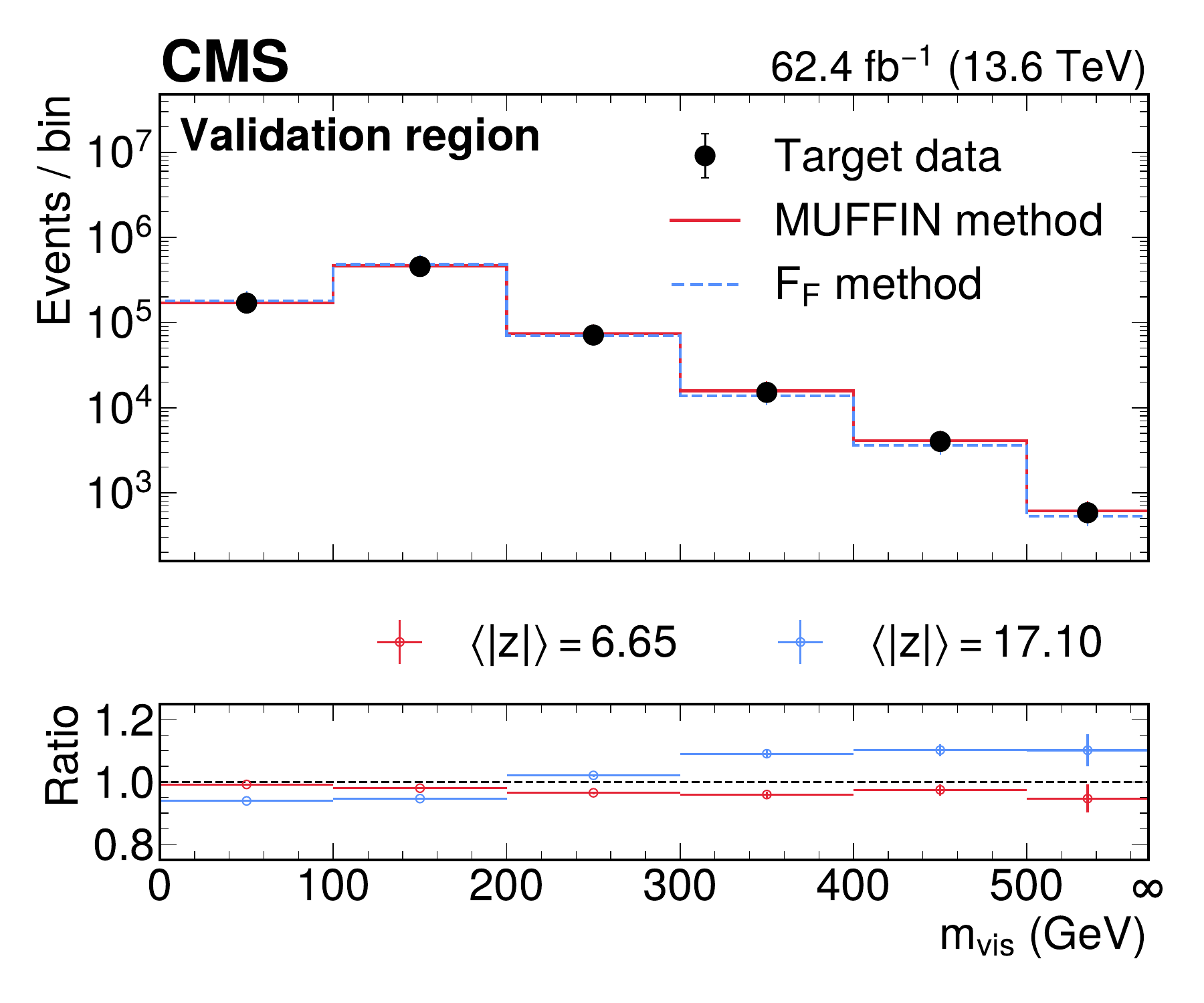}
    \includegraphics[width=\cmsFigWidth]{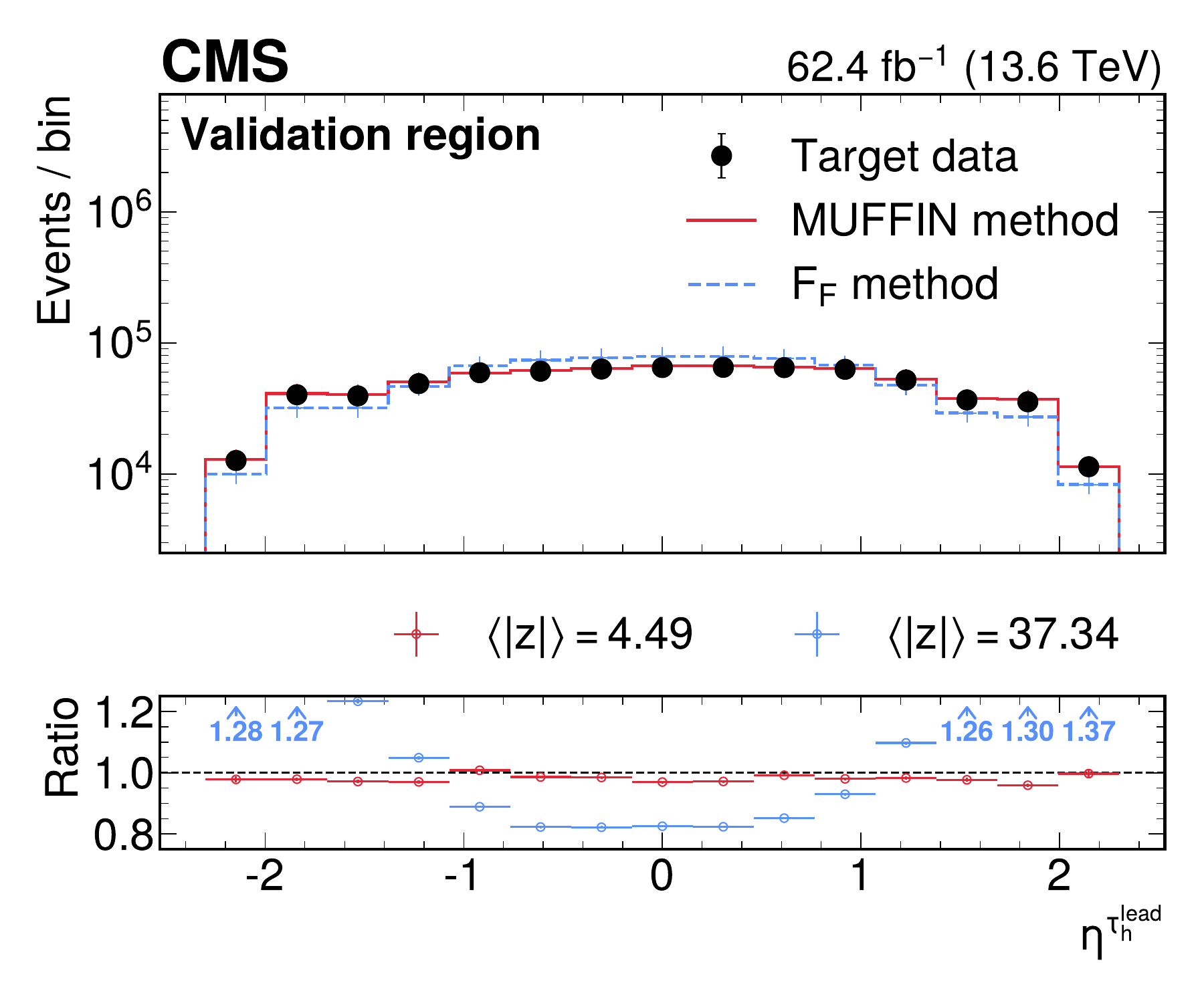}
  \caption{Comparison of the visible mass \mvis and $\eta$ distributions in the same \tautau AR with the MUFFIN (red) and \FF (blue, dashed) predictions. Vertical bars on the points indicate the statistical uncertainty. The ratio panels highlight residual deviations and the corresponding $\langle\abs{z}\rangle$ metric values. Arrows indicate bins outside the plotted range. Overflow is included in the final bin ($\infty$), where applicable.}
  \label{fig:tautau_ar_extra}
\end{figure}

The performance of the MUFFIN method in the \tautau channel is evaluated using the dedicated VR defined following Section~\ref{sec:bdt_ff_dr_ar}. In this channel, the \jtf background arises primarily from the QCD multijet process, with smaller contributions from \Wjets and \ttbar production. Figures~\ref{fig:tautau_ar} and~\ref{fig:tautau_ar_extra} show representative comparisons between distributions obtained with the MUFFIN and \FF methods for the key observables described above. The ratio panels include the $\langle\abs{z}\rangle$ metric values for each method, allowing a quantitative comparison of their performance.

The MUFFIN method is observed to significantly improve the modelling of all distributions compared to the \FF method. 
This demonstrates that the MUFFIN method provides improved modelling of the \jtf background in the \tautau channel, supporting its use in physics analyses targeting this final state.

\subsection{Performance in the \texorpdfstring{\mutau}{mutau}  channel}

The performance of the MUFFIN method is also evaluated in the \mutau channel using dedicated VRs defined following Section~\ref{sec:bdt_ff_dr_ar} and simulated \Wjets and \ttbar events. 
In this channel, the \jtf background modelling is assessed for QCD multijet, \Wjets, and \ttbar production modes in turn. 

\begin{figure}[ht!]
  \centering
    \includegraphics[width=\cmsFigWidth]{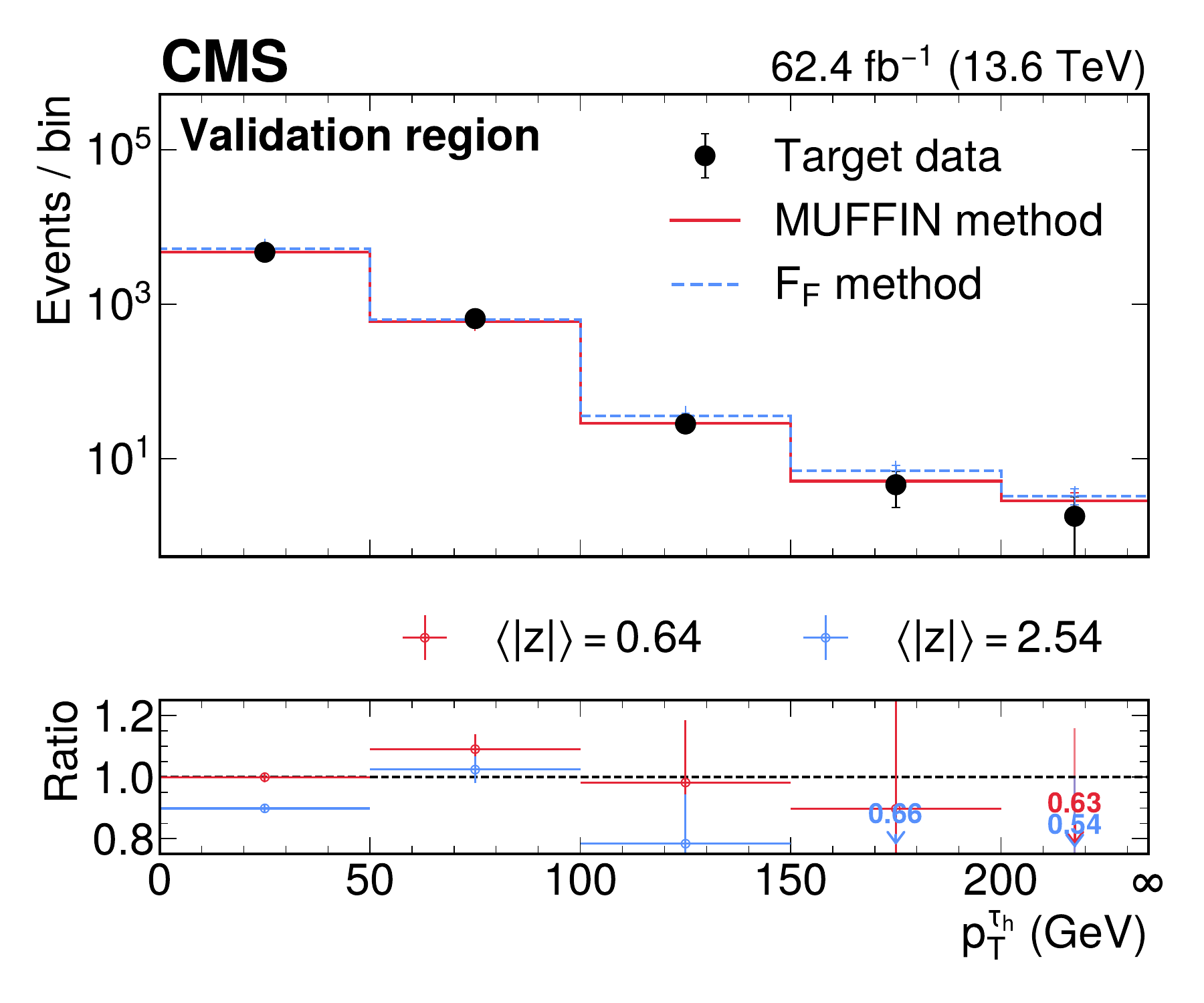}
    \includegraphics[width=\cmsFigWidth]{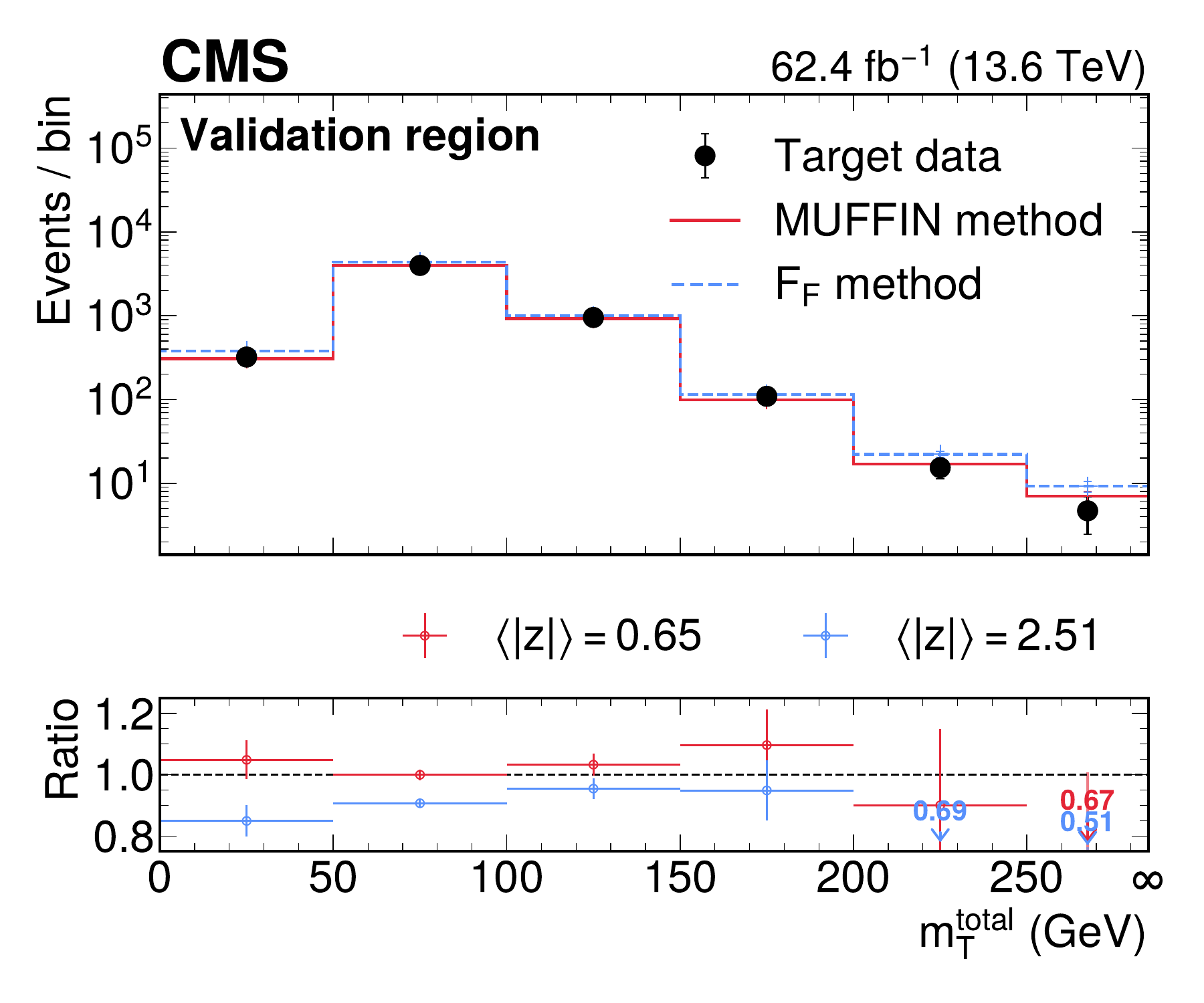}
    \includegraphics[width=\cmsFigWidth]{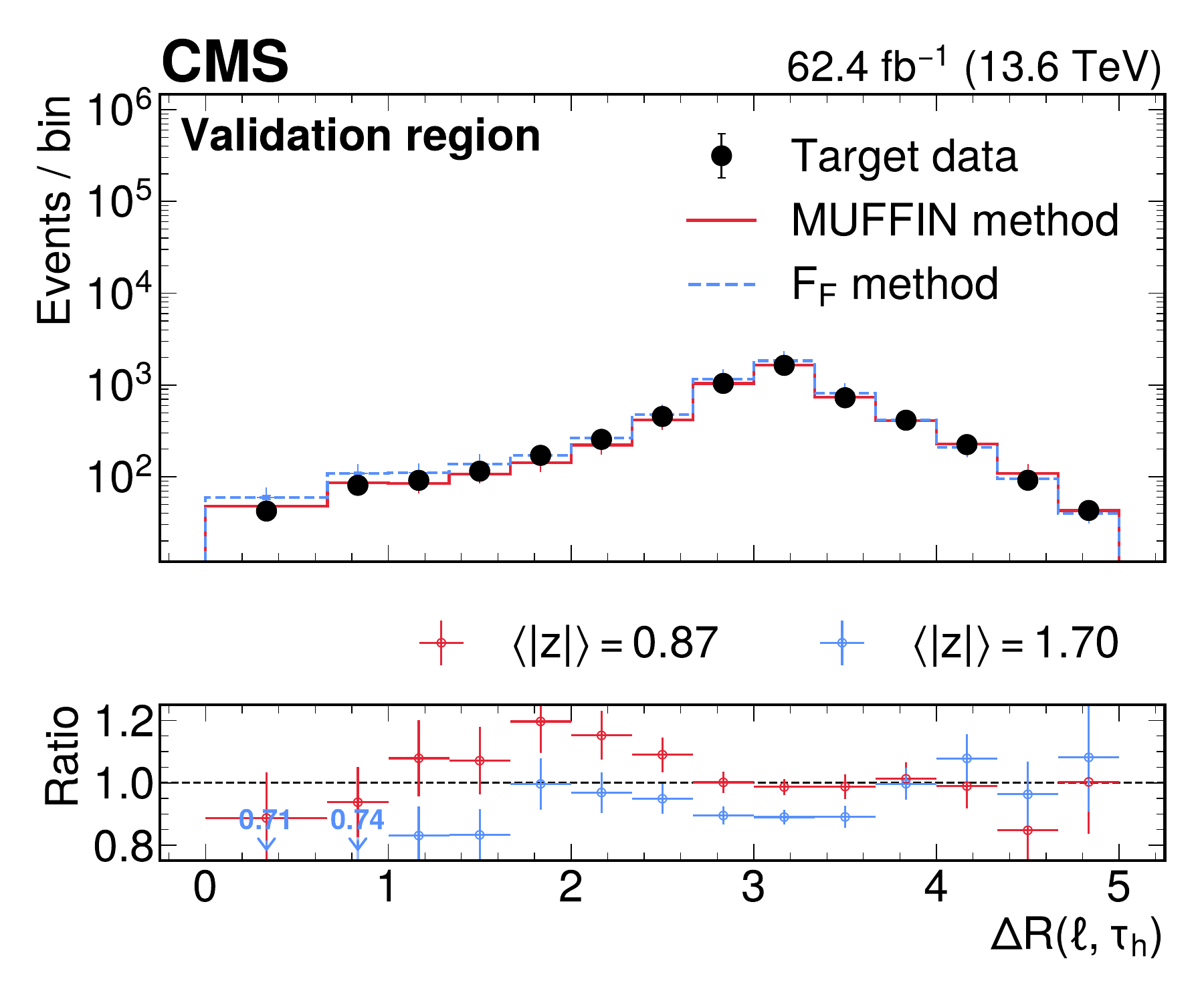}
    \includegraphics[width=\cmsFigWidth]{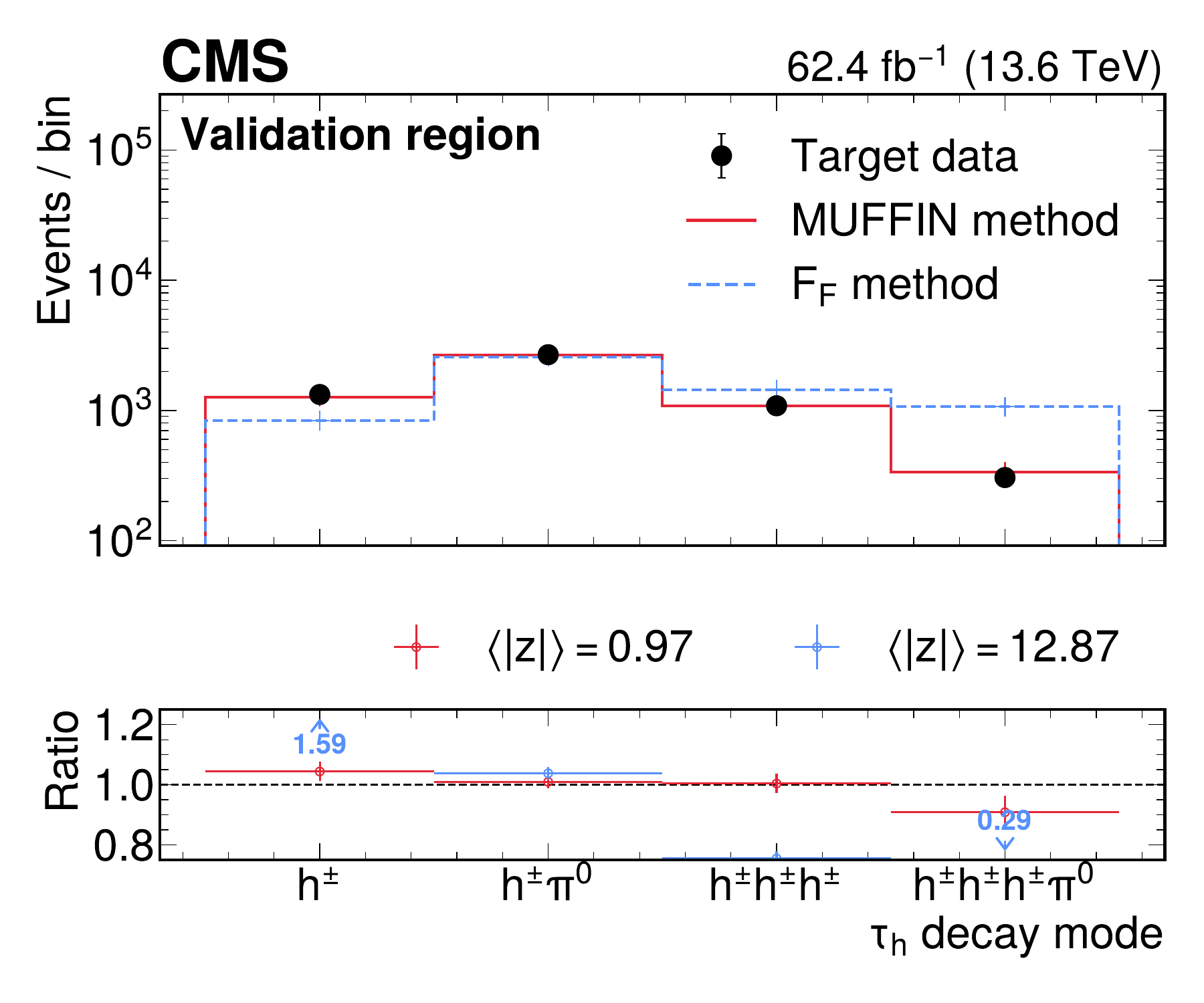}
  \caption{Validation of the QCD multijet background modelling in the \mutau VR. Predictions from MUFFIN (red) and \FF (blue, dashed) are compared to data for \pttauh, $\mT^{\text{total}}$, $\Delta R$, and \tauh decay mode. Vertical bars on the points indicate the statistical uncertainty. The lower panels show the data-to-prediction ratio, with $\langle\abs{z}\rangle$ values indicating the overall level of agreement. Arrows indicate bins outside the plotted range. Overflow is included in the final bin ($\infty$), where applicable.}
  \label{fig:mutau_QCD_ar}
\end{figure}

\begin{figure}[ht!]
  \centering
    \includegraphics[width=\cmsFigWidth]{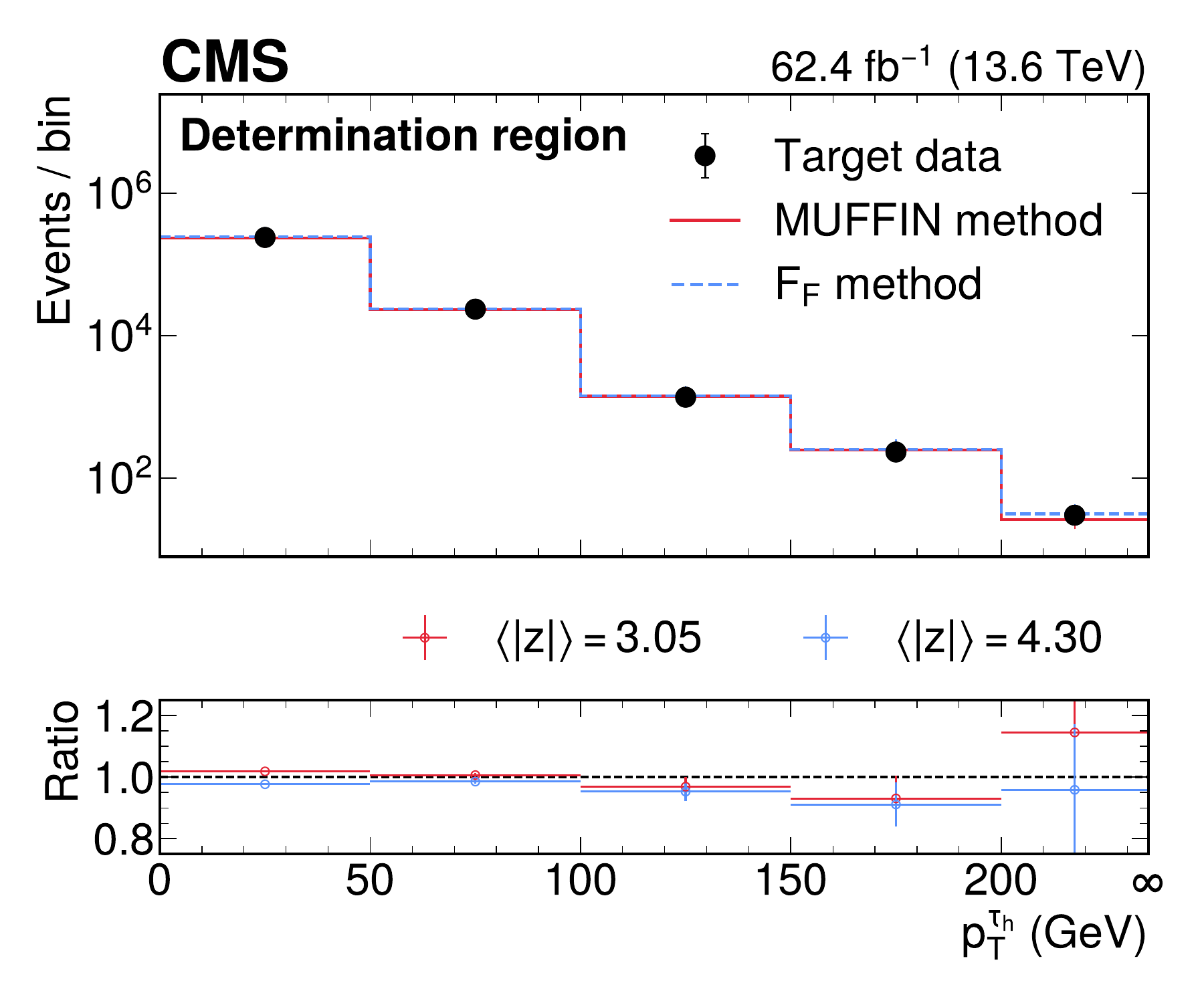}
    \includegraphics[width=\cmsFigWidth]{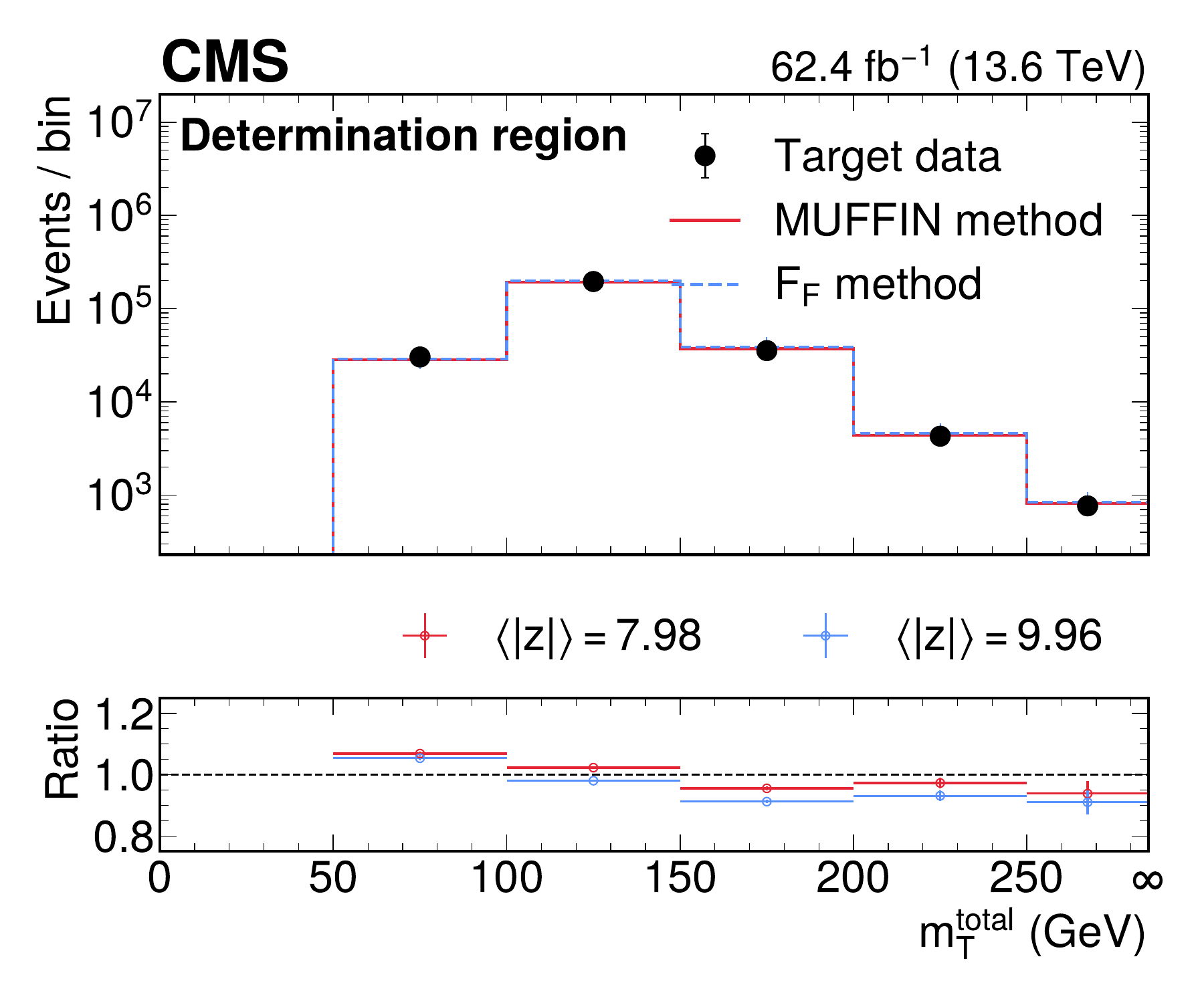}
    \includegraphics[width=\cmsFigWidth]{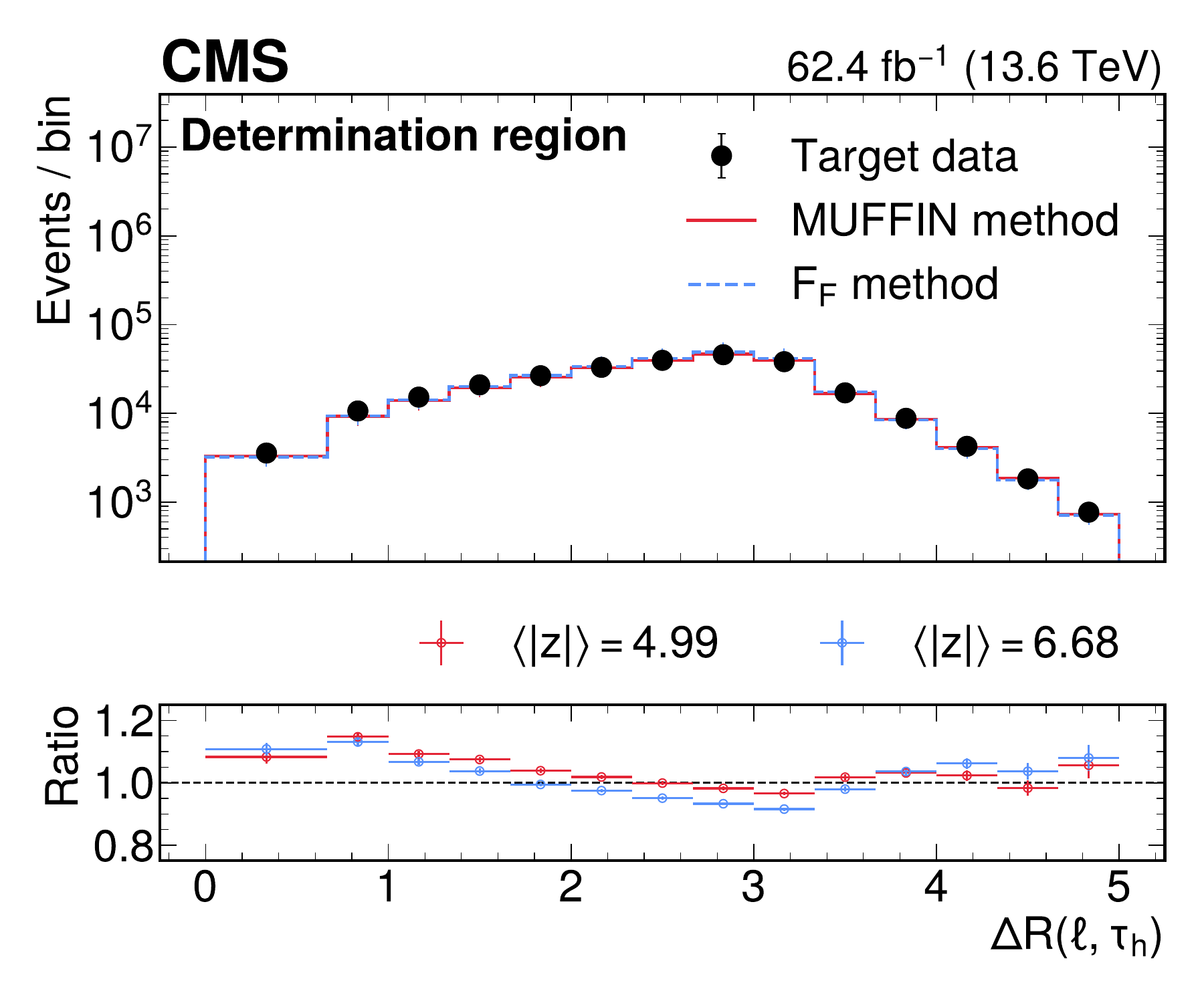}
    \includegraphics[width=\cmsFigWidth]{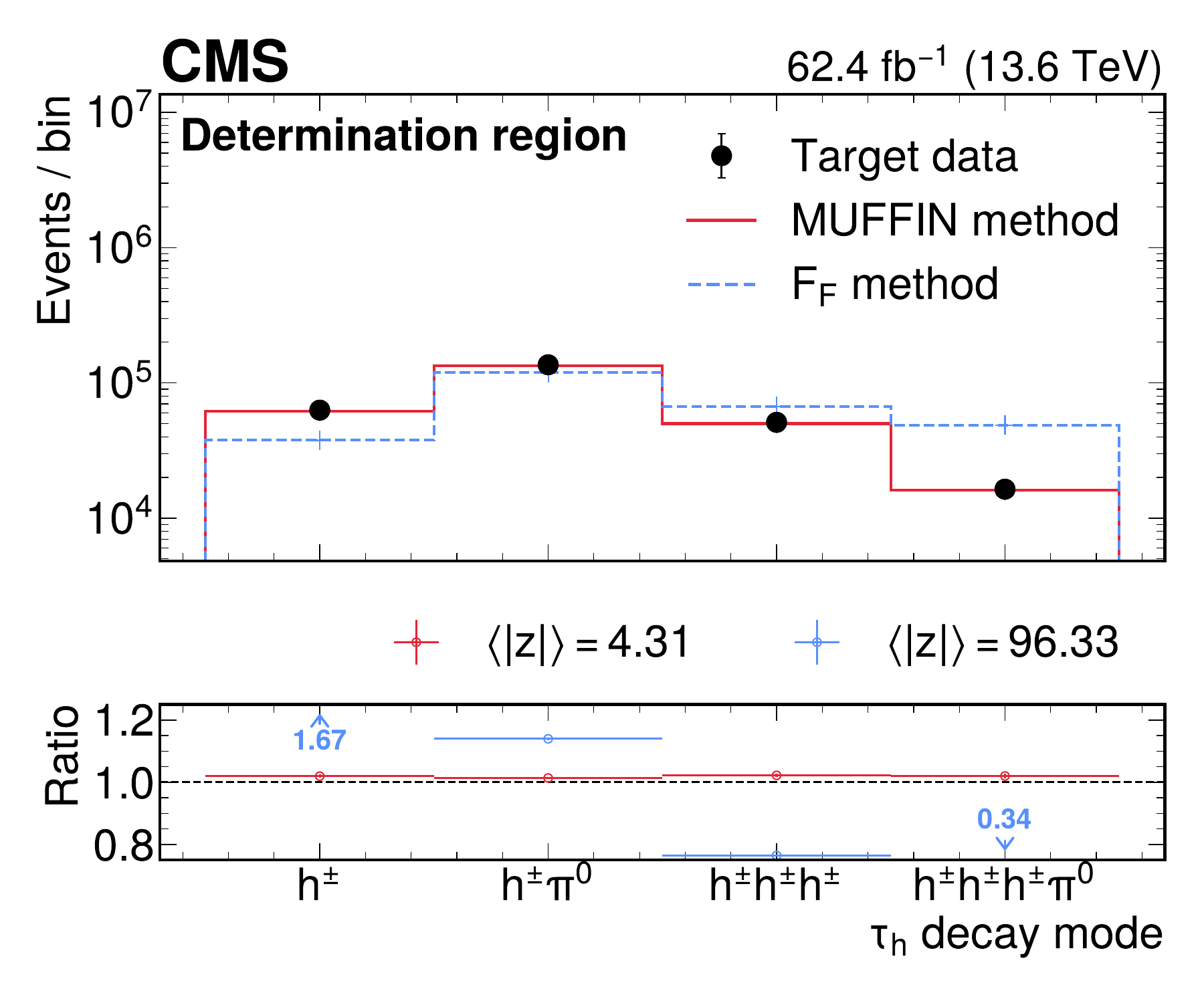}
  \caption{Comparison of \pttauh, $\mT^{\text{total}}$, $\Delta R$, and \tauh decay mode in the \Wjets DR of the \mutau channel, showing MUFFIN (red) and \FF (blue, dashed). Vertical bars on the points indicate the statistical uncertainty. The ratio panels report the $\langle\abs{z}\rangle$ values for each method; arrows mark bins outside the plotted range. Overflow is included in the final bin ($\infty$), where applicable.}
  \label{fig:mutau_Wjets_dr}
\end{figure}

\begin{figure}[ht!]
  \centering
    \includegraphics[width=\cmsFigWidth]{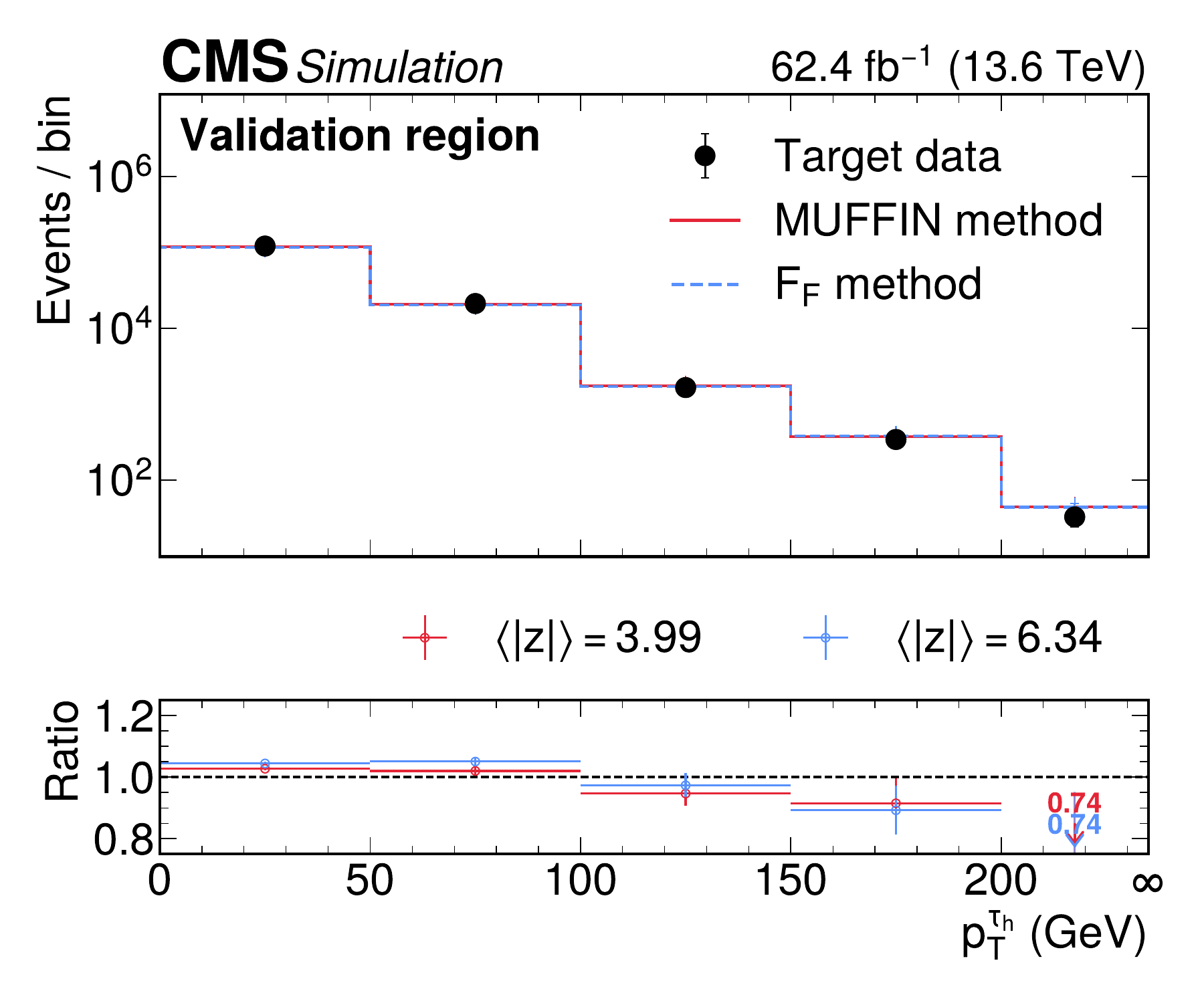}
    \includegraphics[width=\cmsFigWidth]{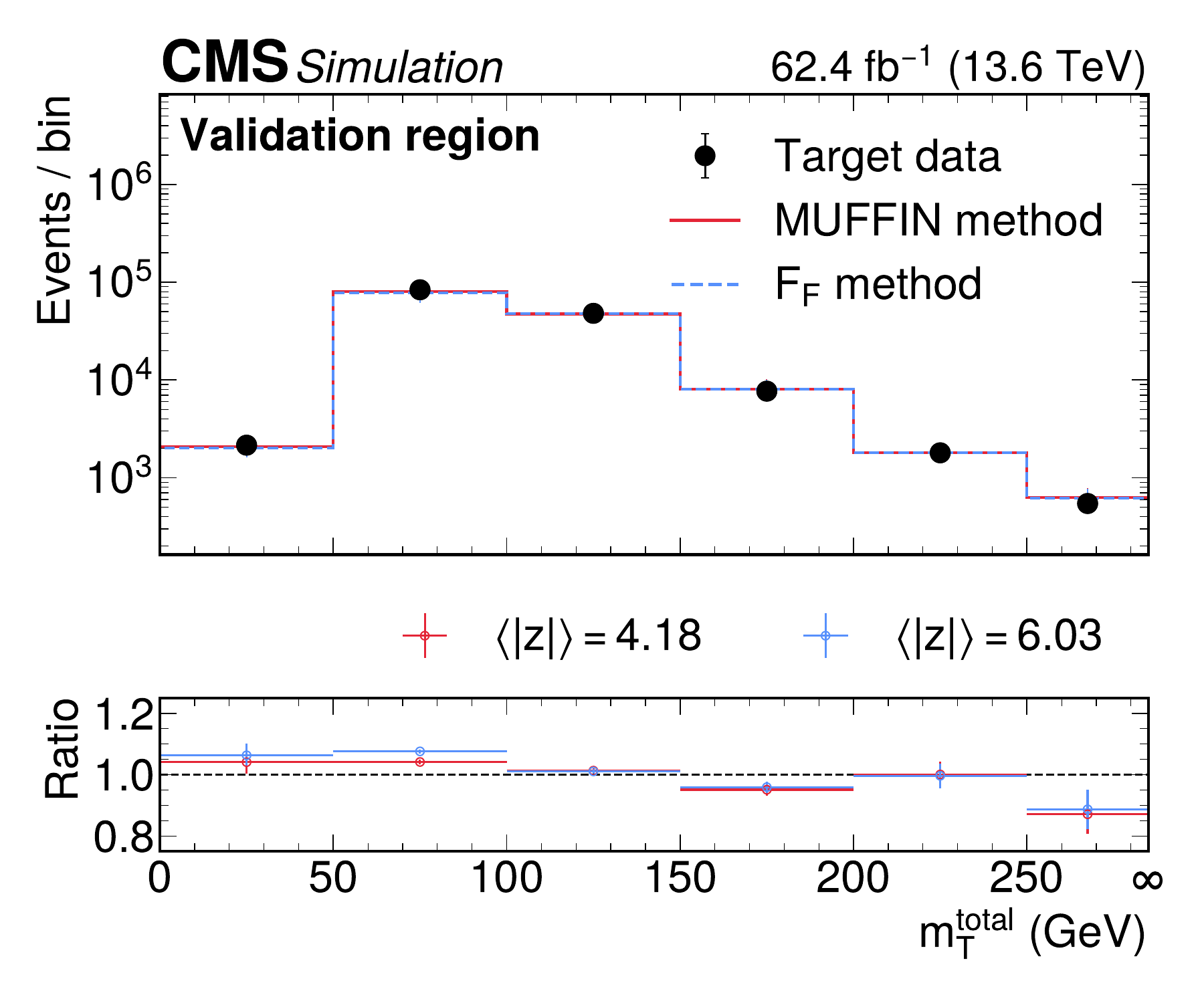}
    \includegraphics[width=\cmsFigWidth]{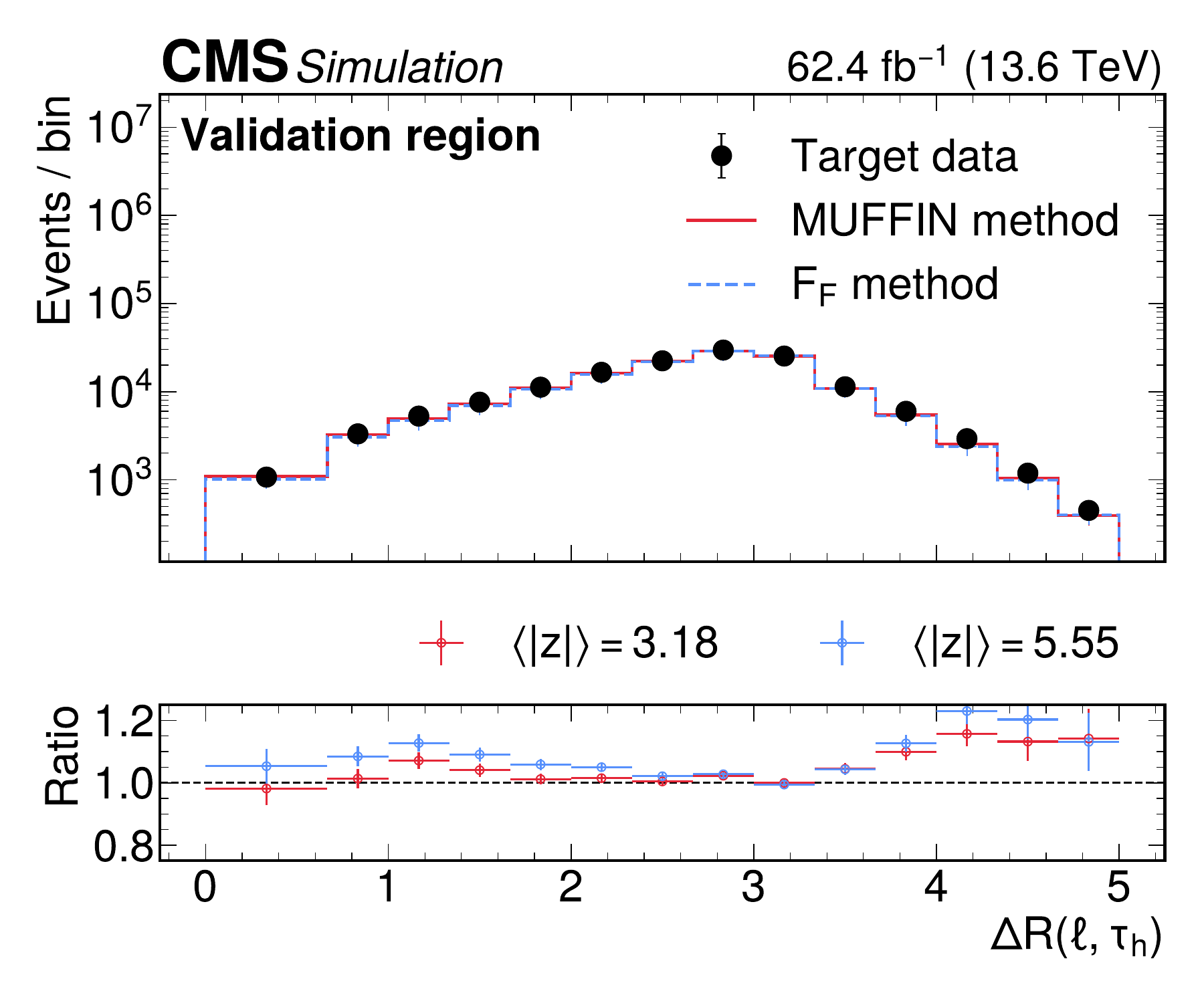}
    \includegraphics[width=\cmsFigWidth]{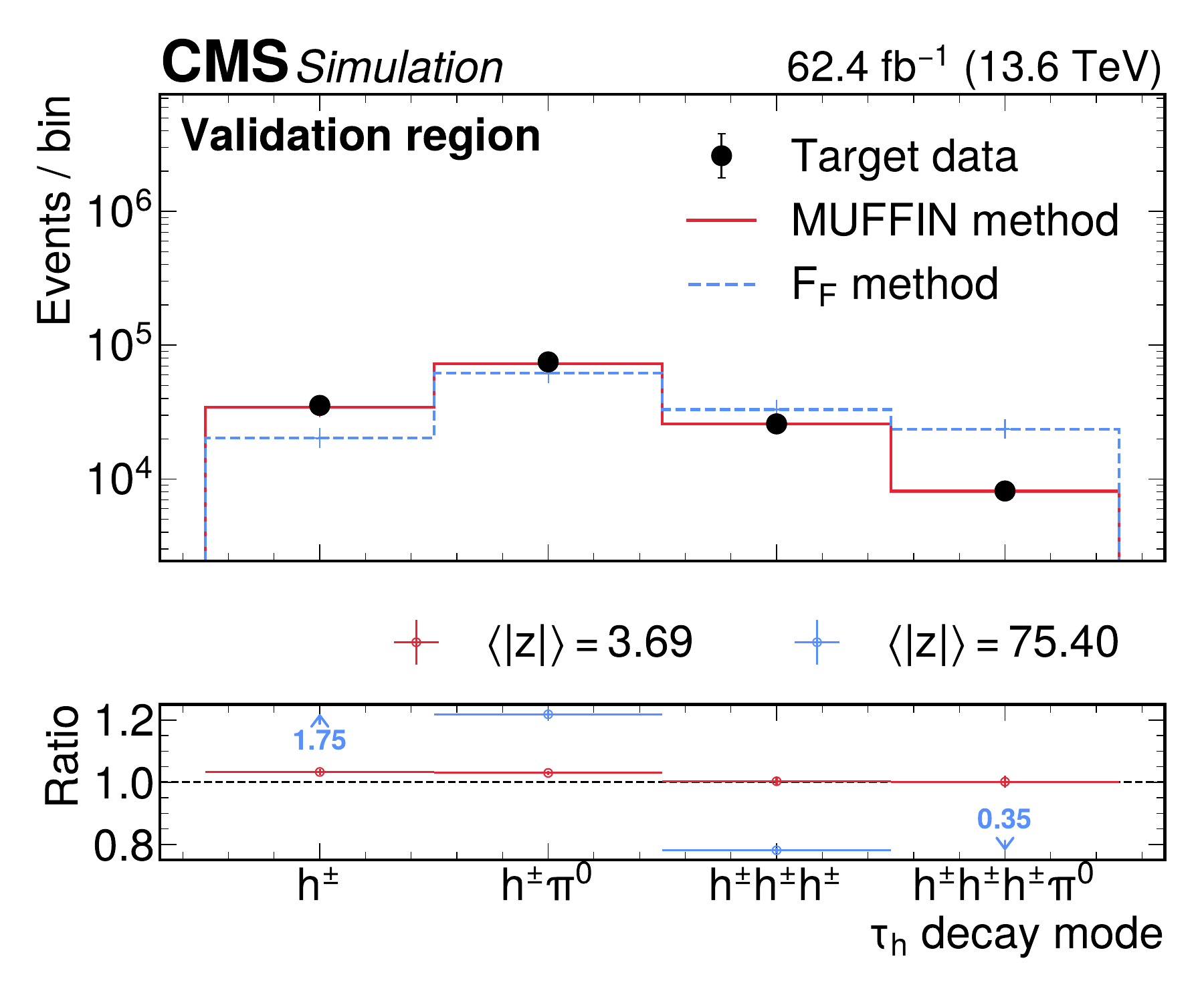}
  \caption{Comparison of \pttauh, $\mT^{\text{total}}$, $\Delta R$, and \tauh decay mode for MUFFIN (red) and \FF (blue, dashed) using simulated \Wjets events. Vertical bars on the points indicate the statistical uncertainty. Ratio panels show agreement with respect to the simulated target and report the corresponding $\langle\abs{z}\rangle$ values. Arrows indicate bins outside the plotted range. Overflow is included in the final bin ($\infty$), where applicable.}
  \label{fig:mutau_WjetsMC_ar}
\end{figure}

\begin{figure}[htbp]
  \centering
    \includegraphics[width=\cmsFigWidth]{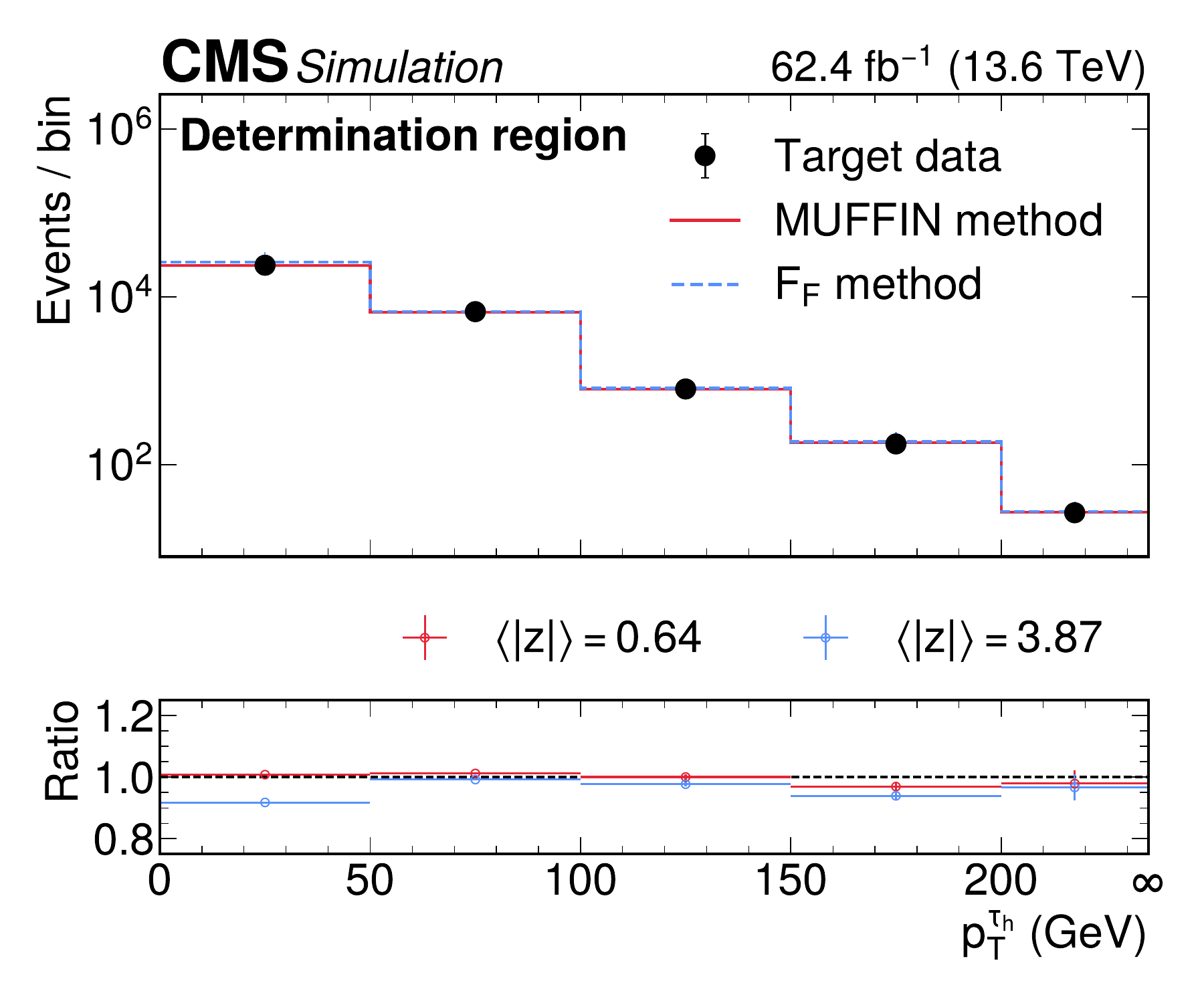}
    \includegraphics[width=\cmsFigWidth]{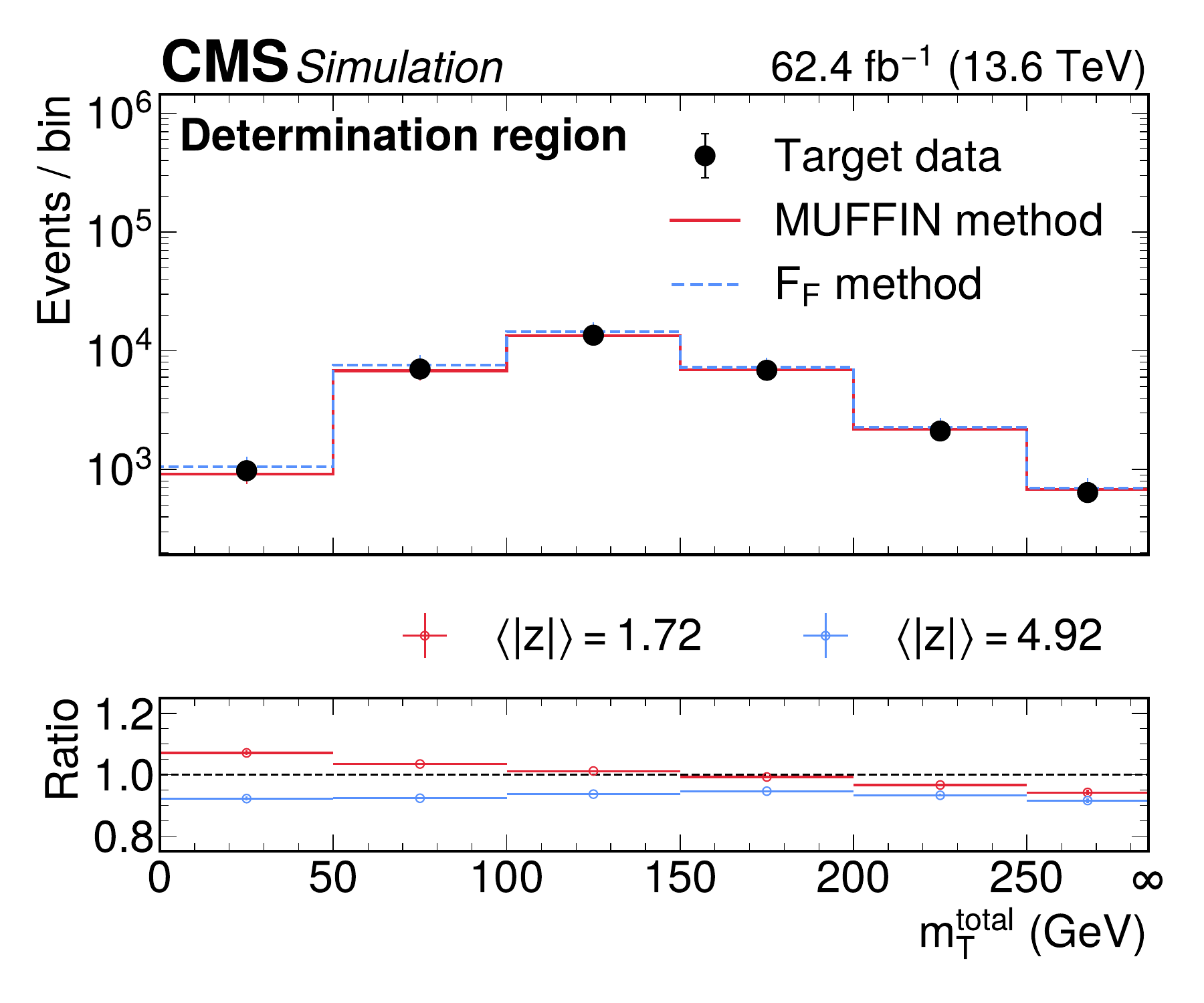}
    \includegraphics[width=\cmsFigWidth]{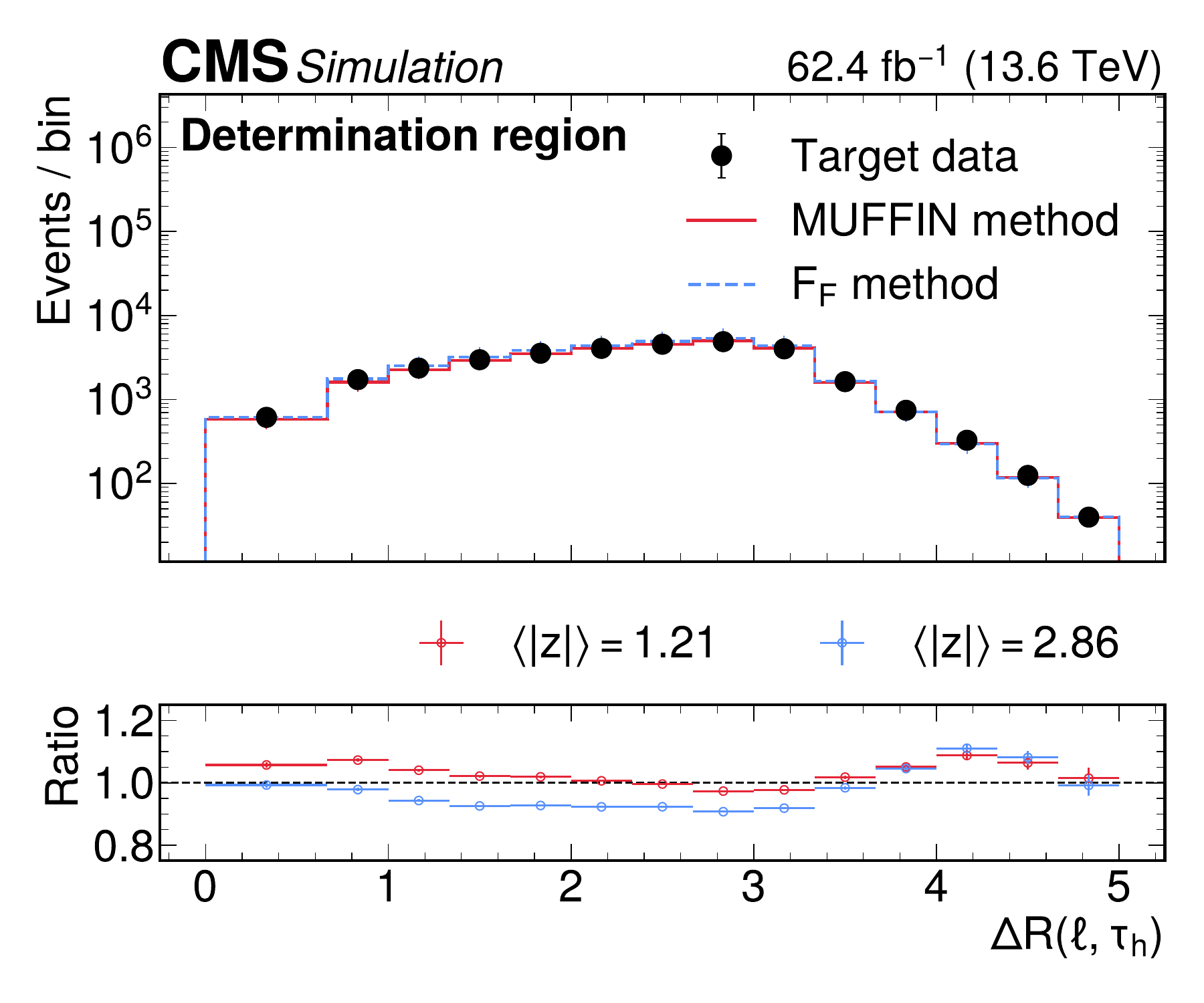}
    \includegraphics[width=\cmsFigWidth]{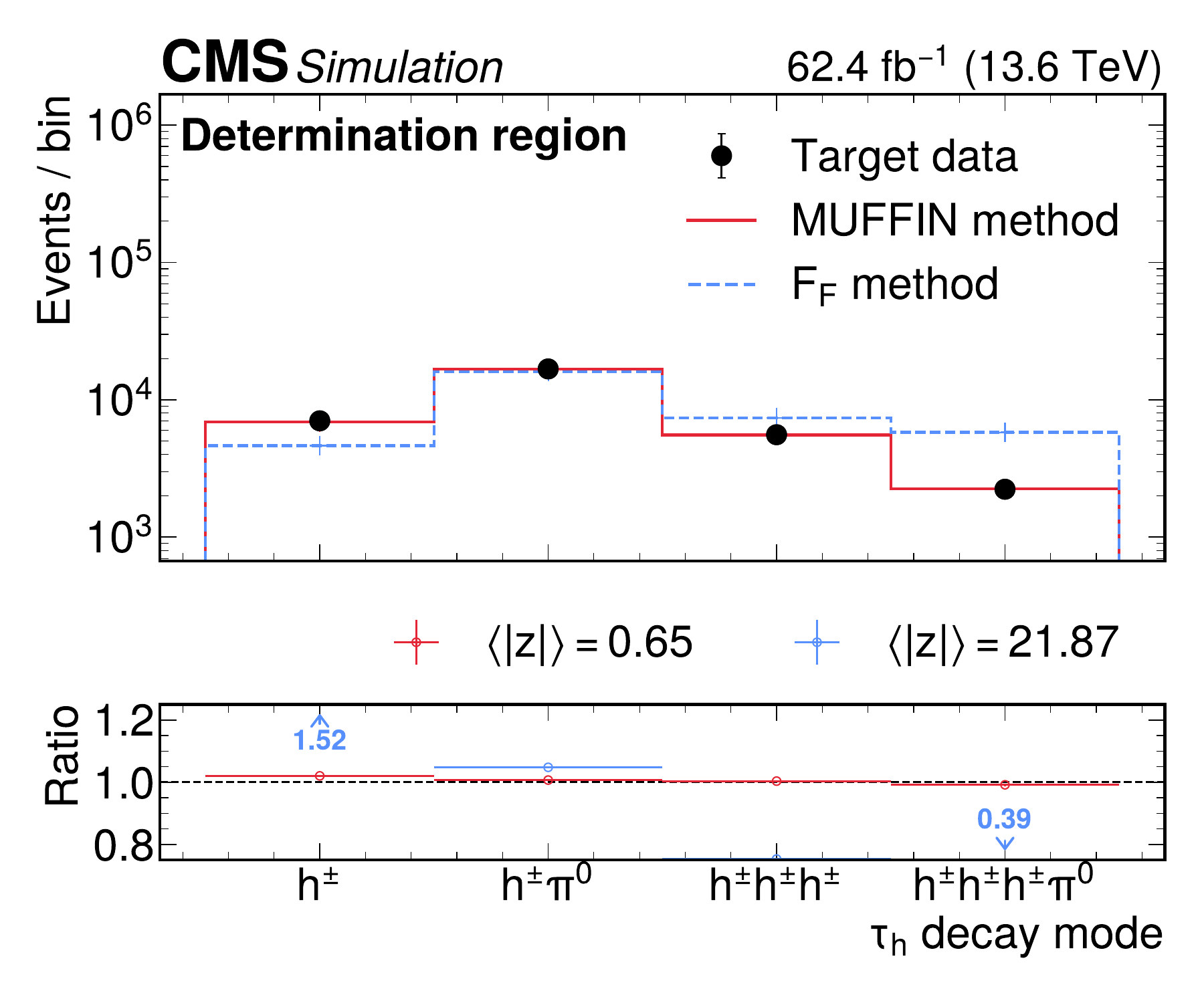}
  \caption{Comparison of \pttauh, $\mT^{\text{total}}$, $\Delta R$, and \tauh decay mode for MUFFIN (red) and \FF (blue, dashed) in simulated
   \ttbar for the \mutau channel. Vertical bars on the points indicate the statistical uncertainty. The ratio panels show agreement with respect to the simulated target and report the corresponding $\langle\abs{z}\rangle$ values. Arrows indicate bins outside the plotted range. Overflow is included in the final bin ($\infty$), where applicable.}
  \label{fig:mutau_ttbar_dr}
\end{figure}

Figures~\ref{fig:mutau_QCD_ar}--\ref{fig:mutau_ttbar_dr} show representative comparisons between distributions obtained with the MUFFIN and \FF methods for the key observables described at the beginning of this section. For brevity, the $\eta$ and \mvis distributions are not shown here, but similar improvements with the MUFFIN method are observed as in the \tautau channel.

Figure~\ref{fig:mutau_QCD_ar} shows comparisons between distributions obtained with the MUFFIN and \FF methods in a representative VR for the QCD multijet background. 
The MUFFIN method provides improved modelling across a wide range of observables, particularly in the $\Delta R$ and decay mode distributions, with the remaining observables also showing improved modelling with the MUFFIN method.

Figure~\ref{fig:mutau_Wjets_dr} shows comparisons between distributions obtained with the MUFFIN and \FF methods in the \Wjets DR. This contains the same events that are used to train MUFFIN and derive the \FF TFs, so good agreement is expected.
Nevertheless, the MUFFIN method again provides improved modelling across all observables considered.
Figure~\ref{fig:mutau_WjetsMC_ar} shows additional comparisons between distributions obtained in a dedicated simulation-only validation procedure for the \Wjets background, defined by applying the methods trained in the high-\mT DR to simulated events with low \mT.
The MUFFIN method again demonstrates improved modelling across all observables. 

Finally, Fig.~\ref{fig:mutau_ttbar_dr} shows comparisons between distributions obtained with the MUFFIN and \FF methods in simulated \ttbar events. 
The MUFFIN method provides improved modelling across all observables considered.

These results demonstrate that the MUFFIN method provides enhanced modelling of the \jtf background in the \mutau channel as well, supporting its broader applicability across different analysis channels. 
This is particularly evident at the high end of the \tauh \pt spectrum, where the MUFFIN method better captures the kinematic dependencies of the misidentification rates, leading to improved background estimates in high-\pt regions critical for new physics searches, as further demonstrated in Section~\ref{sec:physics_impact}. 
This resonates with the findings in the \tautau channel, highlighting the robustness and versatility of the MUFFIN method in addressing the challenges associated with \jtf background estimation, especially in kinematic regions with limited event yields.
A residual shape dependence remains in the $\Delta R$ distributions in the \mutau channel, more pronounced than in the \tautau channel and qualitatively similar to that seen with the \FF method. As $\Delta R$ is deliberately excluded from the BDT training to serve as an independent closure test (Section~\ref{sec:systematic-uncertainties}), this points to an additional phase space dependence that could be captured via the addition of a suitable variable to an extended input variable list for the MUFFIN method.

\section{Uncertainty model}\label{sec:uncertainties}

The uncertainties associated with the MUFFIN method are categorised into statistical and systematic components. The uncertainty model is designed to provide a systematic treatment of the dominant sources of variation affecting the BDT-based reweighting procedure and its application from determination to analysis phase space.

\subsection{Statistical uncertainties} \label{sec:statistical-uncertainties}

The statistical uncertainty of the MUFFIN method arises primarily from the finite size of the training data set used to derive the BDT reweighting factors. To propagate this uncertainty to the final background prediction, an ensemble approach based on bootstrap resampling is employed. An ensemble of $50$ BDT models is trained using independent bootstrapped replicas of the training data. This choice provides a practical balance between computational cost and the stability of the bootstrap-based uncertainty estimate, for which further increases in the number of replicas were found to have only a small impact on the resulting uncertainty.
For each model, the full reweighting procedure is applied, yielding a set of alternative predictions. The spread of these predictions is interpreted as the statistical uncertainty associated with the reweighting factors and is propagated to all observables used in the analysis. 

\subsection{Systematic uncertainties} \label{sec:systematic-uncertainties}

Systematic uncertainties associated with the MUFFIN method are assigned to account for potential mismodelling in the BDT reweighting procedure and its extrapolation from DR to SR. These uncertainties are grouped into four categories, corresponding to distinct sources of potential bias.

\subsubsection{Non-closure uncertainty}
Residual mismodelling can arise in observables that are not explicitly constrained by the BDT training. These include variables orthogonal to the input feature set, such as $\Delta R$ and \ptmiss, which are not included as training inputs and therefore provide an independent test of the modelling performance. The non-closure is evaluated using Eq.~(\ref{eq:stress-nonclosure}) as the bin-by-bin ratio of observed to predicted yields in the VR. A flat uncertainty of 5\% is assigned based on the typical level of residual non-closure observed in these orthogonal observables. This uncertainty covers mismodelling in variables not seen during training and potential non-closure in trained variables when extrapolating from the DR to the SR phase space is separately addressed by the extrapolation uncertainty described in Section~\ref{sec:extrapolation-uncertainty}. This uncertainty is significantly smaller than the non-closures of up to 20\% or more that can be observed with the \FF method, which often necessitate explicit closure corrections.

\subsubsection{Extrapolation uncertainty} \label{sec:extrapolation-uncertainty}
The BDT model is trained in DRs enriched in misidentified jets and subsequently applied in VRs. Differences in phase space between these regions can introduce residual biases
that are not fully constrained by the training. To quantify this effect, the level of non-closure in a key analysis observable is evaluated, as described in Section~\ref{sec:stress-tests}. 
For QCD-enriched DRs such as those considered in Section~\ref{sec:stress-tests}, the resulting uncertainty was found to remain within approximately 10\%, and is significantly smaller over most of the accessible phase space, particularly in regions dominated by non-\jtf backgrounds. Since this uncertainty is derived from the non-closure evaluated in a VR, the VR must be chosen to adequately capture the phase space difference between the DR and the SR, so that the resulting non-closure provides a reliable estimate of the extrapolation uncertainty. Therefore, both the choice of VR and the resulting magnitude of the uncertainty are analysis dependent.

\subsubsection{Modelling uncertainty}
Potential biases related to the ML model itself are assessed by varying the training configuration. These variations include changes to the set of input features, hyperparameter choices, and the inclusion of different combinations of global event variables. For each variation, the full training and reweighting procedure is repeated, and the resulting differences in the predictions are propagated as a systematic uncertainty. 
This uncertainty is treated as dependent on the analysis channel and background process, reflecting the differing sensitivities of each category to modelling choices. The magnitude of this uncertainty is typically at the level of 2\% or less, indicating a robust performance of the BDT-based reweighting procedure against variations in the training configuration.

\subsubsection{Background subtraction uncertainty}
The subtraction of genuine-\tauh contributions in the DRs introduces an associated uncertainty on the estimated TFs. This is propagated by varying the subtracted backgrounds by $\pm20\%$ for the QCD multijet process and $\pm10\%$ for the \Wjets process, reflecting the typical uncertainty in the genuine-\tauh contamination in the respective DRs. The variation is applied in both the numerator and denominator of the weight formula, and the full reweighting procedure is repeated for each variation to propagate the effect to the final background estimate. In rare cases, statistical fluctuations in sparsely populated bins can yield negative TFs; the method remains well-defined in this limit; when the nominal TF is negative, the unsubtracted ratio $w_{\text{MUFFIN}}(\mathbf{z}) = ({P_{\text{pass}}^{\text{data}}(\mathbf{z})})/({P_{\text{fail}}^{\text{data}}(\mathbf{z})})$ is used as both the nominal prediction and the uncertainty bound, ensuring no negative event weights enter the final prediction. The bootstrap ensemble naturally captures the impact of such fluctuations on the final background prediction. This uncertainty is not applicable to simulation-based derivations (\Wjets and \ttbar), where there is no genuine-\tauh contamination.

Together, these uncertainty components provide a comprehensive treatment of the statistical and systematic effects associated with the MUFFIN method, ensuring robust coverage of residual mismodelling, extrapolation effects, and training-related variations. The total relative uncertainty in the \jtf background yield in representative inclusive SR selections for each channel is summarised in Table~\ref{tab:uncertainty_totals}. In both channels, the extrapolation uncertainty is the single largest source of systematic uncertainty, while all other components remain well below the 2\% level.

\begin{table}[htbp]
    \centering
    \topcaption{Total relative uncertainty in the \jtf background yield in representative inclusive SR selections for each channel.}
    \label{tab:uncertainty_totals}
    \begin{tabular}{lcc}
        \hline
        Channel & Dominant background & Total relative uncertainty \\
        \hline
        $\tautau$ & QCD multijet & 6\% \\
        $\mutau$     & QCD multijet, \Wjets, \ttbar & 5\% \\
        \hline
    \end{tabular}
\end{table}

\section{Performance in physics analyses}\label{sec:physics_impact}

\subsection{Application of the MUFFIN method to signal regions} \label{sec:signal_regions}
In this section we present the application of the MUFFIN method in physics analyses at CMS involving \tauh leptons. The results shown correspond to baseline SR selections as described in Section~\ref{sec:selections}, without additional categorisation beyond the nominal event selection. The MUFFIN method is used to estimate the dominant backgrounds arising from jets misidentified as \tauh candidates, encompassing contributions from QCD multijet, \Wjets, and \ttbar processes. In the \tautau channel, the method is applied to estimate the \jtf background from the QCD multijet process. 

\begin{figure}[htbp]
    \centering
    \includegraphics[width=\cmsFigWidth]{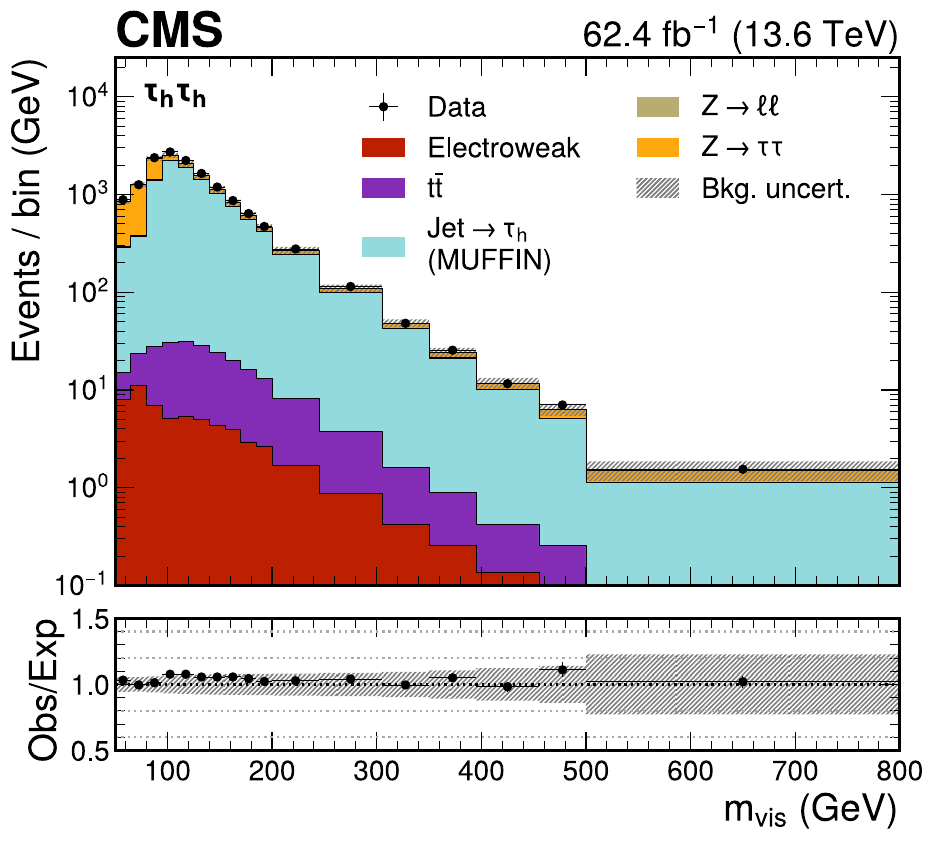}
    \includegraphics[width=\cmsFigWidth]{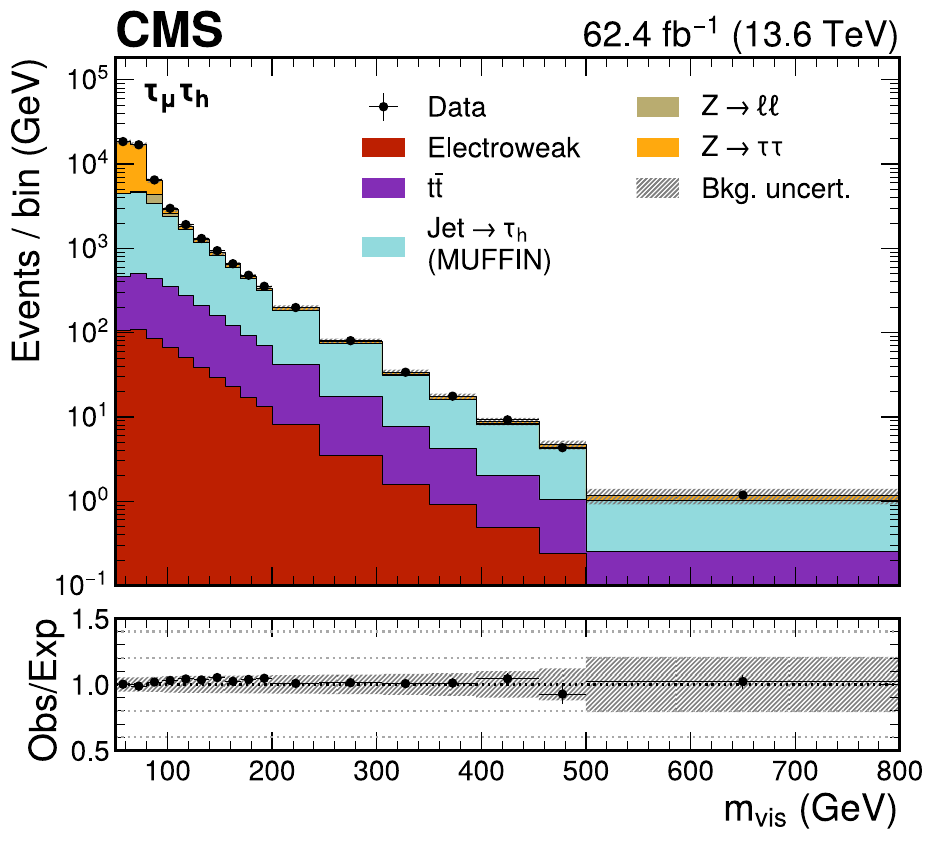}
    \caption{Distributions of the visible mass in the \tautau (left) and the \mutau (right) channels in the SRs, as defined in Section~\ref{sec:selections}, of physics analyses targeting Higgs boson decays to \PGt lepton pairs. The MUFFIN method is used to estimate the \jtf background contributions. Vertical bars represent the statistical uncertainty on the data points. The lower panels show the ratio of data to the total background prediction, with shaded bands representing the combined statistical and systematic uncertainties from the MUFFIN method.}
    \label{fig:physics_impact}
\end{figure}

In the \mutau channel, the method is applied to estimate the combined \jtf background from all relevant processes. This is done by calculating process-specific reweighting factors and combining them according to their relative contributions in the respective analysis categories. The resulting \mvis distributions in the \tautau and \mutau channels are shown for the SRs in Fig.~\ref{fig:physics_impact}, where the MUFFIN-based prediction is found to be in good agreement with the observed data; the corresponding improvement relative to the \FF method is demonstrated in Sections~\ref{sec:performance} and~\ref{sec:stress-tests}.

When the MUFFIN method is applied to SRs, agreement within uncertainties between the predicted background and the observed data is found in the \mvis distributions. A $\chi^2$ goodness-of-fit test~\cite{Cowan:1998}, computed using statistical uncertainties only, yields $\chi^2/\mathrm{dof} = 11.42/17$ ($p = 0.834$) in the \tautau channel and $\chi^2/\mathrm{dof} = 7.14/17$ ($p = 0.982$) in the \mutau channel, where $\mathrm{dof}$ is the number of degrees of freedom, confirming compatibility between the observed data and the prediction. The non-closure systematic uncertainties derived from the VRs, as quantified in Sections~\ref{sec:stress-tests} and~\ref{sec:uncertainties}, are excluded from the $\chi^2$ denominator to avoid circularity, and are shown separately as the shaded band in Fig.~\ref{fig:physics_impact}.

\subsection{Comparison of MUFFIN and \texorpdfstring{\FF}{F\_F} methods in categorised and multivariate analyses} \label{sec:stress-tests}

While overall agreement in the total distribution of a primary fit observable is a necessary validation, it may not fully capture the impact of \jtf modelling once analyses probe extreme kinematics, categorise events into jet multiplicity or high-\pt bins, and/or use multivariate discriminants built from many correlated inputs. To address this, additional closure tests are performed in VRs that are orthogonal to both the DRs used to derive the MUFFIN reweighting and the SRs used to evaluate the final analysis sensitivity, but are kinematically representative of the SRs. In these VRs, the predicted \jtf background is compared to the observed data distribution in a relevant analysis observable, and the resulting residual differences between predicted and observed yields are used directly to estimate the systematic uncertainty associated with the \jtf background estimation with each method. For each analysis setup, the non-closure in the VR is evaluated as a function of the relevant analysis observable. 

The performance of the MUFFIN and \FF methods is evaluated using the non-closure
\begin{equation}
r(\vec{z}) \equiv \frac{N_{\text{obs}}(\vec{z})}{N_{\text{pred}}(\vec{z})},
\label{eq:stress-nonclosure}
\end{equation}
where $\vec{z}$ denotes a bin in the (possibly multidimensional) space of relevant observables, $N_{\text{obs}}$ is the observed yield in the signal-like VR, and $N_{\text{pred}}$ is the corresponding method prediction, computed across the full distribution. The bin-by-bin non-closure is taken directly as the systematic uncertainty in the \jtf background estimate in that bin. In sparsely populated regions where statistical precision does not support fine binning, a constant is fit to the ratio above a threshold, yielding $r_{\text{tail}}$, which is then used as the systematic uncertainty in that region.

The corresponding total uncertainty per bin is constructed as
\begin{equation}
\text{Total uncertainty} =
\sqrt{(1 - r)^2 + \sigma_{r}^{2}},
\label{eq:nonclosure-uncert}
\end{equation}
which combines the statistical component $\sigma_r$ with the observed deviation from unity. Crucially, by evaluating these non-closures in a VR that is kinematically representative of the SR, the resulting per-bin uncertainties provide a data-driven estimate of the residual modelling uncertainty that would be assigned in an actual analysis. A smaller non-closure therefore translates directly into a smaller systematic uncertainty in the \jtf background estimate, and hence improved sensitivity.

Three representative analysis setups are studied, spanning some of the most commonly used cases for \jtf background estimation in analyses involving \tauh decays at CMS. The first setup targets Higgs boson $\PH(125)\to\PGt\PGt$ decays in a \CP-structure measurement~\cite{cp_paper,CMS-HIG-25-012}, where a multivariate discriminant is used to enhance signal purity by separating Higgs boson events from background. The second one focuses on differential cross section measurements in the simplified template cross section (STXS) framework~\cite{CMS:2022kdi}. The third setup searches for new heavy resonances decaying to \tautau and \mutau final states at high mass~\cite{HIG-21-001}. Together, these cover a representative range of analysis strategies, from inclusive SRs to finely categorised and multivariate-discriminant-based selections, and allow the systematic uncertainty reduction achievable with the MUFFIN method to be quantified in each context. 

\subsubsection{The \texorpdfstring{\CP}{CP} structure in \texorpdfstring{$\PH(125)\to\PGt\PGt$}{H(125) to tautau} decays: multivariate discriminant-based analysis}
To probe whether improved modelling propagates to multivariate analyses, closure tests are performed directly on the output of a signal-versus-background BDT discriminator $D(\vec{x})$ used in the $\PH(125)\to\PGt\PGt$ \CP analysis. The bin-by-bin non-closure and integrated bias,
\begin{equation}
b \equiv \frac{\sum_i N_i^{\text{pred}} - \sum_i N_i^{\text{obs}}}{\sum_i N_i^{\text{obs}}},
\label{eq:integrated-bias}
\end{equation}
are evaluated across the full BDT score range, with particular attention to the signal-enriched high-score region where modelling quality has the largest impact on sensitivity.

\begin{figure}[htbp]
  \centering
  \includegraphics[width=\cmsFigWidth]{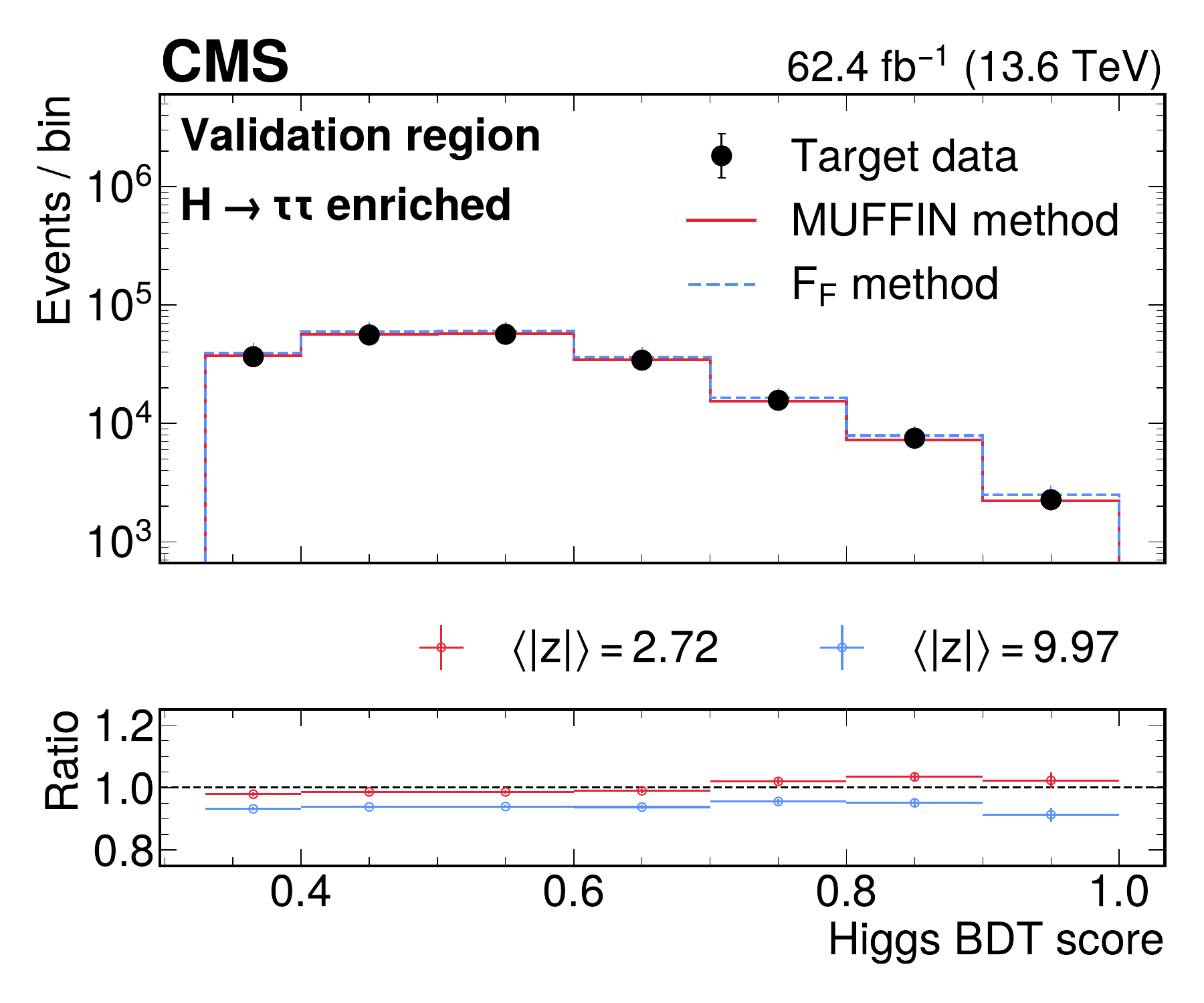}
  \includegraphics[width=\cmsFigWidth]{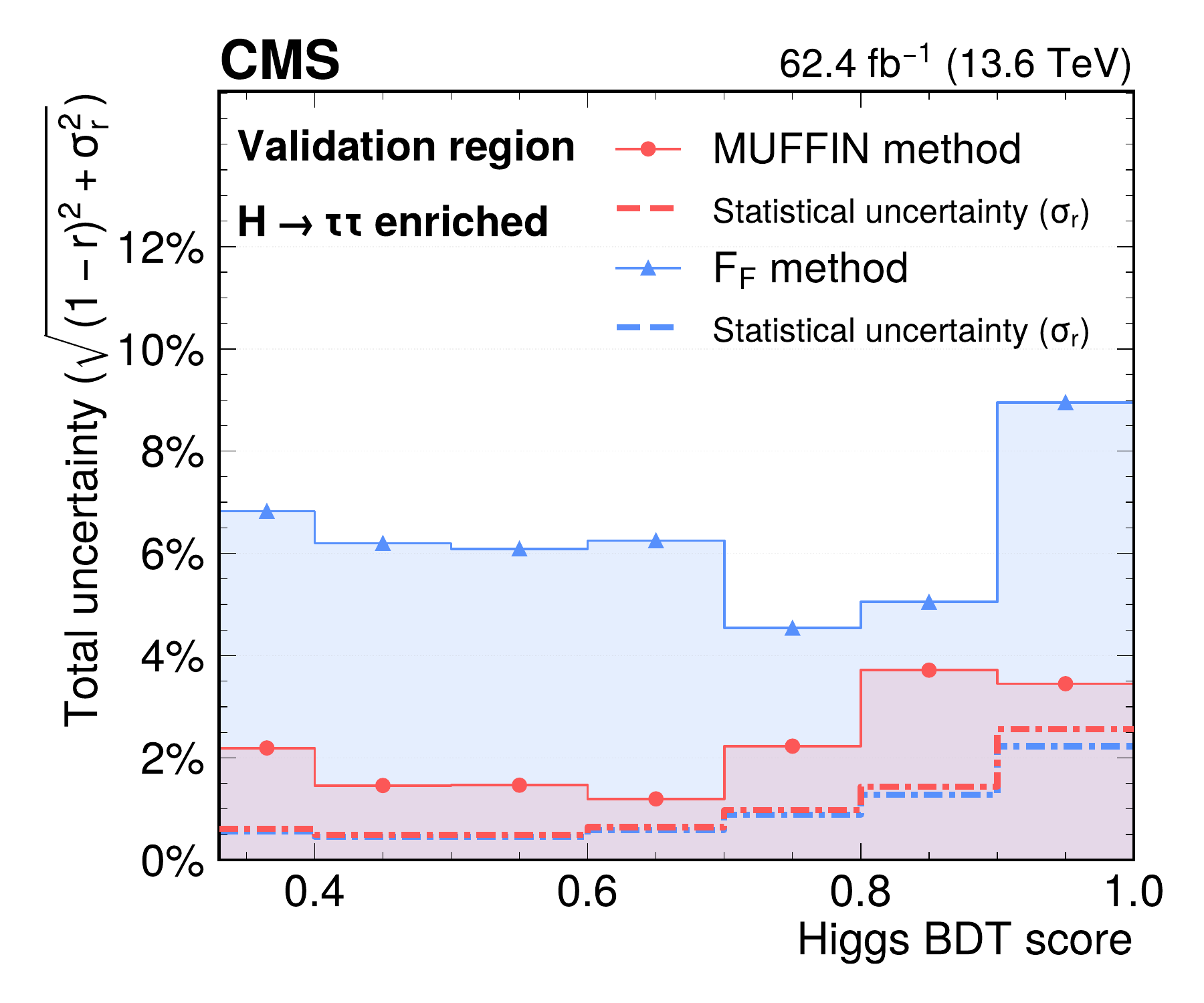}
  \includegraphics[width=\cmsFigWidth]{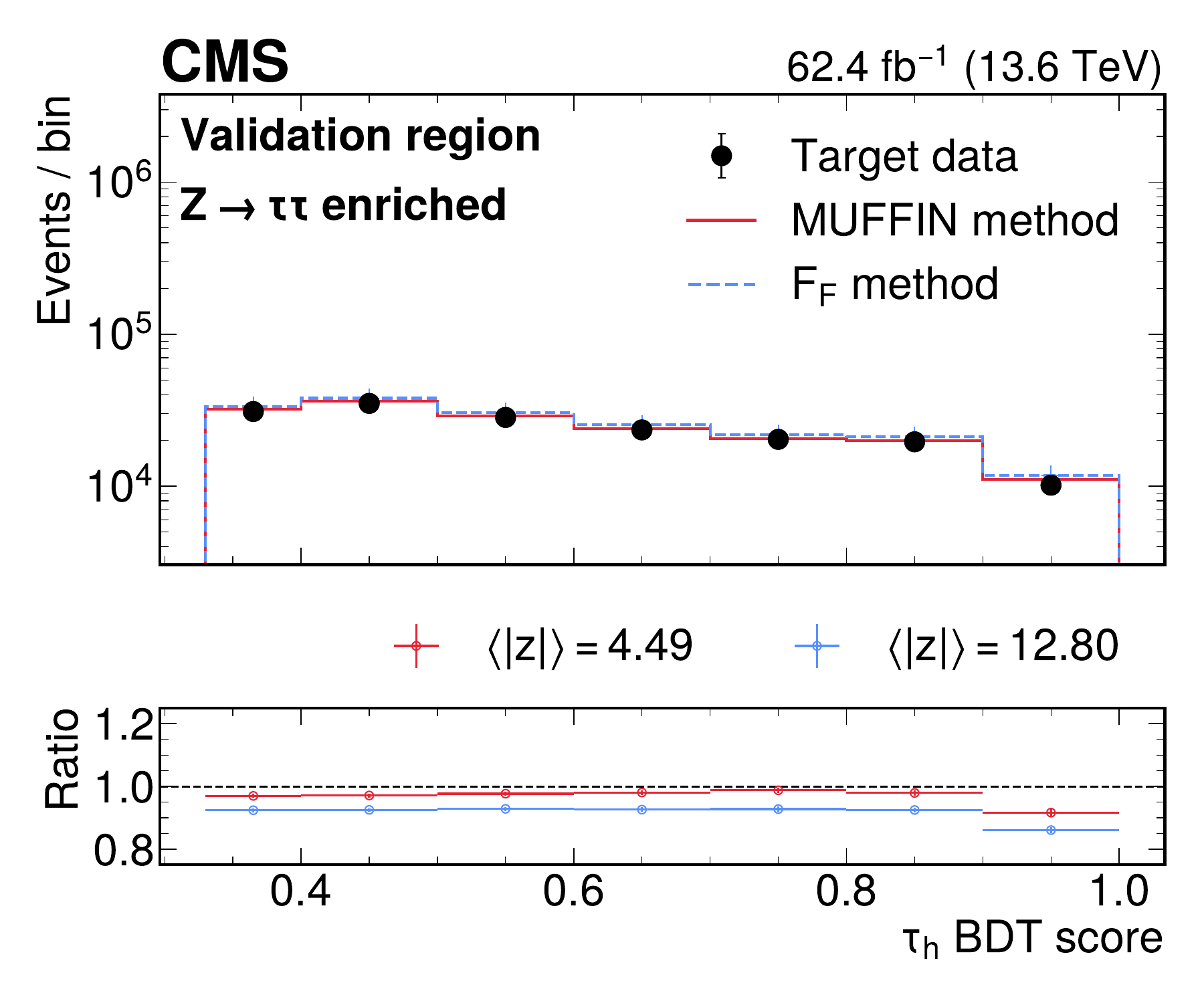}
  \includegraphics[width=\cmsFigWidth]{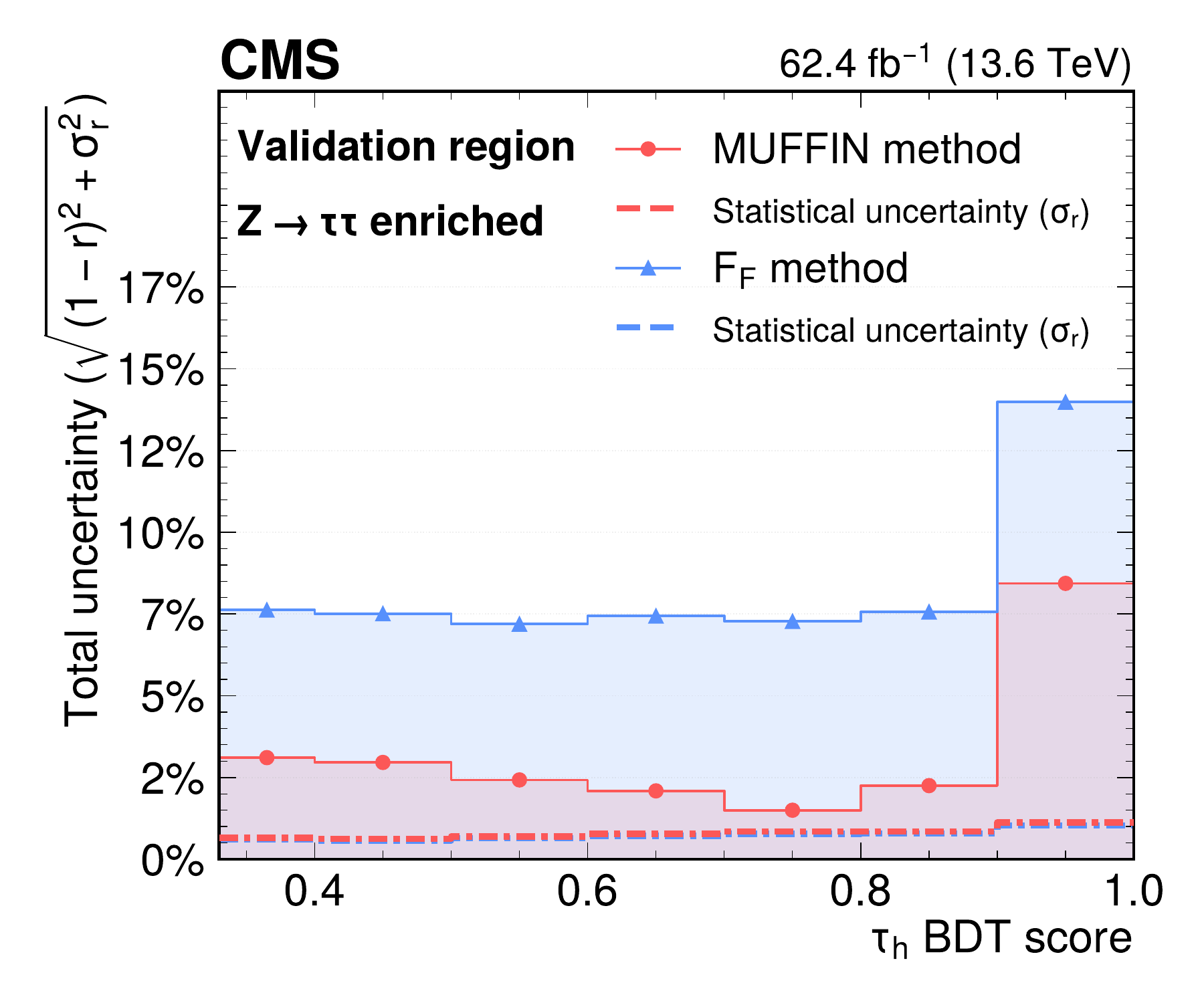}
  \includegraphics[width=\cmsFigWidth]{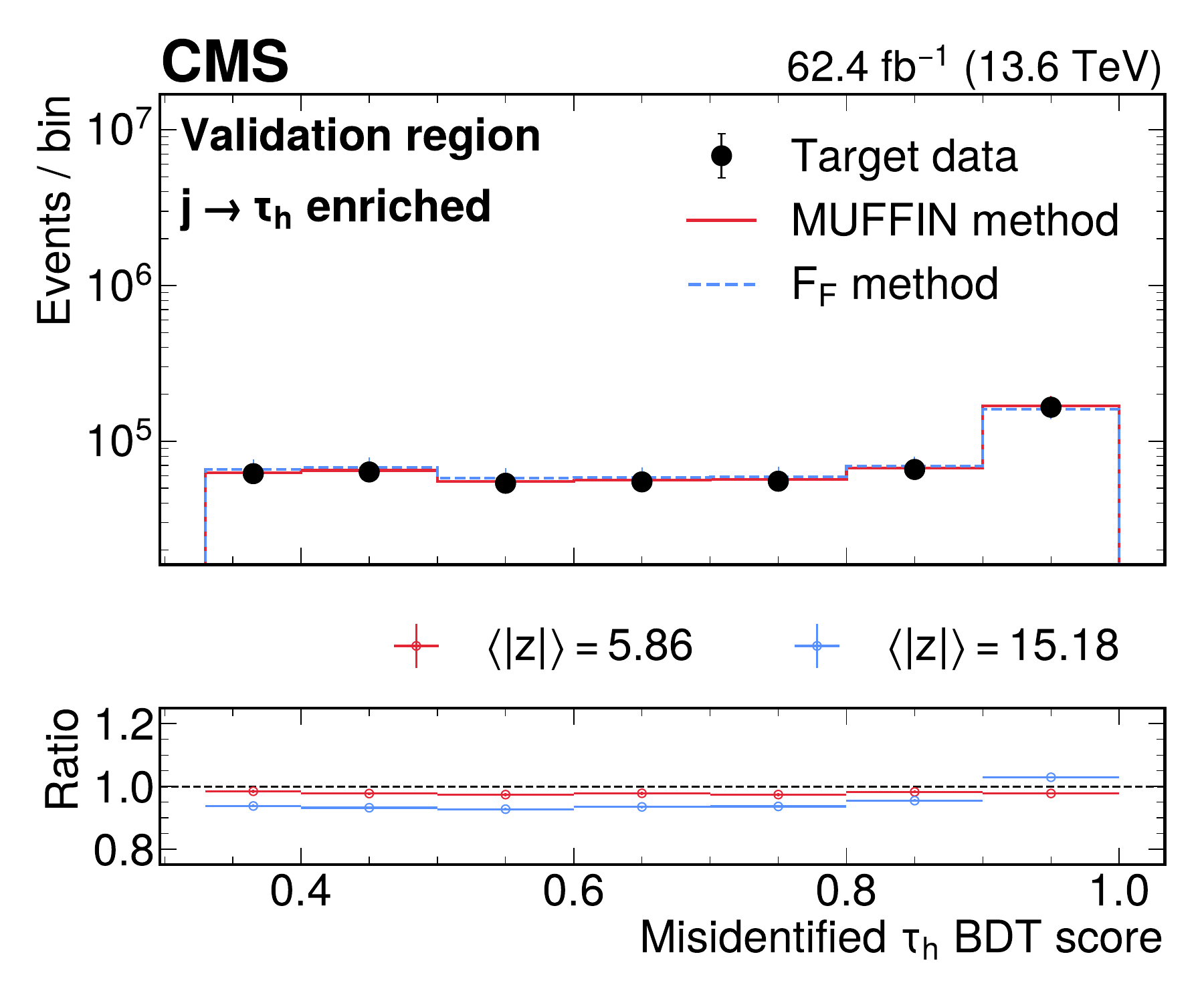}
  \includegraphics[width=\cmsFigWidth]{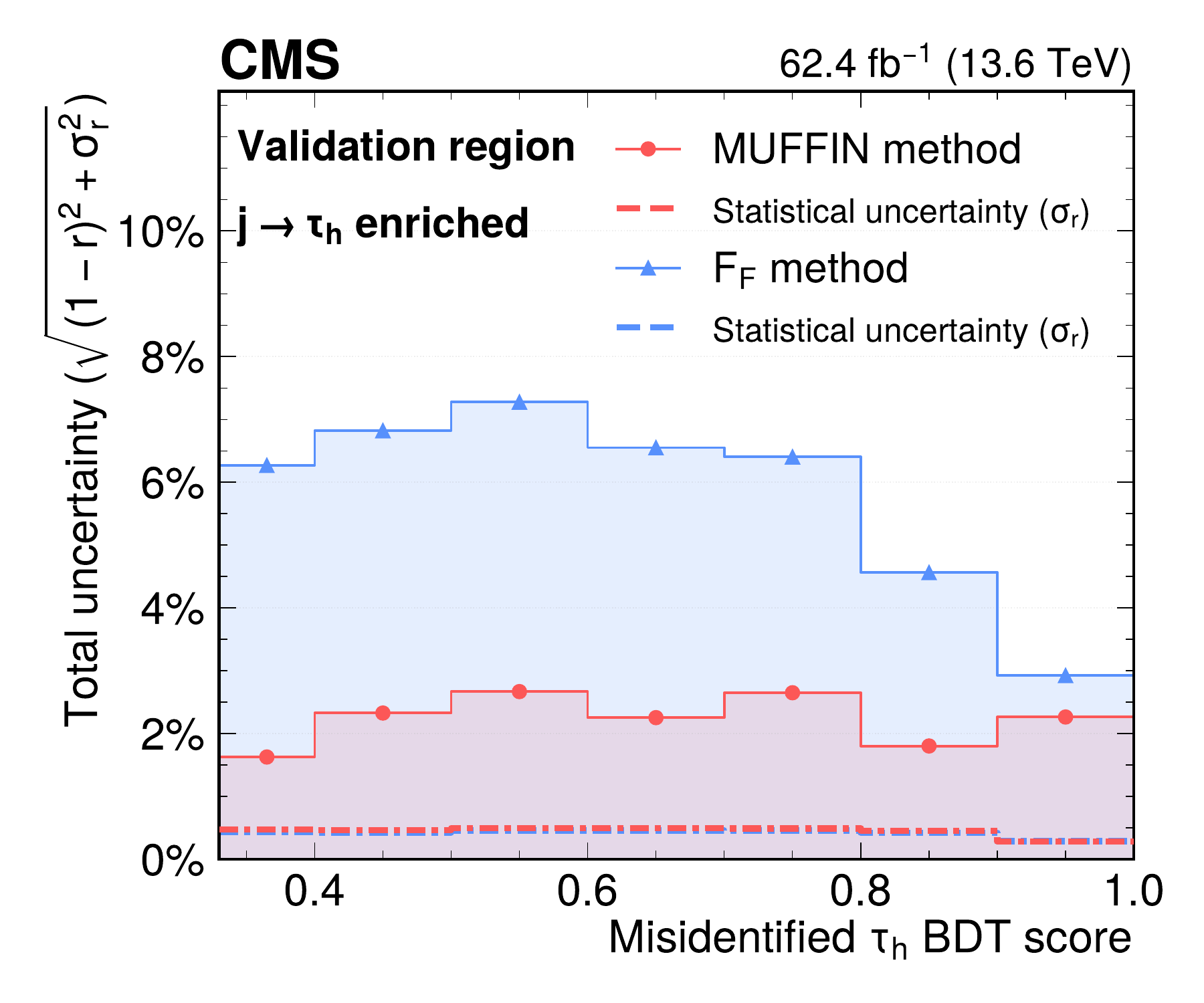}

  \caption{The BDT score validation in a $\PH(125)\to\PGt\PGt$ \CP VR in the \tautau channel for the Higgs boson enriched, \PZ boson enriched, and misidentified jet enriched categories. MUFFIN (red) and \FF (blue, dashed) predictions are compared at the distribution level (left) and through the corresponding non-closure evaluation (right). The statistical uncertainty is indicated by the vertical bars on the data points. In the right panels, the hatched bars indicate the statistical component of the total uncertainty.}
  \label{fig:tt_bdt_scores}
\end{figure}

\begin{figure}[htbp]
  \centering
  \includegraphics[width=\cmsFigWidth]{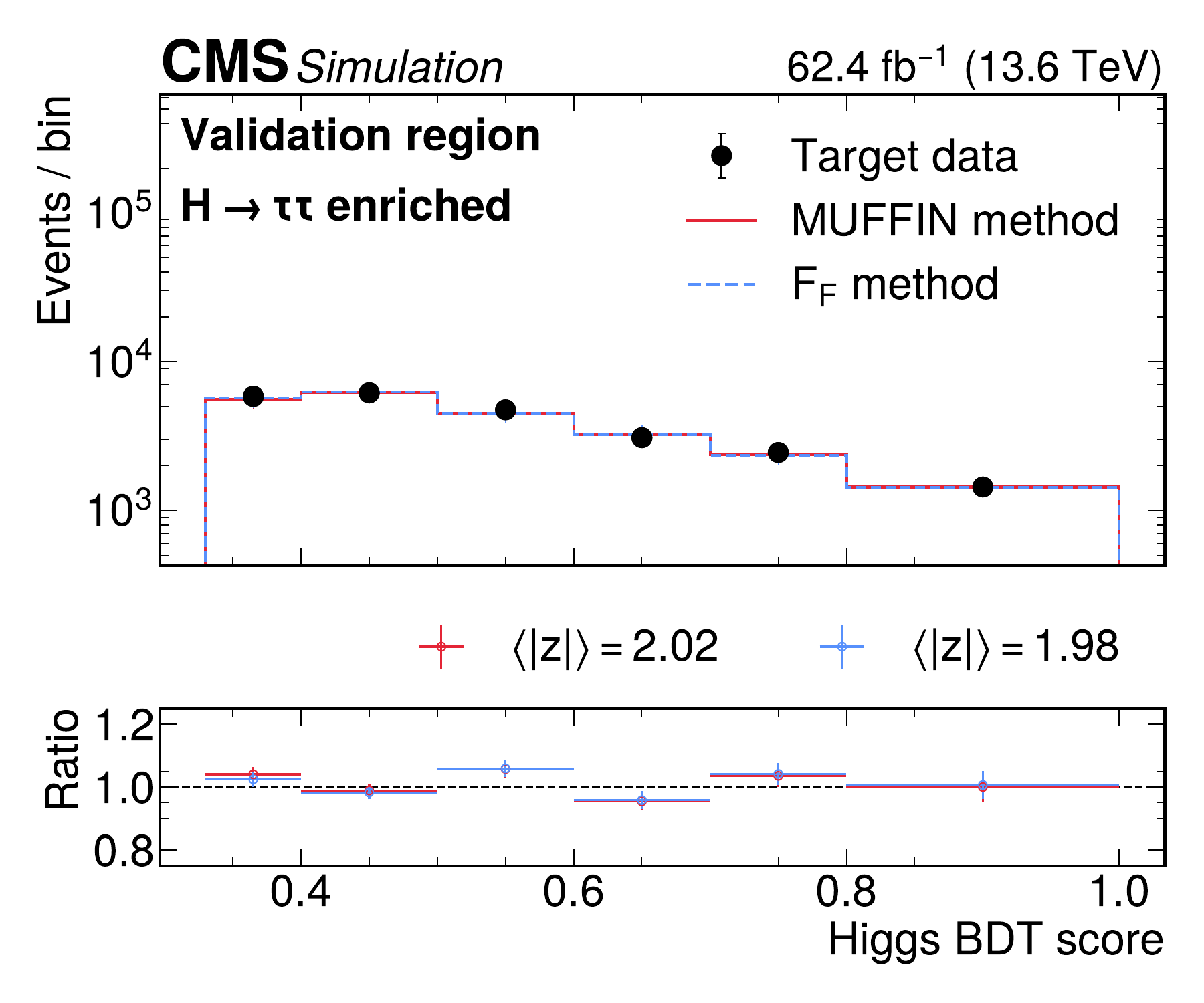}
  \includegraphics[width=\cmsFigWidth]{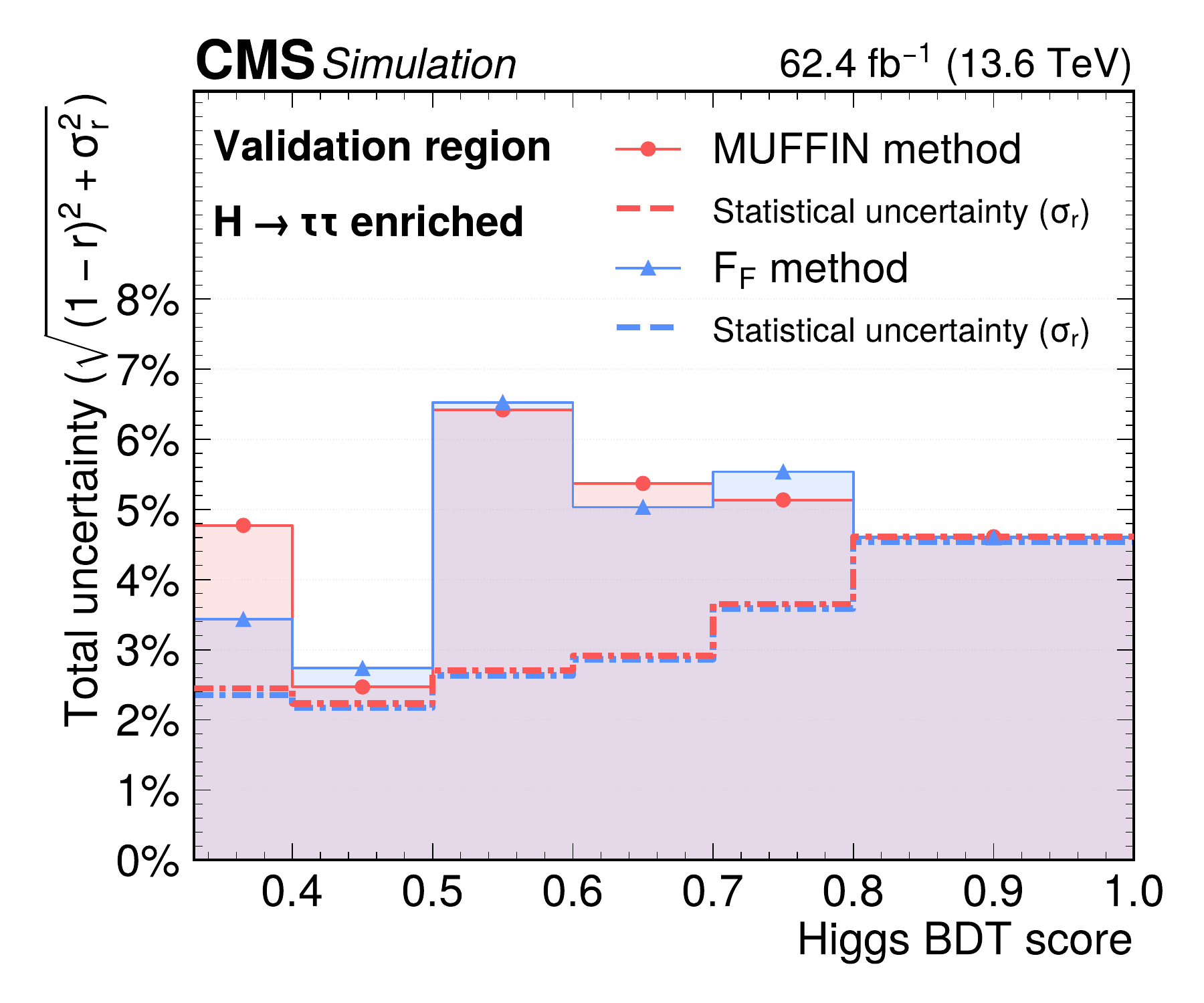}
  \includegraphics[width=\cmsFigWidth]{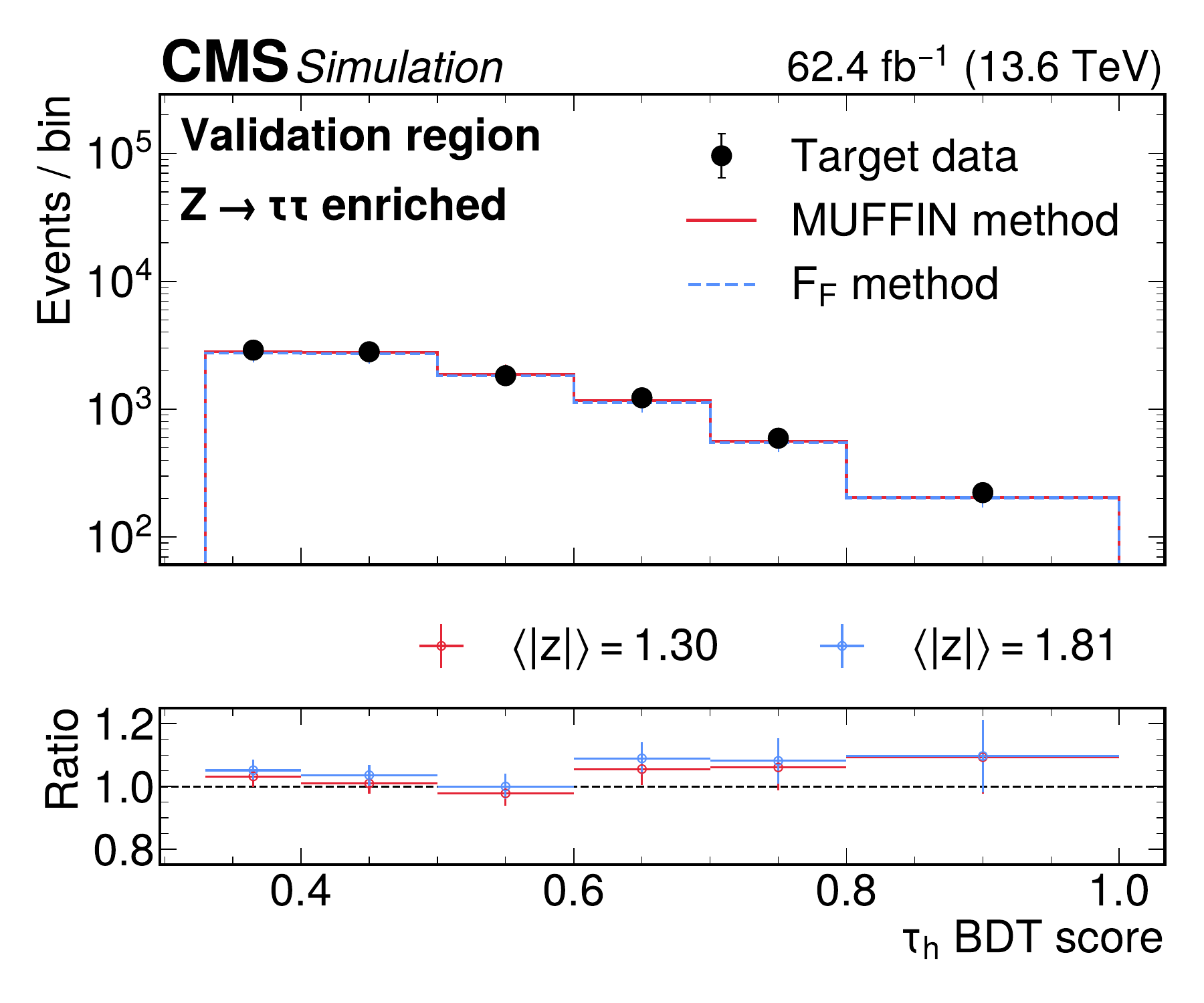}
  \includegraphics[width=\cmsFigWidth]{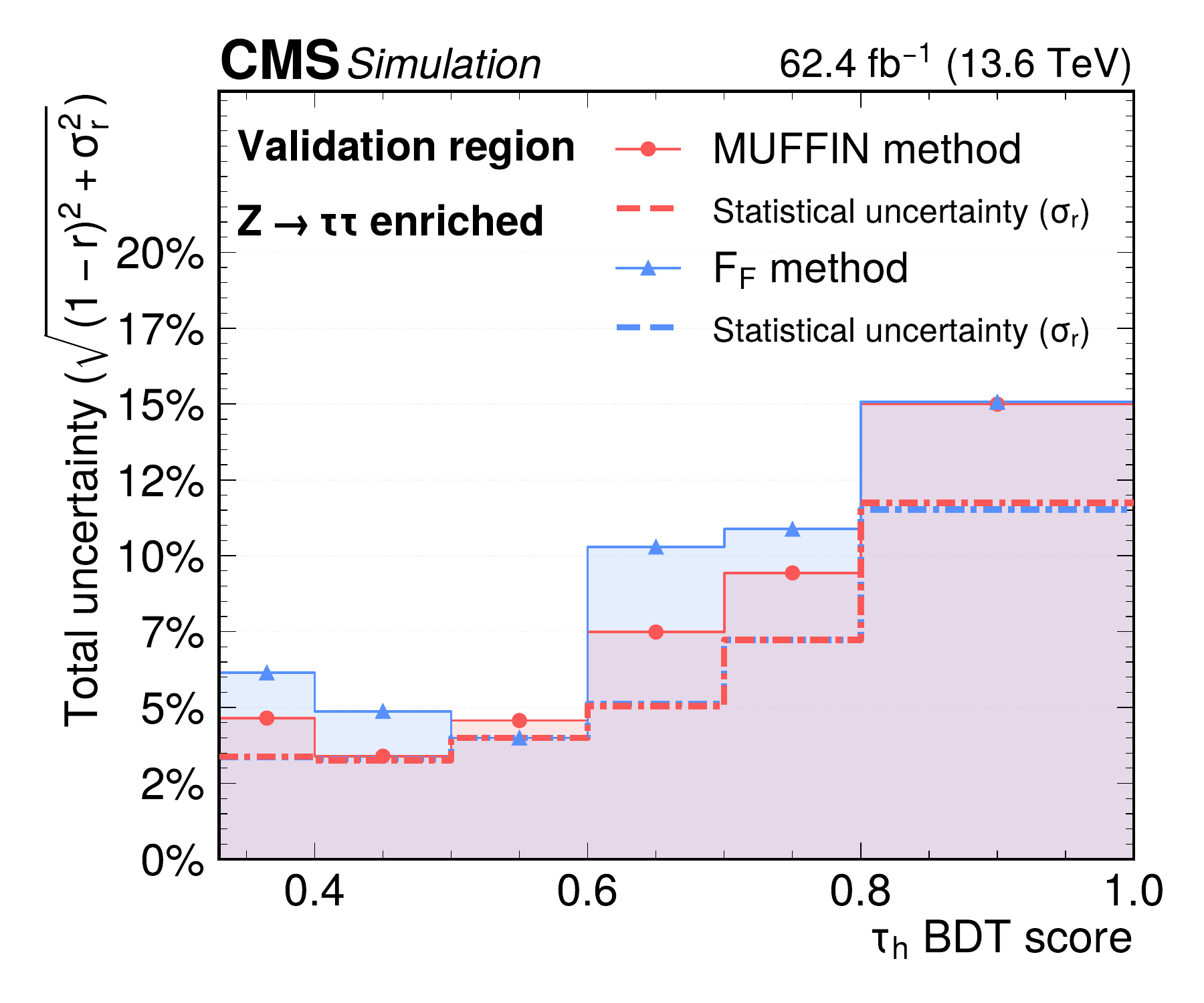}
  \includegraphics[width=\cmsFigWidth]{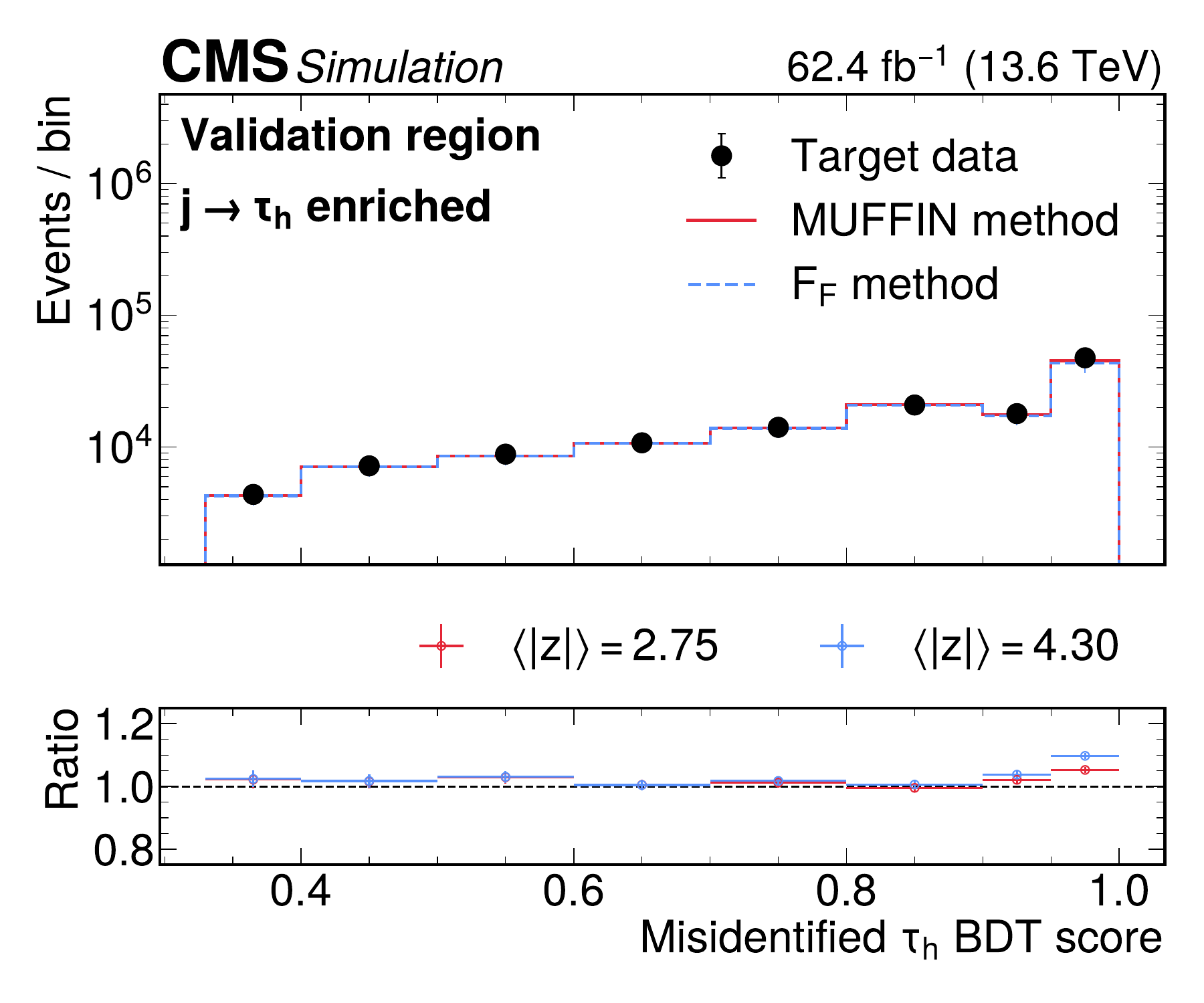}
  \includegraphics[width=\cmsFigWidth]{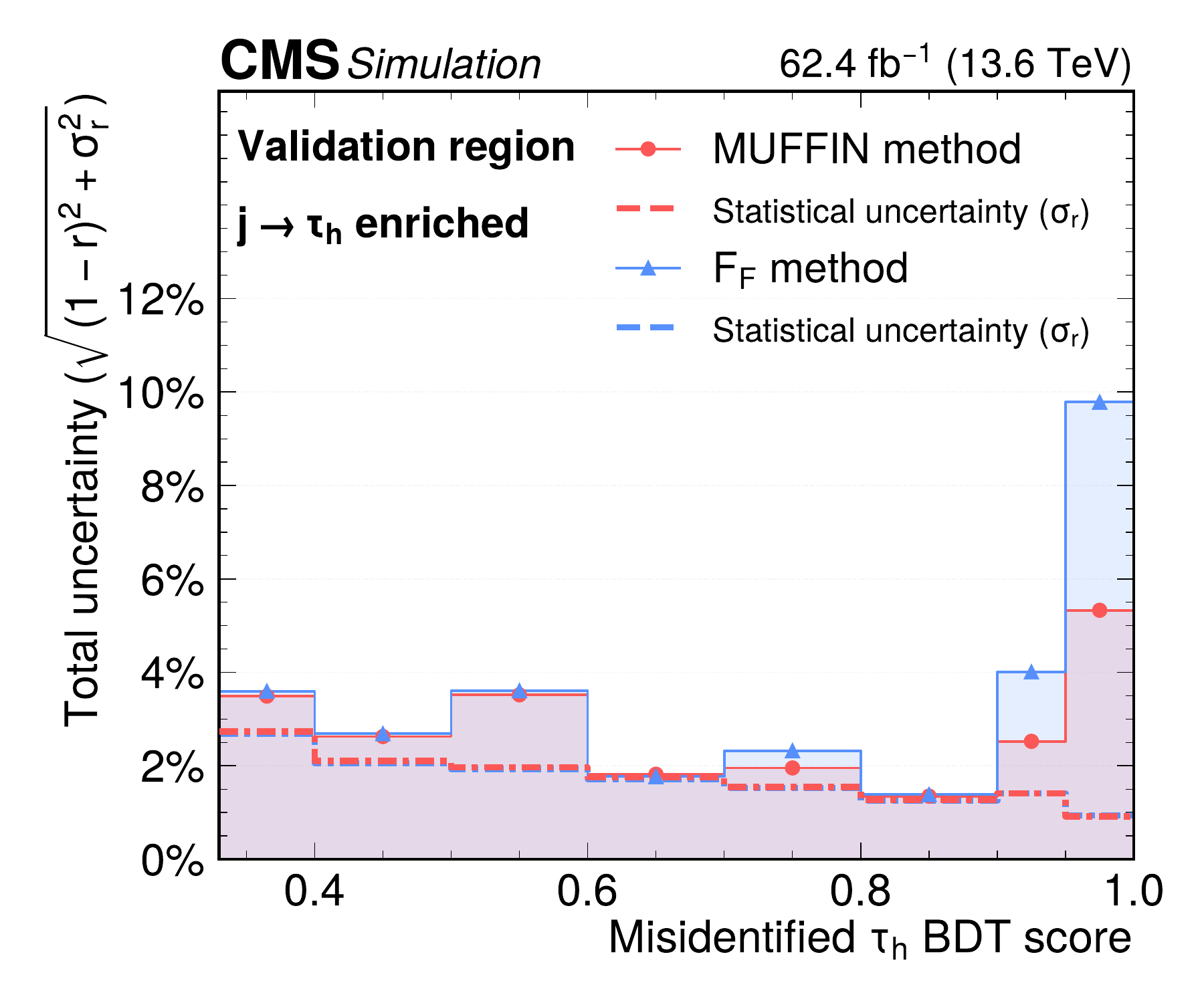}
  \caption{The BDT score validation in a \Wjets-enriched VR in the \mutau channel for the Higgs boson enriched, \PZ boson enriched, and misidentified jet enriched categories. MUFFIN (red) and \FF (blue, dashed) predictions are compared at the distribution level (left) and through the corresponding non-closure evaluation (right). The statistical uncertainty is indicated by the vertical bars on the data points. In the right panels, the hatched bars indicate the statistical component of the total uncertainty.}
  \label{fig:lt_wjets_mc_bdt_scores}
\end{figure}

In the illustrated $\PH(125)\to\PGt\PGt$ BDT example in Fig.~\ref{fig:tt_bdt_scores}, the root-mean-square of the bin-by-bin residuals improves from ${\sim}\,6.6\%$ (\FF) to ${\sim}\,3.4\%$ (MUFFIN), and the integrated bias improves from ${\sim}\,6.5\%$ to ${\sim}\,3.3\%$, demonstrating substantially reduced systematic uncertainties in the signal-enriched (high-score) region. Complementary results in a \Wjets enriched VR in the \mutau channel are shown in Fig.~\ref{fig:lt_wjets_mc_bdt_scores}. In this case the non-closure evaluation is statistically limited at high BDT scores, where the signal-enriched bins are sparsely populated, and the resulting systematic uncertainties are dominated by their statistical component. The non-closure values obtained with both methods are therefore comparable within current statistical precision, and the systematic component is expected to become better constrained as larger data sets become available.

\subsubsection{Differential cross section STXS measurements: jet multiplicity and high-\texorpdfstring{\pt}{pT} categories}
A second robustness check targets the STXS measurement framework, which is widely used in Higgs boson measurements. In this approach, the phase space is divided into exclusive categories based on jet multiplicity and event topology (\eg 0-jet, 1-jet, $\geq$2-jet, and high-\pt regions), in order to probe different Higgs boson production mechanisms. These categories introduce strong kinematic and topological constraints, making them particularly sensitive to mismodelling of backgrounds. Neither method is explicitly tuned to STXS definitions (as provided in Ref.~\cite{CMS:2022kdi}); both are applied without category-specific optimisation to probe robustness under aggressive phase space slicing. The non-closure is evaluated per STXS category by comparing predicted-to-observed yields in a fixed \textsc{FastMTT} mass~\cite{Kalinowski:2025vci} window, which provides an approximate reconstruction of the di-\PGt invariant mass in the presence of neutrinos from \PGt decays, with the category-level non-closures, as defined in Eq.~(\ref{eq:nonclosure-uncert}), used to estimate the per-category systematic uncertainty in the \jtf background.

\begin{figure}[htbp]
  \centering
  \includegraphics[width=\textwidth]{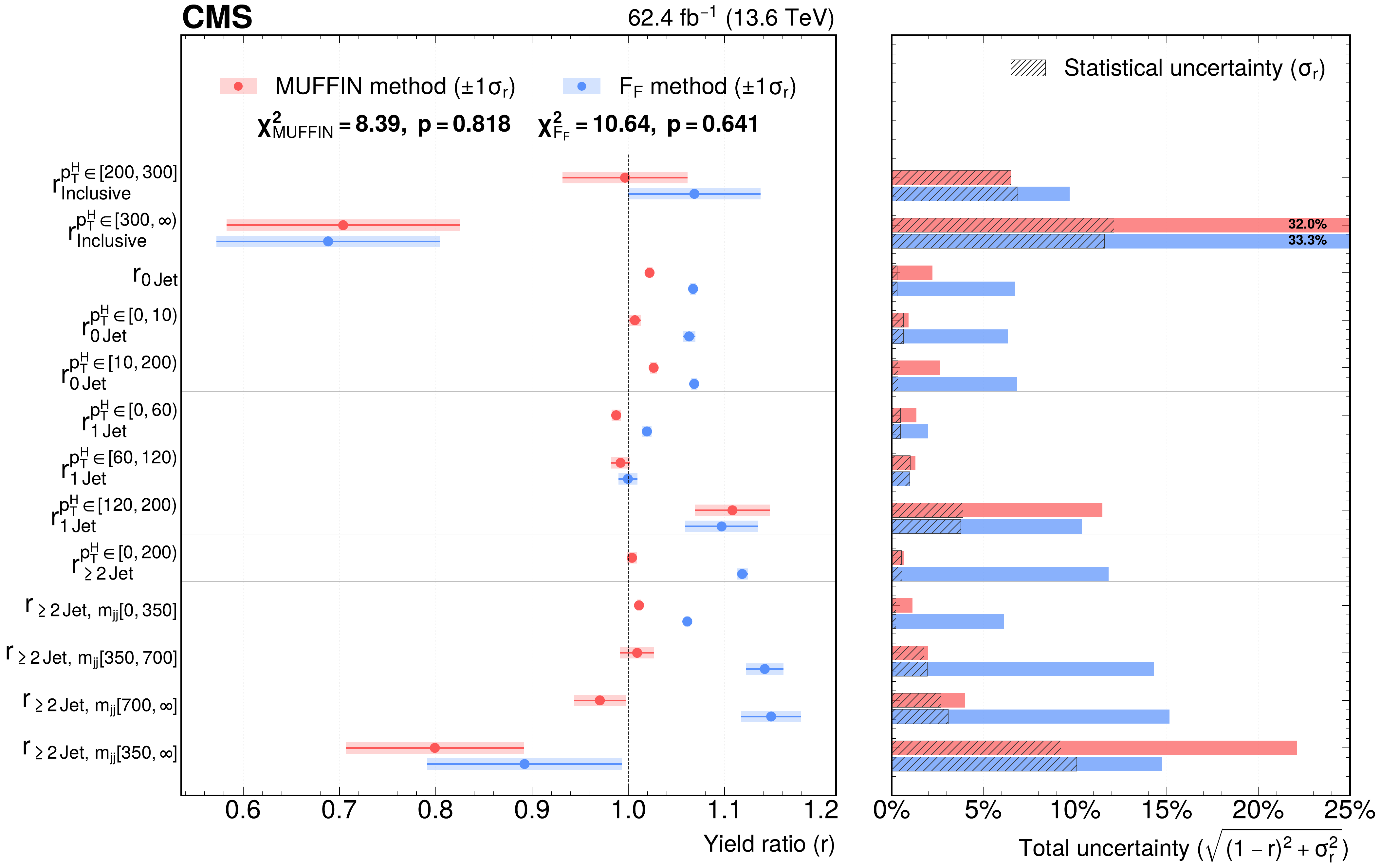}
  \includegraphics[width=\textwidth]{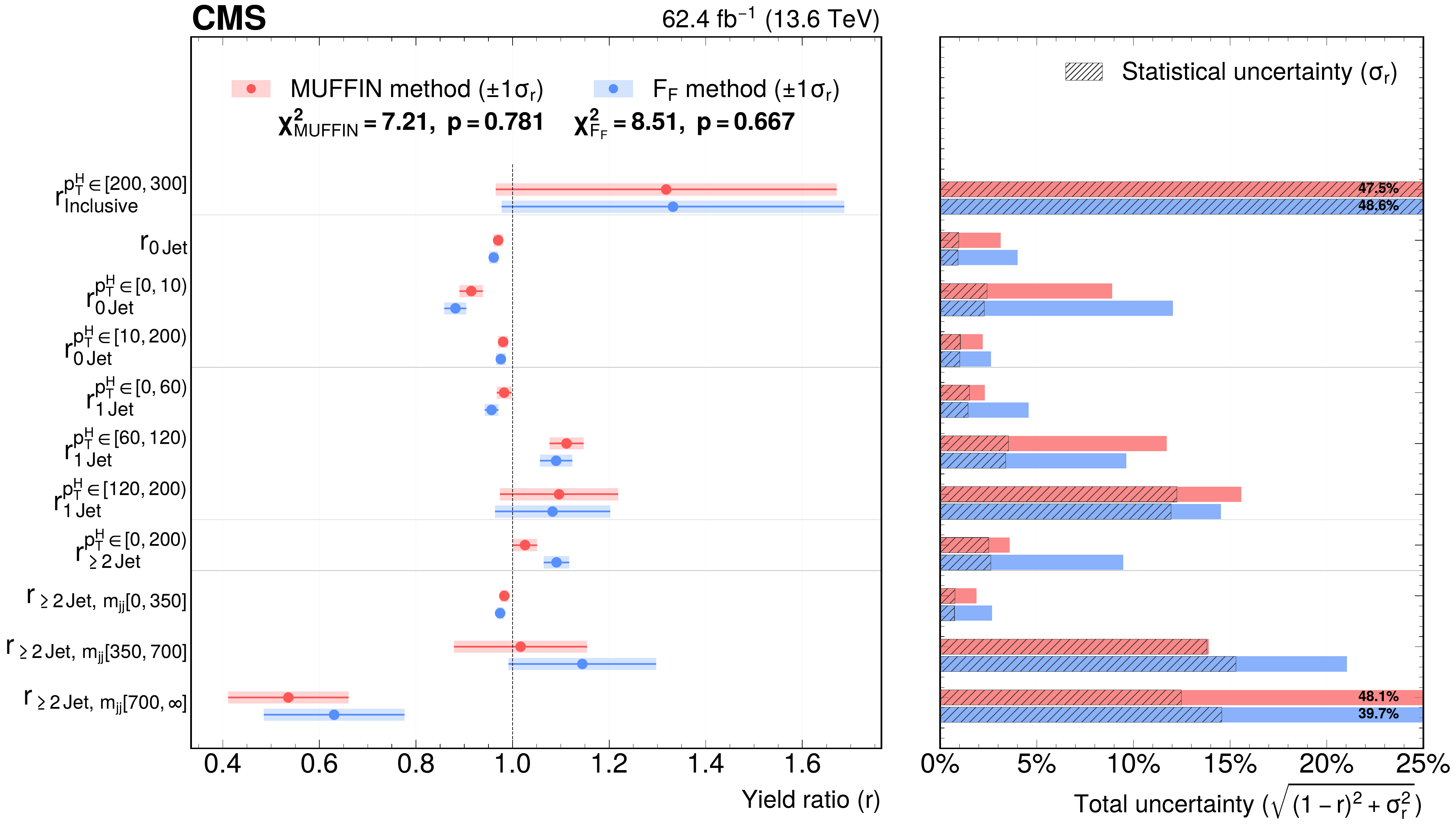}
  \caption{The STXS-inspired category-by-category yield closure for a \tautau (upper) and \mutau (lower) VR. Points show the ratio of observed to predicted yields per STXS category (non-closure), for the MUFFIN (red) and \FF (blue) methods, where deviations from unity are interpreted as systematic uncertainties in the \jtf background estimate. The right panel shows the total uncertainty per category, with the hatched grey bars indicating the statistical component. Only statistical uncertainties are shown in the left panel; the observed non-closure along with the statistical uncertainty define the systematic uncertainty that would be assigned in a real analysis.}
  \label{fig:stxs-closure}
\end{figure}

In a $\tautau$ QCD multijet-dominated VR, the MUFFIN method achieves $\chi^2 = 8.39$ ($p = 0.818$) across all STXS categories, compared to $\chi^2 = 10.64$ ($p = 0.641$) for the \FF method, with the largest improvements observed in the high-\pt and multi-jet categories where the \FF method is most challenged by its inability to capture correlations between jet activity and the \jtf misidentification probability. In a \mutau \Wjets-dominated test, MUFFIN achieves $\chi^2 = 7.21$ ($p = 0.781$) compared to $\chi^2 = 8.51$ ($p = 0.667$) for the \FF method. Categories for which the available sample size is insufficient to obtain a statistically meaningful non-closure estimate are excluded from the comparison. In the majority of categories where a reliable comparison is possible, MUFFIN achieves smaller or comparable deviations from unity than the \FF method, demonstrating that the improvement is systematic and not confined to particular phase space regions. In a small number of individual categories the central non-closure value for MUFFIN is marginally larger than for \FF, but in all such cases the difference is well within the statistical uncertainties of the two methods. Figure~\ref{fig:stxs-closure} shows the category-by-category yield closure and derived systematic uncertainties for both VRs, illustrating the improved per-category agreement and the reduced systematic uncertainties achievable with the MUFFIN method, particularly in the high-\pt and multijet categories. In statistically limited categories, both methods perform comparably, with the non-closure values dominated by their statistical component. As larger data sets become available, the improved modelling of the MUFFIN method is expected to translate into significantly reduced systematic uncertainties in these categories as well.

\subsubsection{High-mass resonance searches: kinematic tails sensitive regions}
For searches targeting BSM heavy resonances decaying to \tautau and \mutau final states, sensitivity often comes from the high-\pt tails of the \tauh kinematic distributions, where the \jtf background systematic uncertainty is typically the limiting factor. The non-closure in these tail regions is evaluated by fitting a constant to the residuals above a \tauh \pt threshold ($\pttauh>200\GeV$ in \tautau and $\pttauh>100\GeV$ in \mutau), yielding $r_{\text{tail}}$ and the corresponding tail systematic uncertainty $\sigma^{\text{tail}}$ as defined in Eq.~(\ref{eq:nonclosure-uncert}).

\begin{figure}[htbp]
  \centering
  \includegraphics[width=\cmsFigWidth]{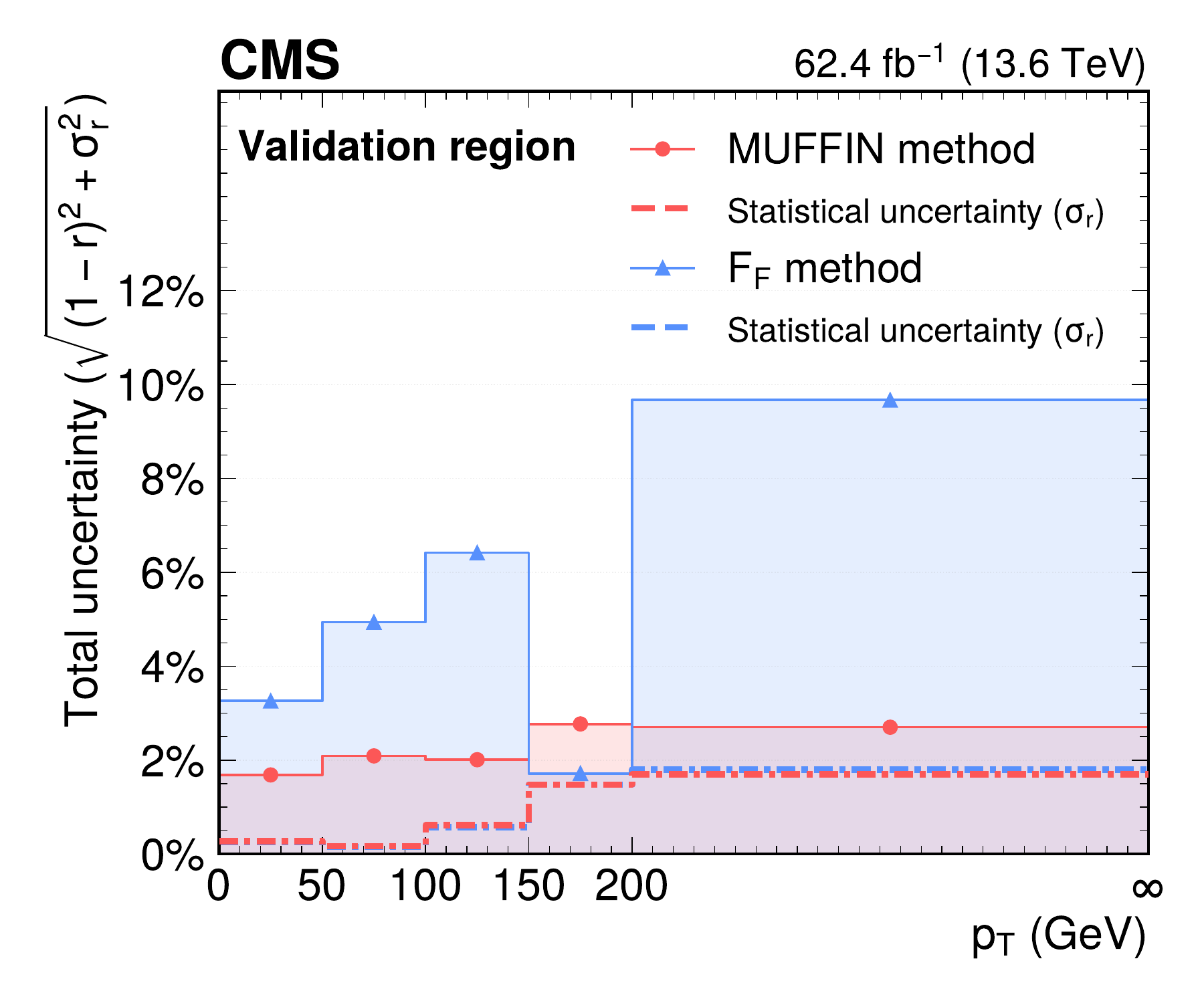}
  \includegraphics[width=\cmsFigWidth]{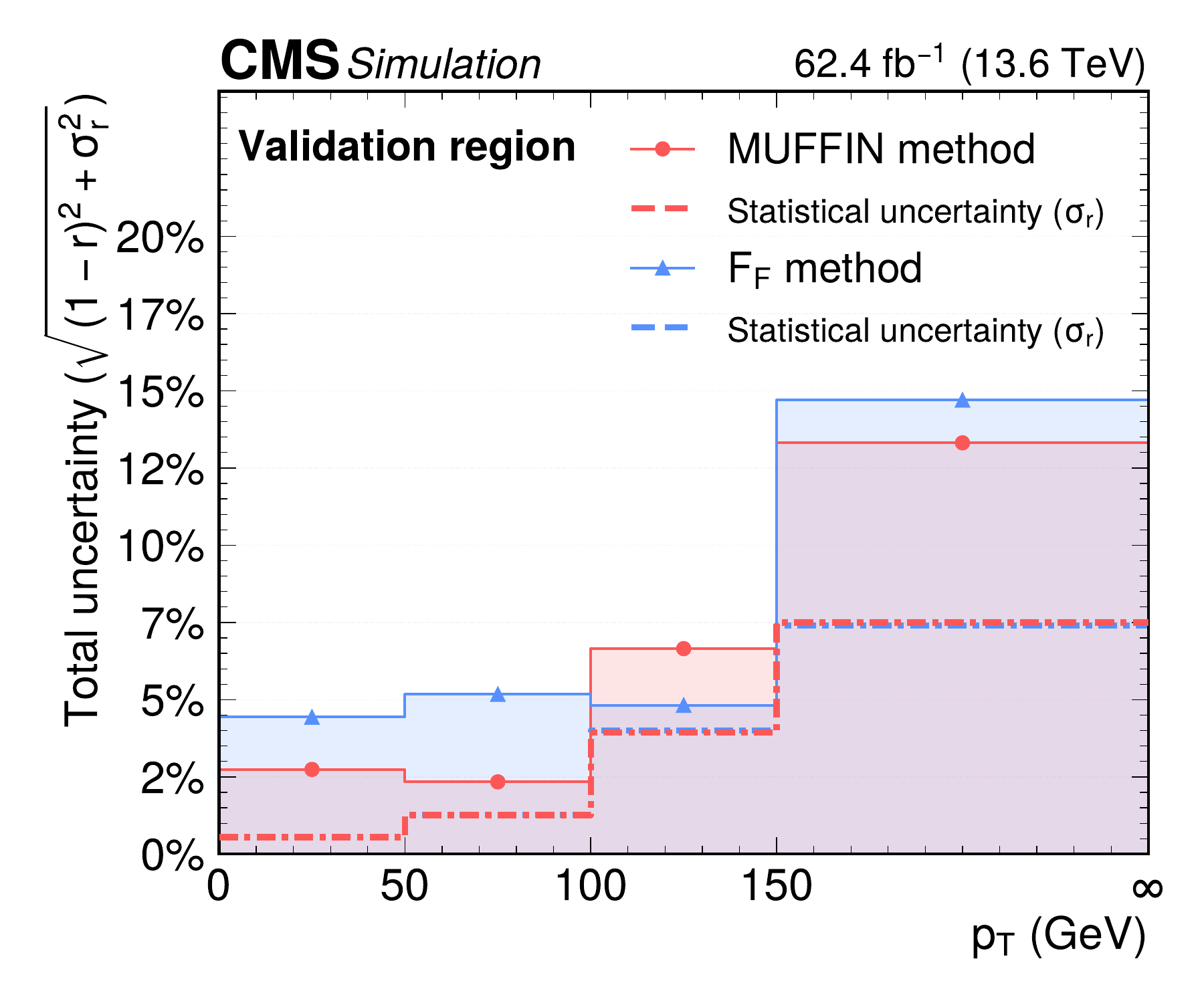}
  \caption{Closure in the high-\pt tails of the kinematic distributions for a \tautau QCD multijet-dominated (left) and \mutau \Wjets (right) VRs. The dashed lines indicate the statistical component of the uncertainty. The ratio in each bin reflects the non-closure, and the fitted constant (solid horizontal line) is derived by fitting over all events above the \pt threshold, providing a more statistically robust estimate of the aggregate tail closure than any individual bin. The MUFFIN method achieves significantly improved agreement in both tails, leading to substantially reduced systematic uncertainties for the \tautau channel and moderate improvements in the \mutau channel. Overflow is included in the final bin ($\infty$).}
  \label{high-pt}
\end{figure}

In the \tautau QCD multijet-dominated VR, the fitted tail residuals are $r_{\text{tail}}^{\FF} = 1.095 \pm 0.018$ and $r_{\text{tail}}^{\text{MUFFIN}} = 1.021 \pm 0.018$ for the \tauh \pt tail, corresponding to tail systematic uncertainties of $\sigma^{\text{tail}}_{\FF}=9.7\%$ and $\sigma^{\text{tail}}_{\text{MUFFIN}}=2.7\%$, as shown in Fig.~\ref{high-pt}. In the \mutau \Wjets-dominated VR, the fitted tail residuals are $r_{\text{tail}}^{\FF} = 0.873 \pm 0.074$ and $r_{\text{tail}}^{\text{MUFFIN}} = 0.890 \pm 0.075$, corresponding to tail systematic uncertainties of $\sigma^{\text{tail}}_{\FF}=14.7\%$ and $\sigma^{\text{tail}}_{\text{MUFFIN}}=13.3\%$. This corresponds to a reduction of ${\sim}\,70\%$ in the \tautau channel. In the \mutau channel, where the evaluation is limited by the available event yields, a modest improvement of ${\sim}\,10\%$ is observed, with both methods yielding comparable uncertainties within statistical precision. These reductions directly translate into reduced systematic uncertainties in the \jtf background estimate in the high-mass SR, and hence into improved sensitivity for BSM physics searches.

Taken together, the three case studies above demonstrate that the improvements in inclusive modelling observed in Section~\ref{sec:performance} translate directly into reduced systematic uncertainties across all major analysis strategies. The MUFFIN method consistently achieves smaller non-closures in BDT score distributions, STXS categories, and kinematic tails, without introducing instabilities in well-modelled regions. The resulting reduction in \jtf systematic uncertainties is most pronounced in the phase space regions of greatest sensitivity to new physics and differential measurements, where the \FF method is most limited by its inability to capture the full multidimensional dependence of the misidentification probability.

\section{Summary and outlook}\label{sec:summary}

In this paper, a machine-learning-based alternative to the ``Fake Factor'' (\FF) method for estimating backgrounds from jets misidentified as hadronically decaying \PGt leptons (\tauh) is presented, referred to as MUltivariate Fake-Factor INference (MUFFIN).
This method is a generalisation of the previous \FF method by replacing low-dimensional transfer factors with multidimensional, per-object reweighting functions learned directly from observed data, while preserving the fully data-driven nature of the background estimation.

The method has been developed and validated using CMS data collected in 2022--2023 and is benchmarked against the \FF method using a series of closure tests. Across the analysis channels and observables considered, the MUFFIN method has demonstrated improved modelling performance, increased stability in regions sensitive to correlations between multiple variables, and a reduced reliance on additional non-closure corrections. At the same time, the method has provided a more unified and reproducible workflow for background estimation, leading to a reduction of up to 70\% in the associated systematic uncertainty (Section~\ref{sec:stress-tests}).

These results indicate that MUFFIN constitutes a robust method for future analyses involving hadronically decaying \PGt leptons. As analyses increasingly exploit fine-grained categorisation and multidimensional phase space information, the method presented here provides a natural framework to improve background modelling precision and to enhance the sensitivity of measurements and searches involving \tauh final states.

Looking forward, further improvement in the performance of the MUFFIN method is expected as larger data sets become available. Training on larger, combined data sets therefore offers a natural path to increasing statistical precision while maintaining robustness against changing experimental conditions. The method also does not depend on the specific \tauh identification algorithm and can be readily deployed with future algorithm developments. More broadly, the general nature of the MUFFIN method makes it readily applicable to other data-driven background estimation problems involving misidentified or non-prompt objects, beyond \tauh candidates.

\begin{acknowledgments}
We congratulate our colleagues in the CERN accelerator departments for the excellent performance of the LHC and thank the technical and administrative staffs at CERN and at other CMS institutes for their contributions to the success of the CMS effort. In addition, we gratefully acknowledge the computing centres and personnel of the Worldwide LHC Computing Grid and other centres for delivering so effectively the computing infrastructure essential to our analyses. Finally, we acknowledge the enduring support for the construction and operation of the LHC, the CMS detector, and the supporting computing infrastructure provided by the following funding agencies: SC (Armenia), BMFWF and FWF (Austria); FNRS and FWO (Belgium); CNPq, CAPES, FAPERJ, FAPERGS, and FAPESP (Brazil); MES and BNSF (Bulgaria); CERN; CAS, MoST, and NSFC (China); MINCIENCIAS (Colombia); MSES and CSF (Croatia); RIF (Cyprus); SENESCYT (Ecuador); ERC PRG and PSG, TARISTU24-TK10 and MoER TK202 (Estonia); Academy of Finland, MEC, and HIP (Finland); CEA and CNRS/IN2P3 (France); SRNSF (Georgia); BMFTR, DFG, and HGF (Germany); GSRI (Greece); MATE and NKFIH (Hungary); DAE and DST (India); IPM (Iran); SFI (Ireland); INFN (Italy); MSIT and NRF (Republic of Korea); MES (Latvia); LMTLT (Lithuania); MOE and UM (Malaysia); BUAP, CINVESTAV, CONACYT, LNS, SEP, and UASLP-FAI (Mexico); MOS (Montenegro); MBIE (New Zealand); PAEC (Pakistan); MSHE, NSC, and NAWA (Poland); FCT (Portugal); MESTD (Serbia); MICIU/AEI and PCTI (Spain); MOSTR (Sri Lanka); Swiss Funding Agencies (Switzerland); NSTC (Taiwan); MHESI (Thailand); TUBITAK and TENMAK (T\"{u}rkiye); NASU (Ukraine); STFC (United Kingdom); DOE and NSF (USA).

\begin{sloppypar}
\setlength\emergencystretch{\hsize}
\hyphenation{Rachada-pisek} Individuals have received support from the Marie-Curie programme and the European Research Council and Horizon 2020 Grant, contract Nos.\ 675440, 724704, 752730, 758316, 765710, 824093, 101115353, 101002207, 101001205, and COST Action CA16108 (European Union); the Leventis Foundation; the Alfred P.\ Sloan Foundation; the Alexander von Humboldt Foundation; the Science Committee, project no. 22rl-037 (Armenia); the Fonds pour la Formation \`a la Recherche dans l'Industrie et dans l'Agriculture (FRIA) and Fonds voor Wetenschappelijk Onderzoek contract No. 1228724N (Belgium); the Beijing Municipal Science \& Technology Commission, No. Z191100007219010, the Fundamental Research Funds for the Central Universities, the Ministry of Science and Technology of China under Grant No. 2023YFA1605804, the Natural Science Foundation of China under Grant No. 12535004, and USTC Research Funds of the Double First-Class Initiative No.\ YD2030002017 (China); the Ministry of Education, Youth and Sports (MEYS) of the Czech Republic; the Shota Rustaveli National Science Foundation (Georgia); the Deutsche Forschungsgemeinschaft (DFG), among others, under Germany's Excellence Strategy -- EXC 2121 ``Quantum Universe" -- 390833306, and under project number 400140256 - GRK2497; the Hellenic Foundation for Research and Innovation (HFRI), Project Number 2288 (Greece); the Hungarian Academy of Sciences, the New National Excellence Program - \'UNKP, the NKFIH research grants K 131991, K 138136, K 143460, K 143477, K 147557, K 146913, K 146914, K 147048, TKP2021-NKTA-64, 2026-4.1.1-MISSZI\'{O}K-2026-00005 and 2025-1.1.5-NEMZ\_KI-2025-00004, and MATE KKP and KKPCs Research Excellence and Flagship Research Groups grants (Hungary); the Council of Science and Industrial Research, India; ICSC -- National Research Centre for High Performance Computing, Big Data and Quantum Computing, FAIR -- Future Artificial Intelligence Research, and CUP I53D23001070006 (Mission 4 Component 1), funded by the NextGenerationEU program, the Italian Ministry of University and Research (MUR) under Bando PRIN 2022 -- CUP I53C24002390006, PRIN PRIMULA 2022RBYK7T (Italy); the Latvian Council of Science; the Ministry of Science and Higher Education, project no. 2022/WK/14, and the National Science Centre, contracts Opus 2021/41/B/ST2/01369, 2021/43/B/ST2/01552, 2023/49/B/ST2/03273, and the NAWA contract BPN/PPO/2021/1/00011 (Poland); the Funda\c{c}\~ao para a Ci\^encia e a Tecnologia (Portugal); the National Priorities Research Program by Qatar National Research Fund; MICIU/AEI/10.13039/501100011033, ERDF/EU, ``European Union NextGenerationEU/PRTR", projects PID2022-142604OB-C21, PID2022-139519OB-C21, PID2023-147706NB-I00, PID2023-148896NB-I00, PID2023-146983NB-I00, PID2023-147115NB-I00, PID2023-148418NB-C41, PID2023-148418NB-C42, PID2023-148418NB-C43, PID2023-148418NB-C44, PID2024-158190NB-C22, RYC2021-033305-I, RYC2024-048719-I, CNS2023-144781, CNS2024-154769 and Plan de Ciencia, Tecnolog{\'i}a e Innovaci{\'o}n de Asturias, Spain; the Chulalongkorn Academic into Its 2nd Century Project Advancement Project, the National Science, Research and Innovation Fund program IND\_FF\_68\_369\_2300\_097, and the Program Management Unit for Human Resources \& Institutional Development, Research and Innovation, grant B39G680009 (Thailand); the Eric \& Wendy Schmidt Fund for Strategic Innovation through the CERN Next Generation Triggers project under grant agreement number SIF-2023-004; the Kavli Foundation; the Nvidia Corporation; the SuperMicro Corporation; the Welch Foundation, contract C-1845; and the Weston Havens Foundation (USA).
\end{sloppypar}
\end{acknowledgments}\section*{Data availability} Release and preservation of data used by the CMS Collaboration as the basis for publications is guided by the  \href{https://doi.org/10.7483/OPENDATA.CMS.1BNU.8V1W}{CMS data preservation, re-use and open access policy}.

\bibliography{auto_generated}
\cleardoublepage \appendix\section{The CMS Collaboration \label{app:collab}}\begin{sloppypar}\hyphenpenalty=5000\widowpenalty=500\clubpenalty=5000\cmsinstitute{Yerevan Physics Institute, Yerevan, Armenia}
{\tolerance=6000
A.~Gevorgyan\cmsorcid{0000-0003-2751-9489}, A.~Hayrapetyan, V.~Makarenko\cmsorcid{0000-0002-8406-8605}, A.~Tumasyan\cmsAuthorMark{1}\cmsorcid{0009-0000-0684-6742}
\par}
\cmsinstitute{Marietta Blau Institute for Particle Physics, Vienna, Austria}
{\tolerance=6000
W.~Adam\cmsorcid{0000-0001-9099-4341}, L.~Benato\cmsorcid{0000-0001-5135-7489}, T.~Bergauer\cmsorcid{0000-0002-5786-0293}, M.~Dragicevic\cmsorcid{0000-0003-1967-6783}, S.~Gundacker\cmsorcid{0000-0003-2087-3266}, A.K.~Guven\cmsorcid{0009-0004-5670-5138}, P.S.~Hussain\cmsorcid{0000-0002-4825-5278}, M.~Jeitler\cmsAuthorMark{2}\cmsorcid{0000-0002-5141-9560}, N.~Krammer\cmsorcid{0000-0002-0548-0985}, A.~Li\cmsorcid{0000-0002-4547-116X}, D.~Liko\cmsorcid{0000-0002-3380-473X}, M.~Matthewman, J.~Schieck\cmsAuthorMark{2}\cmsorcid{0000-0002-1058-8093}, R.~Sch\"{o}fbeck\cmsAuthorMark{2}\cmsorcid{0000-0002-2332-8784}, M.~Shooshtari\cmsorcid{0009-0004-8882-4887}, M.~Sonawane\cmsorcid{0000-0003-0510-7010}, N.~Van~Den~Bossche\cmsorcid{0000-0003-2973-4991}, W.~Waltenberger\cmsorcid{0000-0002-6215-7228}, C.E.~Wulz\cmsAuthorMark{2}\cmsorcid{0000-0001-9226-5812}
% ...
\par}
$^{1}$Also at Yerevan State University, Yerevan, Armenia\\
$^{2}$Also at Technische Universit\"{a}t Wien, Vienna, Austria\\
% ...
\end{sloppypar}
\end{document}